\documentclass[twocolumn]{aastex62}

\newcommand\aastex{AAS\TeX}

\received{January 1, 2018}
\revised{January 7, 2018}
\accepted{\today}
\submitjournal{ApJ}

\shorttitle{\aastex\ Distance, Depth and Kinematics of the Taurus Star-Forming Region}
\shortauthors{Galli et al.}

\begin{document}

\title{The Gould's Belt Distances Survey (GOBELINS). \\IV. Distance, Depth and Kinematics of the Taurus Star-Forming Region}

\correspondingauthor{Phillip Galli}
\email{phillip.galli@iag.usp.br}

\author[0000-0003-2271-9297]{Phillip A. B. Galli}
\affil{Instituto de Astronomia, Geof\'isica e Ci\^encias Atmosf\'ericas, Universidade de S\~ao Paulo, Rua do Mat\~ao, 1226, Cidade Universit\'aria, 05508-900, S\~ao Paulo - SP, Brazil
}

\author{Laurent Loinard}
\affiliation{Instituto de Radioastronom\'ia y Astrof\'isica, Universidad Nacional Aut\'onoma de M\'exico, Apartado Postal  3-72, Morelia 58089, M\'exico}
\affiliation{Instituto de Astronom\'ia, Universidad Nacional Aut\'onoma de M\'exico, Apartado Postal 70-264, 04510 Ciudad de M\'exico, M\'exico}

\author{Gisela N. Ortiz-L\'eon}
\affiliation{Max Planck Institut f{\"u}r Radioastronomie, Auf dem H{\"u}gel 69, D-53121, Bonn, Germany}

\author{Marina Kounkel}
\affiliation{Department of Physics and Astronomy, Western Washington University, 516 High St, Bellingham, WA 98225, USA}

\author[0000-0001-6010-6200]{Sergio A. Dzib}
\affiliation{Max Planck Institut f{\"u}r Radioastronomie, Auf dem H{\"u}gel 69, D-53121, Bonn, Germany}

\author{Amy J. Mioduszewski}
\affiliation{National Radio Astronomy Observatory, Domenici Science Operations Center, 1003 Lopezville Road, Socorro, NM 87801, USA}

\author{Luis F. Rodr\'iguez}
\affiliation{Instituto de Radioastronom\'ia y Astrof\'isica, Universidad Nacional Aut\'onoma de M\'exico, Apartado Postal  3-72, Morelia 58089, M\'exico}

\author{Lee Hartmann}
\affiliation{Department of Astronomy, University of Michigan, 1085 S. University st., Ann Arbor, MI
48109, USA}

\author{Ramachrisna Teixeira}
\affiliation{Instituto de Astronomia, Geof\'isica e Ci\^encias Atmosf\'ericas, Universidade de S\~ao Paulo, Rua do Mat\~ao, 1226, Cidade Universit\'aria, 05508-900, S\~ao Paulo - SP, Brazil
}

\author{Rosa M. Torres}
\affiliation{Centro Universitario de Tonal\'a, Universidad de Guadalajara, Avenida Nuevo Perif\'erico No. 555, Ejido San Jos\'e Tatepozco, C.P. 48525, Tonal\'a, Jalisco, M\'exico}

\author{Juana L. Rivera}
\affiliation{Instituto de Radioastronom\'ia y Astrof\'isica, Universidad Nacional Aut\'onoma de M\'exico, Apartado Postal  3-72, Morelia 58089, M\'exico}

\author{Andrew F. Boden}
\affiliation{Division of Physics, Math and Astronomy, California Institute of Technology, 1200 East California Boulevard, Pasadena, CA 91125, USA}

\author{Neal J. Evans II}
\affiliation{Department of Astronomy, The University of Texas at Austin, 2515 Speedway, Stop C1400, Austin, TX 78712-1205, USA}
\affiliation{Korea Astronomy and Space Science Institute, 776 Daedeokdaero, Daejeon 305-348, Korea}
\affiliation{Humanitas College, Global Campus, Kyung Hee University, Yongin-shi 17104, Korea}

\author{Cesar Brice\~no}
\affiliation{Cerro Tololo Interamerican Observatory, Casilla 603, La Serena, Chile}

\author{John J. Tobin}
\affiliation{Homer L. Dodge Department of Physics and Astronomy, University of Oklahoma, 440 W. Brooks Street, Norman, OK 73019, USA}

\author{Mark Heyer}
\affiliation{Department of Astronomy, University of Massachusetts, Amherst, MA 01003, USA}

\begin{abstract}

We present new trigonometric parallaxes and proper motions of young stellar objects in the Taurus molecular cloud complex from observations collected with the Very Long Baseline Array as part of the Gould's Belt Distances Survey (GOBELINS). We detected 26 young stellar objects and derived trigonometric parallaxes for 18 stars with an accuracy of 0.3$\%$ to a few percent. We modeled the orbits of six binaries and determined the dynamical masses of the individual components in four of these systems (V1023~Tau, T~Tau~S, V807~Tau and V1000~Tau). Our results are consistent with the first trigonometric parallaxes delivered by the Gaia satellite and reveal the existence of significant depth effects. We find that the central portion of the dark cloud Lynds~1495 is located at $d=129.5\pm 0.3$~pc while the B~216 clump in the filamentary structure connected to it is at $d=158.1\pm 1.2$~pc. The closest and remotest stars in our sample  are located at $d=126.6\pm 1.7$~pc  and $d=162.7\pm 0.8$~pc yielding a distance difference of about 36~pc. We also provide a new distance estimate for HL~Tau that was recently imaged. Finally, we compute the spatial velocity of the stars with published radial velocity and investigate the kinematic properties of the various clouds and gas structures in this region.

\end{abstract}



\section{Introduction} \label{section1}

Since the discovery of the first T~Tauri star \citep{Joy1945} the nearby Taurus-Auriga molecular cloud complex (or simply ``Taurus")  has become one of the most studied regions of low-mass star formation \citep[see e.g.][for the most recent review]{Kenyon2008}. Taurus hosts more than 300 known young stellar objects (YSO) including pre-main sequence stars and brown dwarfs which are spread over several star-forming clouds and clumps \citep{Luhman2009,Joncour2017}. 

Taurus is composed of multiple filaments \citep{Schneider1979,Hartmann2002,Schmalzl2010,Panopoulou2014} and the spatial distribution of the YSOs in the plane of the sky shows that they are clustered in small groups in or around the different star-forming clouds \citep{Gomez1993}. The question then arises whether the various groups and substructures are bound and have a common origin. Although the morphology of the molecular clouds has been well characterized in previous studies based e.g. on CO surveys and extinction maps \citep{Ungerechts1987,Cambresy1999,Dame2001,Dobashi2005}, little progress has been made so far  to constrain the three-dimensional structure of the complex. Distances to individual stars are urgently required to accurately determine the most fundamental properties of YSOs (luminosity, mass and age), and they could also provide important clues to unravel the history of star formation in this region. 

The distance to Taurus is commonly accepted to be 140~pc  \citep{Elias1978} based on several estimates using a wide variety of techniques. The first results obtained by \citet{Greenstein1937} and \citet{McCuskey1939} using star counts returned a distance of 145~pc and 142~pc, respectively. \citet{Racine1968} determined a shorter distance of $135\pm10$~pc from the photometry of bright stars associated with reflection nebulae. On the other hand, \citet{Gottlieb1969} obtained a somewhat larger distance of 150~pc  based on the reddening turn-on method. Later, \citet{Straizys1980} investigated the area around the dark clouds L~1538, L~1528, L~1521 and L~1495 from the Lynds catalog \citep{Lynds1962} using a similar approach, and concluded that they extend from 140 to 175~pc. In a companion study, \citet{Meistas1981} found that the front-edge of the southern clouds in the complex (L~1551, L~1546 and L~1543) seems to be located at about 140~pc. More recently, \citet{Kenyon1994} derived the canonical distance of $140\pm10$~pc for the northern portion of the Taurus clouds based on the method of spectroscopic parallaxes which is commonly used in the literature. The latter result confirmed previous distance estimates to the Taurus region and supported the idea of a common distance to the various clouds of the complex. 

It is important to note that the measurements listed here refer to the mean distance to Taurus based on indirect methods while the individual distances to most YSOs still remain poorly constrained. Trigonometric parallaxes in the \textit{Hipparcos} catalog \citep{ESA1997} exist only for 17 stars with a median error of about $27\%$. \citet{Bertout1999} divided them into 3 groups (L1495 region, Auriga region and south Taurus) and calculated the mean distance. Using only single stars they concluded that the 3 subgroups are located, respectively, at $125^{+21}_{-16}$, $140^{+16}_{-13}$ and $168^{+42}_{-28}$~pc. These values are statistically compatible between themselves, but they suggest important distance differences among the various clouds in the complex. The mean parallax of all (single) stars together yields a distance of $139^{+10}_{-9}$~pc which confirms previous measurements but at the same time blurs the existence of possible depth effects in this region. 

In a different study, \citet{Bertout2006} applied a variant of the convergent point method and derived the kinematic parallaxes for 67 cluster members using proper motions and radial velocities. The derived parallaxes have a typical error of $20\%$. They investigated the distances of the YSO subclasses in their sample, and concluded that the classical T~Tauri stars are at distances between 126 and 173~pc while the weak emission-line T~Tauri stars can be found on both sides of the molecular clouds between 106 and 259~pc. These results are indicative of real distance differences among Taurus stars which will be tested when more trigonometric parallaxes become available. 

Very Long Baseline Interferometry (VLBI) has been used in recent years to deliver trigonometric parallaxes of nearby stars with an accuracy often better than $1\%$ \citep[see e.g.][]{Melis2014,Forbrich2016}. The first VLBI trigonometric parallax in Taurus was reported by \citet{Lestrade1999} who targeted the weak-line T~Tauri star V773~Tau which was found to be at $148\pm5$~pc. Later, \citet{Torres2012} presented an improved solution for its distance which takes into account the orbital motion of the binary system placing it at $132.8\pm2.3$~pc. The second trigonometric parallax obtained from VLBI radio observations in Taurus was obtained by \citet{Loinard2007}. They measured the distance of  $147.6\pm0.6$~pc for T~Tau~Sb in the well-known T~Tau triple system in the southern clouds of the complex. Subsequently, \citet{Torres2007} measured the trigonometric parallaxes of Hubble~4 and HDE~283572 in the central portion of the complex, and derived the distances of $132.8\pm0.5$~pc and $128.5\pm0.6$~pc, respectively. In a companion study, \citet{Torres2009} measured a somewhat larger distance of $161.2\pm0.9$~pc for HP~Tau~G2 making it the most distant star with known trigonometric parallax in Taurus. In summary, V773~Tau, Hubble~4 and HDE~283572 which are associated with the most prominent dark cloud L~1495 are found to be at the same distance while T~Tau~Sb and HP~Tau~G2 are located at larger distances. There is growing evidence of significant dispersion along the line of sight, but the small number of cluster members with measured trigonometric parallaxes is still insufficient to construct a precise three-dimensional map of this region. 

This paper is one in a series dedicated to measuring stellar distances based on VLBI observations as part of the Gould's Belt Distances Survey \citep[GOBELINS,][]{Loinard2011}. Previous papers of this project investigated the Ophiuchus \citep{OrtizLeon2017a}, Orion \citep{Kounkel2017} and Serpens \citep{OrtizLeon2017b} star-forming regions. Here, we present new trigonometric parallaxes and proper motions for YSOs in Taurus. It is organized as follows. In Section~\ref{section2} we describe our sample and observations. The methodology used to fit the astrometry for both single stars and binaries is explained in Section~\ref{section3}. In Section~\ref{section4} we present our results and comment on individual targets. In Section~\ref{section5} we compare our results derived in this work with the trigonometric parallaxes delivered by the first data release of the Gaia space mission \citep[Gaia-DR1,][]{GaiaDR1} for the targets in common, and complement our sample with the Gaia stars not included in our observing program to construct the most complete, precise and accurate picture of three-dimensional structure of the Taurus region to date. In this section, we also discuss the kinematic properties of the stars and molecular clouds in this region (see Table~\ref{tab10}). Finally, we summarize our results and conclusions in Section~\ref{section6}. 

\section{Sample and Observations} \label{section2}

In a recent study, \citet{Dzib2015} reported on multi-epoch radio observations of the Taurus complex using the Karl G. Jansky Very Large Array (VLA). They detected 59 sources related to YSOs, 18 field stars and another 46 unidentified sources whose radio properties are consistent with YSOs. However, only 56\% of the young stars identified in their study exhibit properties compatible with non-thermal radio emission that can be detected with VLBI observations. These sources constitute the starting point of our sample for the GOBELINS project in the Taurus region. 

\begin{figure*}[ht!]
\plotone{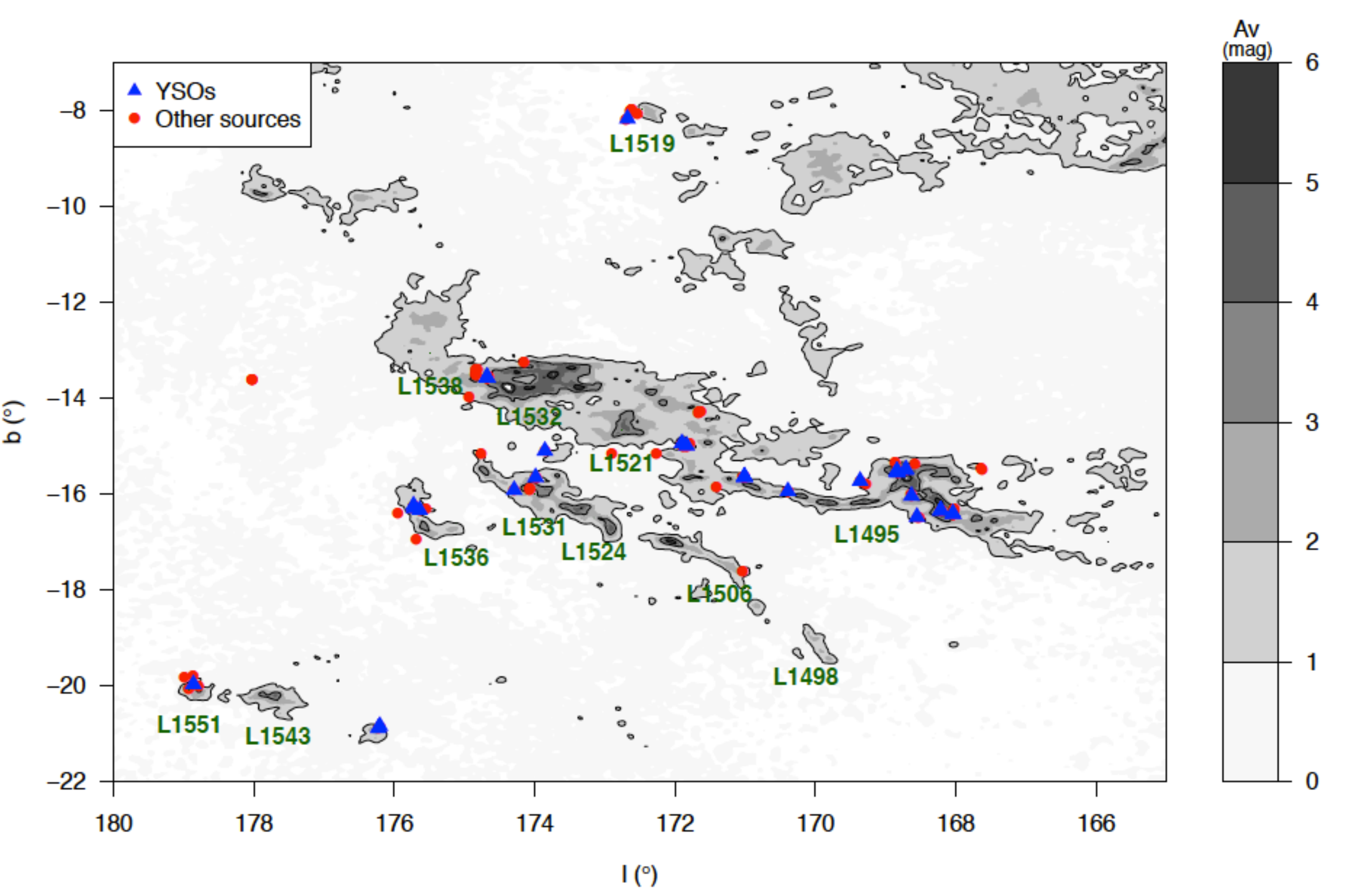}
\caption{Location of our targets in the Taurus star-forming complex overlaid on the extinction map from \citet{Dobashi2005} in Galactic coordinates. The most prominent Lynds dark clouds \citep{Lynds1962} in this region are identified in this diagram. \bigskip
\label{fig1}}
\end{figure*}

The observations presented in this paper were obtained with the Very Long Baseline Array (VLBA) near the equinoxes of every year between August 2012 and October 2017. The data were recorded in dual polarization mode with a bandwidth of 256~MHz centered at $\nu=5.0$ or 8.4~GHz (C- and X-bands, respectively). We observed with the X-band during the first 3~years of our observing program, and then switched to the C-band which reduced significantly the noise in our observations. The VLBA was pointed at the position of our targets that have been accommodated in 52 different fields. Table~\ref{tab1} lists the observed epochs, bands, pointing positions and calibrators for each field. In some of these fields two or more targets are observed simultaneously.  We included additional phase centers within the primary beam to observe other sources reported by \citet{Dzib2015} independently of their nature (YSO candidates, field stars or extragalactic sources) as part of our observing strategy \citep[see][for more details]{OrtizLeon2017a}. Doing so, we observed a total of 86 sources (or stellar systems) during our observing campaign of which 45 are known YSOs in the literature. As shown in Fig.~\ref{fig1} our targets are spread over the various molecular clouds in Taurus. 

Our observations produced 164 different projects under the code BL175 (see Table~\ref{tab1}). Each observing session consisted of cycles alternating between the target(s) and the main phase calibrator. The secondary calibrators were observed every $\sim$50~minutes. In addition, we also observed blocks of geodetic calibrators over a wide range of elevations at the beginning and end of each session. The typical integration time in each cycle was $\sim$2~minutes for science targets and $\sim$1~minute for calibrators. The total integration time for projects that observed with the X-band and C-band were, respectively, 1.1~hours and 1.5~hours.  The calibrators used in our observations are extragalactic sources and the positions of all sources in each project are referenced  to the corresponding primary calibrator. The typical angular separation between the primary calibrator and the YSOs in our sample ranges from 0.9$^{\circ}$ to 3.7$^{\circ}$.


\startlongtable
\begin{deluxetable*}{cccccl}
\tablecaption{Observed fields in Taurus with the VLBA.  \label{tab1}}
\tablehead{
\colhead{Project Code} & 
\colhead{Date} & 
\colhead{Band}& 
\colhead{$\alpha$} &
\colhead{$\delta$} &
\colhead{Calibrators}\\
\colhead{} & 
\colhead{} & 
\colhead{} & 
\colhead{(h:m:s)} & 
\colhead{($^{\circ}$ $^\prime$ $^\prime$$^\prime$)} & 
\colhead{}
}
\startdata
BL175C1	&	20-aug-2012	&	X	&	04	22	02.20	&	19	33	27.0	&	J0412+1856,	J0428+1732,	J0426+2327,	J0412+2305 	\\
BL175C2	&	30-aug-2012	&	X	&	04	31	25.13	&	18	16	16.6	&	J0431+1731,	J0428+1732,	J0438+2153,	J0440+1437	\\
BL175C3	&	31-aug-2012	&	X	&	04	31	40.09	&	18	13	56.7	&	J0431+1731,	J0428+1732,	J0438+2153,	J0440+1437	\\
BL175C4	&	10-sep-2012	&	X	&	04	32	14.58	&	18	20	14.6	&	J0431+1731,	J0428+1732,	J0438+2153,	J0440+1437	\\
BL175C5	&	30-sep-2012	&	X	&	04	31	34.15	&	18	08	04.6	&	J0431+1731,	J0428+1732,	J0438+2153,	J0440+1437	\\
	&		&		&	04	32	29.47	&	18	14	00.3	&					\\
BL175C6	&	30-nov-2012	&	X	&	04	33	26.35	&	22	28	32.0	&	J0438+2153,	J0450+2249,	J0426+2350,	J0431+1731	\\
BL175C7	&	02-dec-2012	&	X	&	04	35	13.27	&	22	59	20.0	&	J0438+2153,	J0426+2327,	J0435+2532,	 J0450+2249	\\
\enddata
\tablecomments{We provide for each field the NRAO project code, epoch of observation, observed band, coordinates of the field center and the calibrators (the main calibrator is the first source in the list). }
\end{deluxetable*}

The VLBA observations were correlated with the DiFX software correlator \citep{Deller2011}. Then, the data were edited and calibrated using the Astronomical Image Processing System \citep[AIPS,][]{Greisen2003} following the standard prescription for VLBA data as described in \citet{OrtizLeon2017a}. We applied the same calibration to all sources in the field when multiple targets were observed in the same session. The calibrated visibilities were imaged with a pixel size of 50-100~$\mu as$ and resulted in a typical angular resolution of 3~mas~$\times$~1~mas at $\nu=5.0$~GHz and  2~mas~$\times$~0.8~mas at $\nu=8.4$~GHz. The mean noise level in the calibrated images were 26~$\mu$Jy/beam and 42~$\mu$Jy/beam, respectively, at the C- and X-bands. Finally, the source position (and the corresponding errors) were obtained from a two-dimensional Gaussian fitting using the AIPS task JMFIT. 

\begin{figure}[ht!]
\plotone{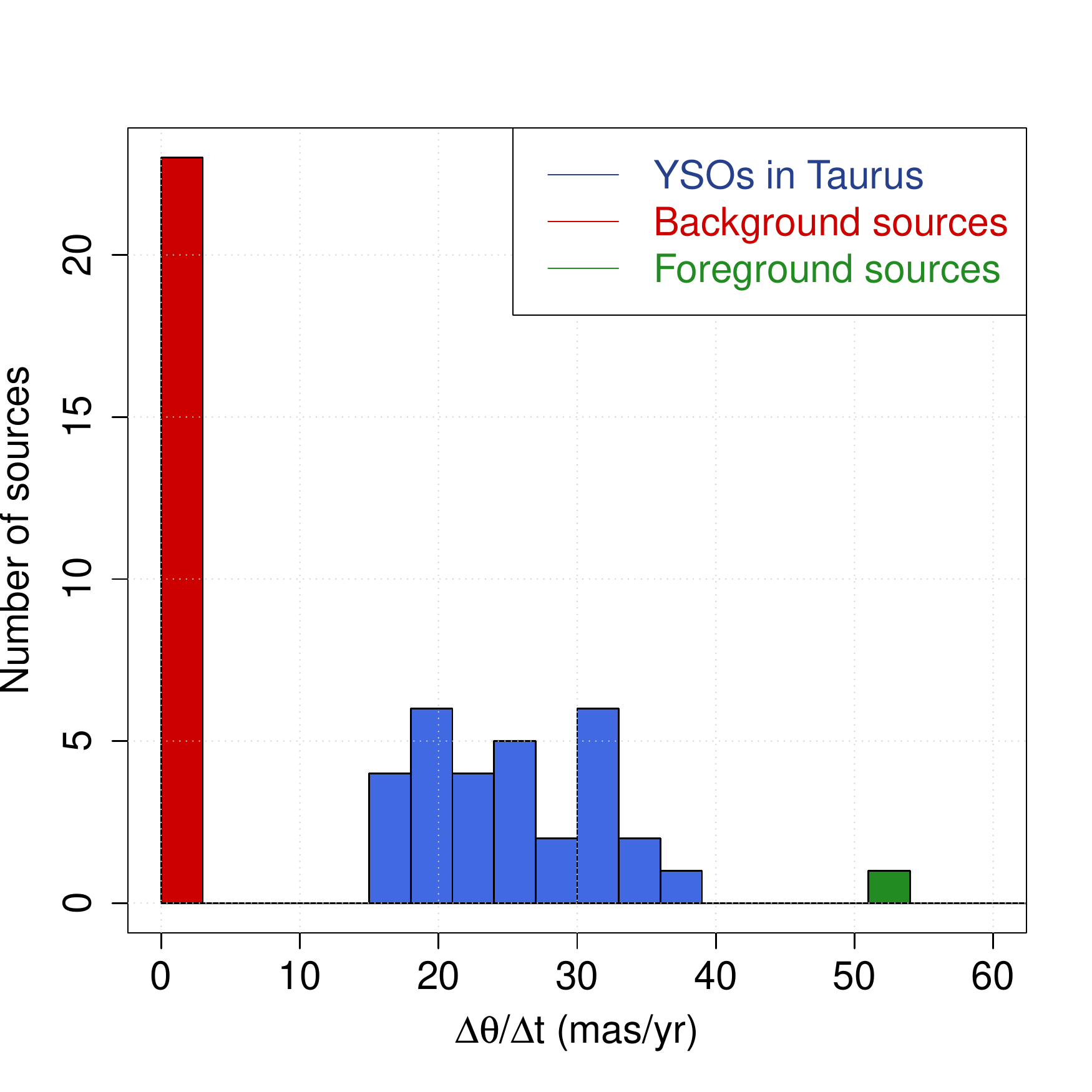}
\caption{Histogram of the position change rate for all sources (including binaries) observed in our campaign with a minimum of two detections (see Table~\ref{tab2}). 
\label{fig2}}
\end{figure}

As mentioned before, we observed a total of 86 targets in Taurus for the GOBELINS project, but only 52 sources could be detected in our observations. Table~\ref{tab2} lists the sources that have been observed in our campaign with the VLBA. We provide the minimum and maximum flux density measured at both frequencies, the brightness temperature $T_{B}$, the number of detections and observations for each source. In some cases we provide an upper flux density limit of $3\sigma$ (based on the noise level of the image) when the source was observed but could not be detected. In this context, it is important to mention that some of the YSOs targeted in this study are highly variable. For binaries and multiple systems that could be resolved in our observations we present the results for each component separately. We note that most sources in Table~\ref{tab2} exhibit a brightness temperature of $T_{B}>10^{6}$~K which is consistent with non-thermal radio emission. Thirty-four sources in our initial target list were not detected, and we interpret the non detections with the VLBA as evidence that their radio emission is thermal. We consider a single detection to be valid when the flux density of the source is above a $5\sigma$ threshold where $\sigma$ is the corresponding noise level of the image. Our effective sample of detected sources contains 26 stars that have been confirmed as YSOs in previous studies, and another 3 sources that require further monitoring to investigate their membership in the Taurus region. The remaining 23 sources that have been detected in our observations are likely to be background contaminants that are not related to the Taurus molecular clouds. As shown in Fig.~\ref{fig2} their proper motion, which are estimated at this stage from the position change rate of the source, are consistent with zero (within the astrometric errors) while the typical proper motion of YSOs in Taurus is about 22~mas/yr \citep[see e.g.][]{Bertout2006}. 

It is important to mention that some sources in our sample had been observed with the VLBA in the past. Thus, we searched the National Radio Astronomy Observatory's (NRAO) data archive for additional information on our targets. We collected a total of 65 projects at different epochs that are listed in Table~\ref{tab3}. These observations were performed between September 2003 and March 2009 which allows us to extend the time base of our analysis to more than 10~years of observations for some targets. This is particularly useful to investigate the orbital motion of binaries and multiple systems in our sample as discussed in the forthcoming sections of this paper. Although the source positions obtained from these observations have already been published in previous studies, we recalibrated these data using the same procedure that was adopted for the GOBELINS project to better combine the different data sets. The typical noise level and angular resolution that we obtain in the re-calibrated images from archival data are, respectively,  100~$\mu$Jy/beam and 2~mas~$\times$~0.8~mas.

\section{Astrometry Fitting} \label{section3}

\subsection{Single Stars}\label{section3.1}

The displacement of a single star in the plane of the sky is the combination of its proper motion $(\mu_{\alpha},\mu_{\delta})$ and the trigonometric parallax $(\pi)$. In this case, the stellar coordinates as a function of time $(t)$ are given by 
\begin{equation}
\alpha(t) = \alpha_{0} + \mu_{\alpha}t + \pi f_{\alpha}(t)\, ,
\end{equation}
\begin{equation}
\delta(t) = \delta_{0} + \mu_{\delta}t + \pi f_{\delta}(t)\, ,
\end{equation}
where $(\alpha_{0},\delta_{0})$ are the coordinates of the star at a given epoch $(t_{0})$. The projections of the parallactic ellipse $(f_{\alpha},f_{\delta})$ are given by \citep[see e.g.][]{Seidelmann1992}
\begin{equation}
f_{\alpha}=(X\sin\alpha - Y\cos\alpha)/\cos\delta\, ,
\end{equation}
\begin{equation}
f_{\delta}=X\cos\alpha\sin\delta + Y\sin\alpha\sin\delta - Z\cos\delta\, , 
\end{equation}
where $(X,Y,Z)$ are the barycentric coordinates of the Earth (in units of A.U.) computed from the planetary ephemerides DE405 using the \texttt{novas} package in Python.

\subsection{Binaries}\label{section3.2}

Binaries and multiple systems are very common in the Taurus star-forming region \citep{Duchene1999} and our sample contains many such systems that require a dedicated analysis. Since both stars in a binary system move around their common centre of gravity, their orbital motion projected onto the plane of the sky has to be taken into account to accurately determine the parallaxes and proper motions of the individual components. 

At this stage it is important to distinguish between (i) binaries with orbital periods much longer than the monitoring time of our observing campaign, and (ii) binaries with short or intermediate periods where our observations cover a significant fraction of the orbital period. In the first case, it is possible to assume an uniform acceleration \citep[see e.g.][]{Loinard2007} leading to
\begin{equation}
\alpha(t) = \alpha_{0} + \mu_{\alpha}t + \frac{1}{2}a_{\alpha}t^{2} + \pi f_{\alpha}(t)\, ,
\end{equation}
\begin{equation}
\delta(t) = \delta_{0} + \mu_{\delta}t + \frac{1}{2}a_{\delta}t^{2} + \pi f_{\delta}(t)\, ,
\end{equation}
where $a_{\alpha}$ and $a_{\delta}$ are the acceleration terms. In the second case, the effects of the binary motion and the existence of a non-uniform acceleration need to be taken into account. It is therefore necessary to fit for the full orbital motion and the astrometric parameters simultaneously. The orbital elements to be considered in this case are the semi-major axis $a$, the orbital period $P$, the eccentricity $e$, the epoch of periastron passage $T_{p}$, the argument of the ascending node $\Omega$, the longitude of the periastron $\omega$ and the inclination $i$ of the orbital plane. Thus, the equations for the primary component with semi-major axis $a_{1}$ can be written in the form  
\begin{equation}
\alpha(t) = \alpha_{0} + \mu_{\alpha}t + \pi f_{\alpha}(t) + a_{1}Q_{\alpha}(t)\, ,
\end{equation}
\begin{equation}
\delta(t) = \delta_{0} + \mu_{\delta}t + \pi f_{\delta}(t) + a_{1}Q_{\delta}(t)\, ,
\end{equation}
where $Q_{\alpha}$ and $Q_{\delta}$ are the orbital factors. They are given by \citep[see e.g.][]{VanDeKamp1967}
\begin{equation}
Q_{\alpha}=B'x(t)+G'y(t)\, ,
\end{equation}
\begin{equation}
Q_{\delta}=A'x(t)+F'y(t)\, ,
\end{equation}
where the orientation factors $B',A',G',F'$ are related to the Thiele-Innes constants as follows
\begin{equation}
B'= -\cos\omega\sin\Omega - \sin\omega\cos\Omega\cos i\, ,
\end{equation}
\begin{equation}
A'= -\cos\omega\cos\Omega + \sin\omega\sin\Omega\cos i\, ,
\end{equation}
\begin{equation}
G'= +\sin\omega\sin\Omega - \cos\omega\cos\Omega\cos i\, ,
\end{equation}
\begin{equation}
F'= +\sin\omega\cos\Omega + \cos\omega\sin\Omega\cos i\, .
\end{equation} 
The elliptical rectangular coordinates $x(t)$ and $y(t)$ are given by
\begin{equation}
x(t)=\cos E(t)-e\, ,
\end{equation}
\begin{equation}
y(t)=\sin E(t)\sqrt{1-e^{2}}\, ,
\end{equation}
where $E(t)$ is the eccentric anomaly given by Kepler's transcendental equation. Similar equations hold for the secondary component. In this case, we replace $a_{1}$ by the semi-major axis $a_{2}$ of the secondary which is scaled by the mass ratio $q$ of the two components. 

To simplify the forthcoming discussion we denote the methodology described above to solve simultaneously for the astrometric parameters and orbital elements of the binary system using absolute positions by the name ``full model''. Some binaries in our sample also exhibit relative astrometry of the two components from optical and near-infrared observations (NIR) published in previous works which have been incorporated to our study. In this case, our analysis is limited to the right-hand term of Eqs.~(7) and (8) to derive the orbital elements of the binary system. The resulting semi-major axis $a$ refers to the relative orbit of the two components. We denote this solution using relative positions by the name ``relative model". Finally, we combine the absolute positions measured in this work and relative positions from the literature to perform a ``joint fit" solution.  

\subsection{Solving the system of equations}\label{section3.3}

The source positions obtained from the JMFIT task in AIPS were used to derive the astrometric and orbital parameters of the stars in our sample. Table~\ref{tab4} lists the measured positions for each source in our sample. However, it is important to mention that the positional uncertainties provided by JMFIT, which roughly represent the expected astrometric precision delivered by the VLBA, do not include various systematic errors that may significantly affect the accuracy of the computed astrometric parameters \citep[see e.g.][]{Pradel2006}. To overcome this problem, additional errors were added quadratically to the positional errors given by JMFIT to adjust the reduced $\chi^{2}$ value in the astrometric fit to unity \citep[see also][]{Menten2007,OrtizLeon2017a}. These additional errors range in most cases from 0.05 to 0.30~mas for both single stars and binaries  (the few exceptions are discussed in Sect.~\ref{section4}).

To solve for the astrometric and orbital elements of the star (or stellar system) we developed our own routine in the Python programming language based on the \texttt{emcee} package \citep{Foreman2012} which implements the Markov chain Monte Carlo (MCMC) method proposed by \citet{Goodman2010}. The algorithm was adapted to our purposes and applied to the general problem of computing the astrometric and orbital parameters from the astrometry fitting. Briefly, the method that we use in this work explores the parameter space using a number of walkers and iteration steps to solve the system of equations presented in Sects.~\ref{section3.1} and \ref{section3.2} via Bayesian inference. The walkers move around the $n$-dimensional parameter space and take tentative steps towards the lowest value of $\chi^{2}$. This produces a distribution of the individual solutions given by the ensemble of walkers. We run the MCMC method typically with 200 walkers to sample the distribution of each parameter with a significant number of individual solutions. Then, we take the mean and standard deviation of the distribution of each parameter as our final result.  

\begin{figure}[ht!]
\plotone{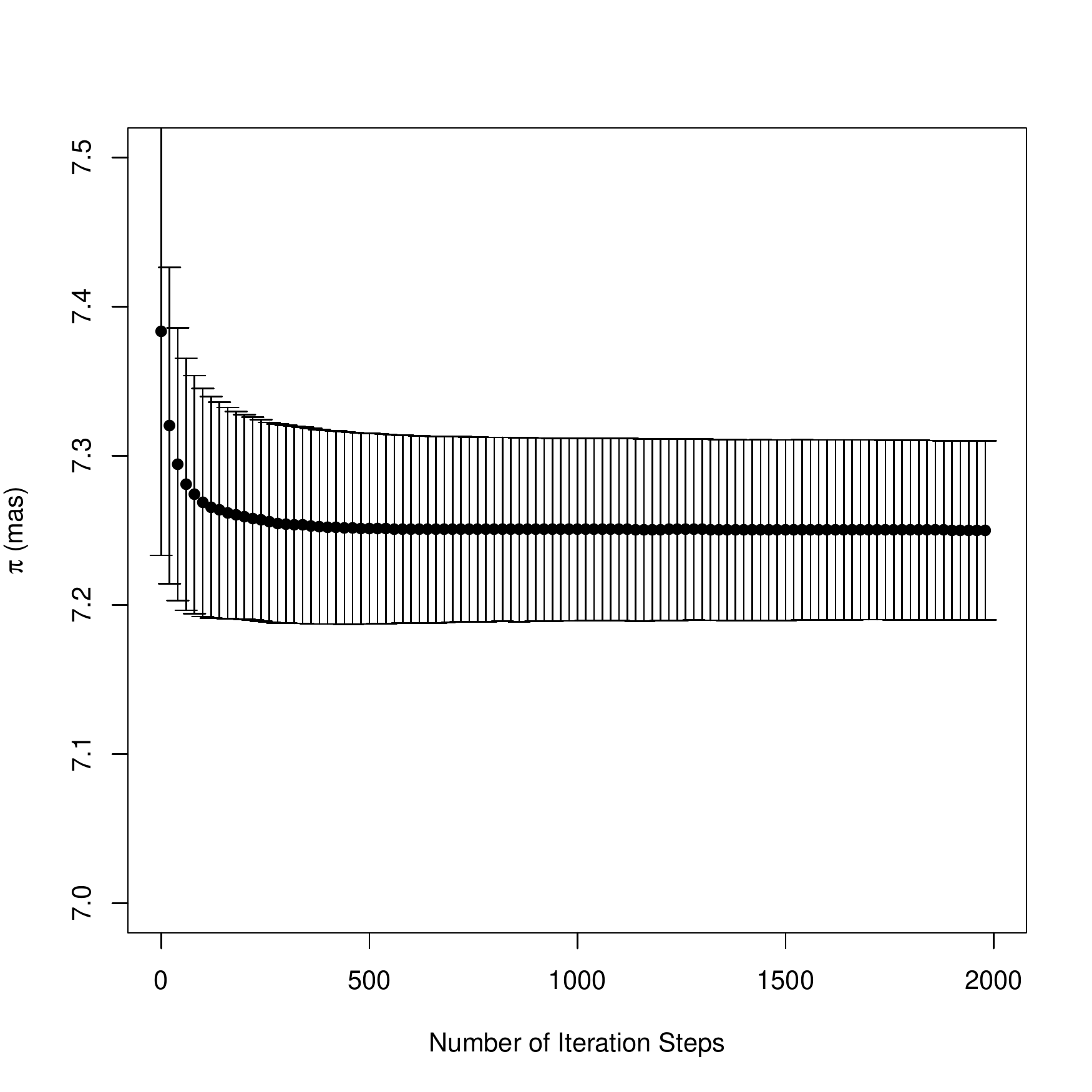}
\caption{Convergence of the trigonometric parallax obtained for LFAM~8 based on the MCMC method implemented in this work. Each point is the average of the trigonometric parallax obtained by 200 walkers for each iteration step, and the error bars indicate the scatter of the solution given by the ensemble of walkers.  
\label{fig3}}
\end{figure}

Our methodology based on the MCMC method was first applied to the sources previously investigated by our team in the Ophiuchus region \citep{OrtizLeon2017a} for calibration purposes. We varied the number of iterations for each walker from 1 to 2000 steps, and verified that convergence of the Markov-chains of the ensemble of walkers is attained after $\sim 100$ iterations steps where the mean (and median) of the computed parameters are clearly bounded by the variance of the sample. Figure~\ref{fig3} illustrates the convergence of our results for the Ophiuchus source LFAM8 as an example. We find a trigonometric parallax of $\pi=7.242\pm 0.060$~mas based on the MCMC methodology described in this section which is in good agreement with the result of $\pi=7.246\pm 0.088$~mas reported by \citet{OrtizLeon2017a}. Thus, the solutions based on the MCMC method presented in this paper are calculated using 500 iteration steps as a conservative threshold to ensure convergence of the Markov chains for both single stars and binaries in our sample.

\begin{longrotatetable}
\startlongtable
\begin{deluxetable}{llccccccc}
\tablecaption{VLBA detections and non-detections in Taurus. \label{tab2}}
\tablehead{
\colhead{GBS-VLA\tablenotemark{a}} & 
\colhead{Other}&
\colhead{Minimum Flux}&
\colhead{Maximum Flux}&
\colhead{Minimum Flux}&
\colhead{Maximum Flux}&
\colhead{$\log T_{B}$}&
\colhead{Type\tablenotemark{b}}&
\colhead{Number}\\
\colhead{Identifier} & 
\colhead{Identifier} &
\colhead{at 5.0~GHz (mJy)}&
\colhead{at 5.0~GHz (mJy)}&
\colhead{at 8.4~GHz (mJy)}&
\colhead{at 8.4~GHz (mJy)}&
\colhead{}&
\colhead{}&
\colhead{Det./Obs.}
}
\colnumbers
\startdata
J041327.23+281624.4 & V1096~Tau~A & $ 0.71 \pm 0.05 $&$ 9.2 \pm 0.07 $& $ 0.38 \pm 0.08 $&$ 0.55 \pm 0.08 $& 8.03 & YSO & 4/8 \\
J041327.23+281624.4 & V1096~Tau~B & $ 0.20 \pm 0.04 $&$ 0.36 \pm 0.05 $& \multicolumn{2}{c}{\nodata}&         6.75 & YSO & 3/8 \\
J041412.93+281211.9 & V773~Tau~Aa & $ 2.77 \pm 0.07 $&$ 11.89 \pm 0.07 $& $ 0.36 \pm 0.08 $&$ 22.23 \pm 0.20 $& 8.66 & YSO & 7/7 \\
J041412.93+281211.9 & V773~Tau~Ab & \multicolumn{2}{c}{$ 1.56 \pm 0.07 $}&     $ 0.54 \pm 0.11 $&$ 1.57 \pm 0.10 $& 7.45 & YSO & 3/7 \\
J041448.00+275234.4 & V1098~Tau & $ 0.22 \pm 0.06 $&$ 0.29 \pm 0.06 $& \multicolumn{2}{c}{$ 1.41 \pm 0.1 $}&     7.48 & YSO & 3/8 \\
J041628.11+280735.4 & V1068~Tau & \multicolumn{2}{c}{$ 11.19 \pm 0.09 $}&     \multicolumn{2}{c}{$ 0.40 \pm 0.13 $}&     8.34 & YSO & 2/7 \\
J041829.10+282618.8 & 2MASS~J04182909+2826191 & $ 0.54 \pm 0.05 $&$ 1.79 \pm 0.04 $& \multicolumn{2}{c}{$ 1.31 \pm 0.1 $}&     7.47 & YSO & 3/9 \\
J041831.12+282715.9 & HD~283518, V410~Tau & $ 1.21 \pm 0.04 $&$ 3.59 \pm 0.06 $& $ 0.72 \pm 0.09 $&$ 13.62 \pm 0.22 $& 8.44 & YSO & 9/9 \\
\enddata
\tablenotetext{a}{ The GBS-VLA (Gould's Belt Very Large Array Survey) identifier is defined by \citet{Dzib2015}.}
\tablenotetext{b}{ Type of source: ``YSO"= Young Stellar Object, ``Star?''= field star and ``B"= background source (Galactic or extragalactic).}

\end{deluxetable}
\end{longrotatetable}

\startlongtable
\begin{deluxetable*}{lccl}
\tablecaption{VLBA data in Taurus from the NRAO archive.   \label{tab3}}
\tablehead{
\colhead{Project Code} & 
\colhead{Date} & 
\colhead{Source}& 
\colhead{Calibrators \tablenotemark{a}}
}
\startdata
BM198A	&	11-mar-2004	&	V773~Tau A	&	J0403+2600, J0408+3032	\\
BM198B	&	12-mar-2004	&	V773~Tau A	&	J0403+2600, J0408+3032	\\
BM198C	&	13-mar-2004	&	V773~Tau A	&	J0403+2600, J0408+3032	\\
BM198D	&	14-mar-2004	&	V773~Tau A	&	J0403+2600, J0408+3032	\\
BM198E	&	15-mar-2004	&	V773~Tau A	&	J0403+2600, J0408+3032	\\
BM198F	&	16-mar-2004	&	V773~Tau A	&	J0403+2600, J0408+3032	\\
BM198G	&	17-mar-2004	&	V773~Tau A	&	J0403+2600, J0408+3032	\\
BL128AA	&	08-sep-2005	&	V773~Tau A	&	J0408+3032, J0403+2600, J0429+2724, J0356+2903	\\
\enddata

\tablenotetext{a}{ The main calibrator is the first source in the list.}
\end{deluxetable*}


\startlongtable
\begin{deluxetable*}{lccccccccc}
\tablecaption{VLBA measured positions for the sources in our sample.  \label{tab4}}
\tablehead{
\colhead{Project Code} & 
\colhead{Julian Day} & 
\multicolumn{3}{c}{$\alpha$}&
\multicolumn{3}{c}{$\delta$}&
\colhead{$\sigma_{\alpha}$}&
\colhead{$\sigma_{\delta}$}\\
\colhead{}&
\colhead{}&
\multicolumn{3}{c}{(h:m:s)}&
\multicolumn{3}{c}{(d:m:s)}&
\colhead{(sec)}&
\colhead{(arcsec)}
}
\startdata
\hline																			
\multicolumn{10}{c}{V1096~Tau}\\																	
\hline																		\multicolumn{10}{l}{First source:}\\														
BL175E8	&	2457123.49009	&	04	&	13	&	27.23089125	&	28	&	16	&	24.2804961	&	0.00000214	&	0.0000826	\\
BL175HL	&	2457304.99447	&	04	&	13	&	27.23164300	&	28	&	16	&	24.2750503	&	0.00000318	&	0.0001228	\\
BL175IE	&	2457508.42722	&	04	&	13	&	27.23100894	&	28	&	16	&	24.2641215	&	0.00000214	&	0.0000568	\\
BL175IU	&	2457635.08049	&	04	&	13	&	27.23210035	&	28	&	16	&	24.2605681	&	0.00000030	&	0.0000086	\\
\multicolumn{10}{l}{Second source:}\\														
BL175IE	&	2457508.42722	&	04	&	13	&	27.23206682	&	28	&	16	&	24.2607435	&	0.00000470	&	0.0001528	\\
\enddata
\end{deluxetable*}

\section{Results} \label{section4}

We present trigonometric parallaxes for 18 stars in our sample with a minimum of 3 detections during our observing campaign. The trigonometric parallaxes and proper motions derived in this study are listed in Table~\ref{tab5}. In addition, we derived the orbital elements of six binary and multiple systems in our sample which are given in Table~\ref{tab6}.  The best fit solution of our results are collectively illustrated in Figures~\ref{fig4} and \ref{fig5}. In Table~\ref{tab7} we compare the astrometry derived from the full model and joint fit for the binaries with measured relative positions in the literature. The relative astrometry model for these sources is illustrated in Figure~\ref{relFIT}. In the following we comment on the properties of the individual stars (and stellar systems) that have been investigated. We present our discussion about individual sources  in the order that they appear in Table~\ref{tab2}. 

\begin{figure*}[]
\gridline{
\fig{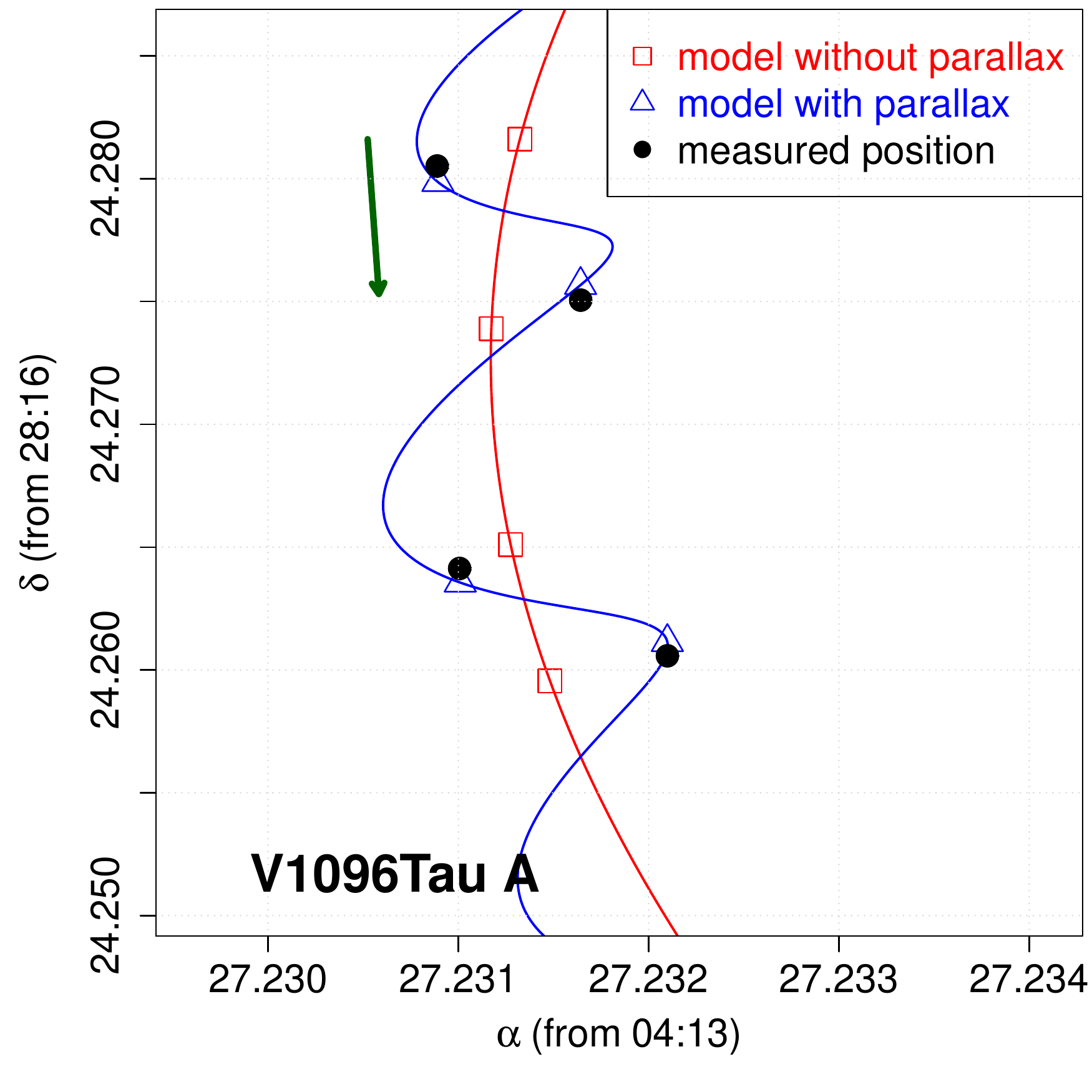}{0.33\textwidth}{}
\fig{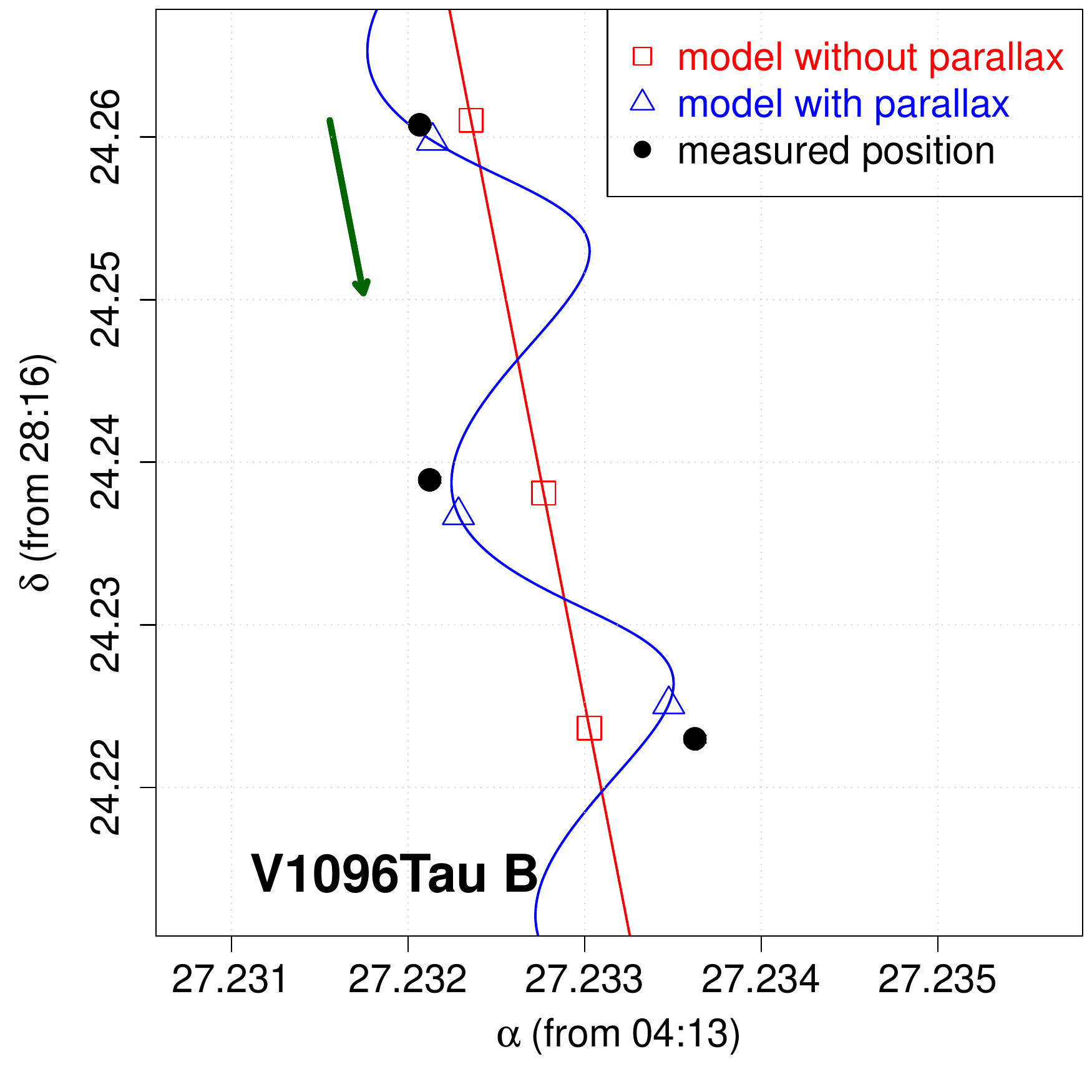}{0.33\textwidth}{}
\fig{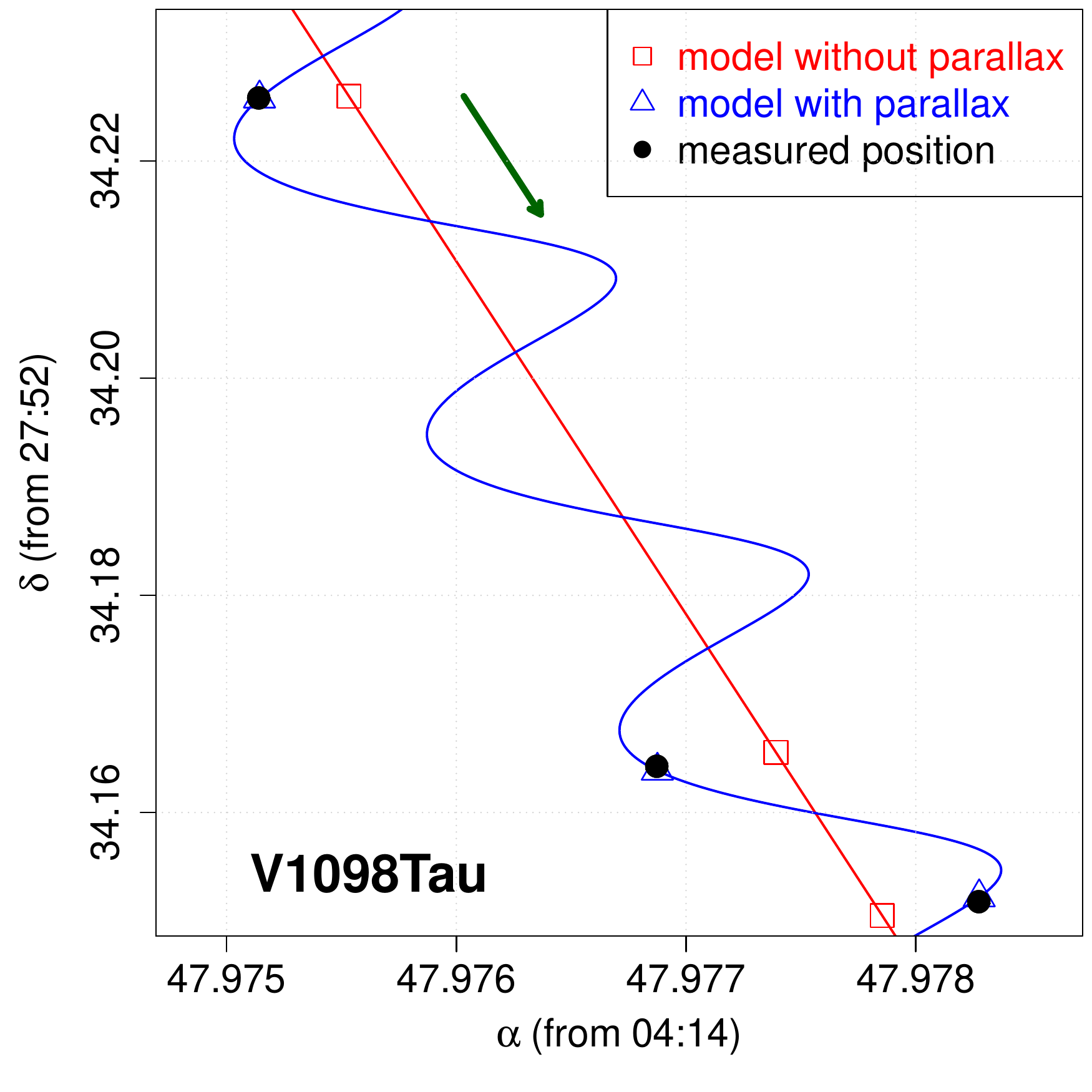}{0.33\textwidth}{}
}
\gridline{
\fig{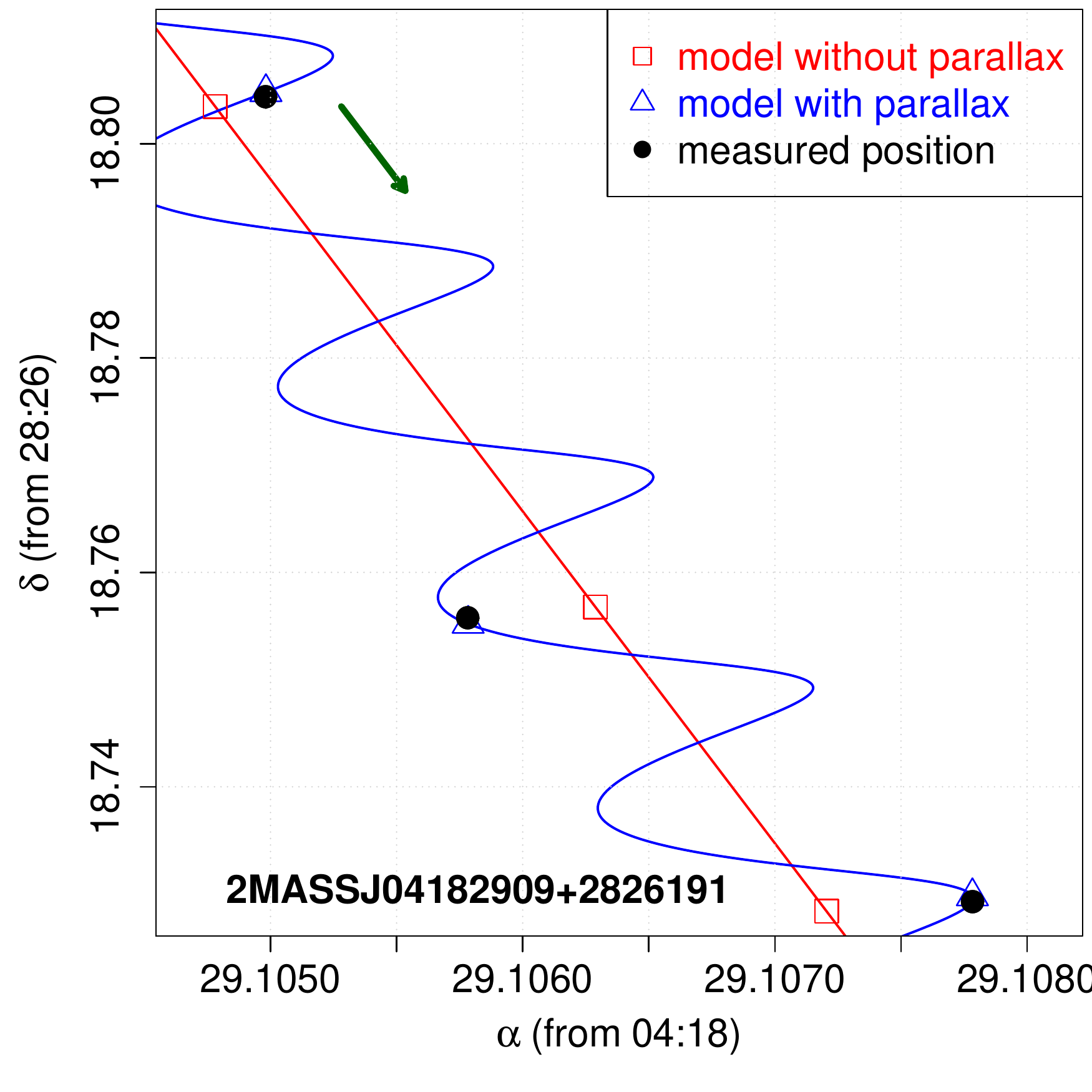}{0.33\textwidth}{}
\fig{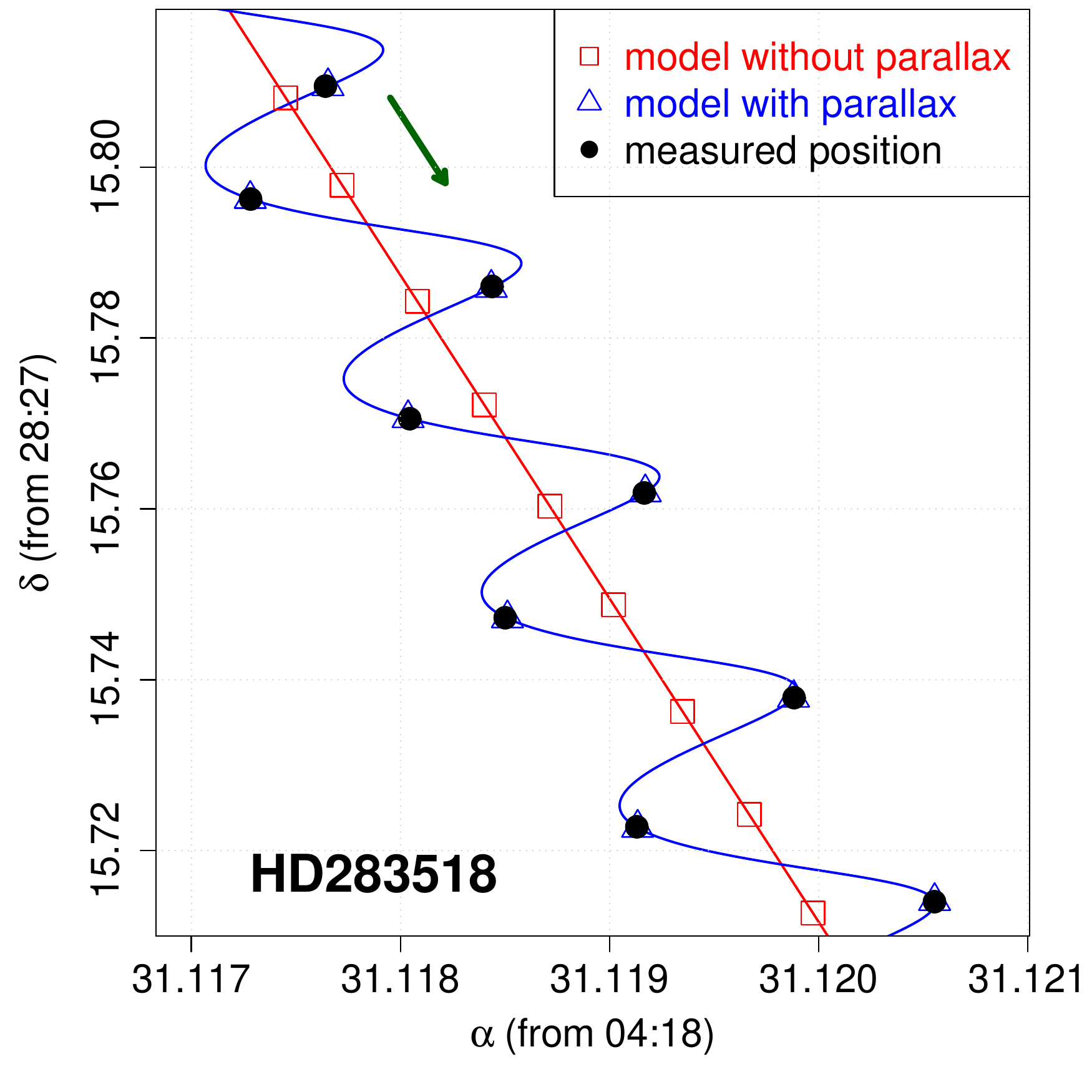}{0.33\textwidth}{}
\fig{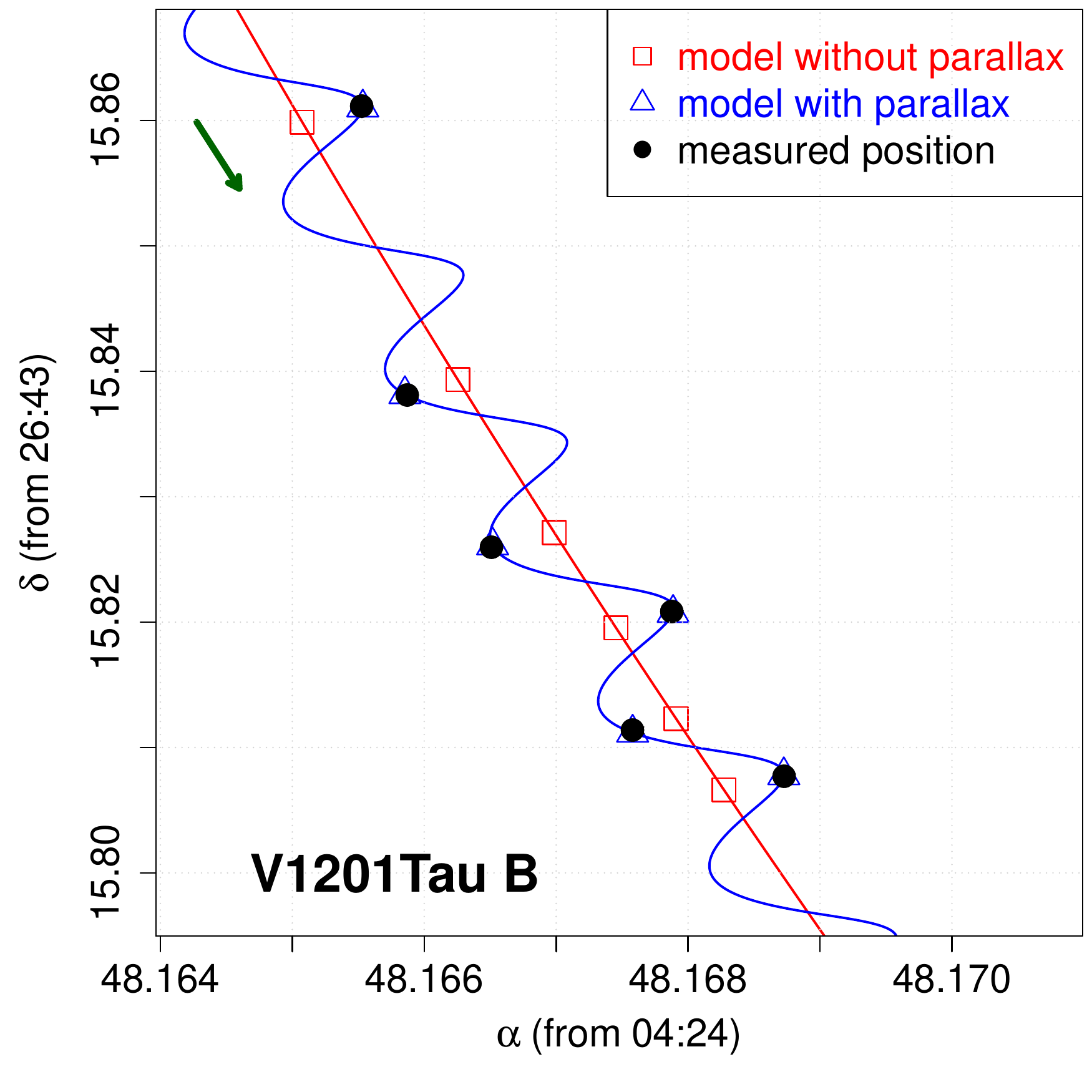}{0.33\textwidth}{}
}
\gridline{
\fig{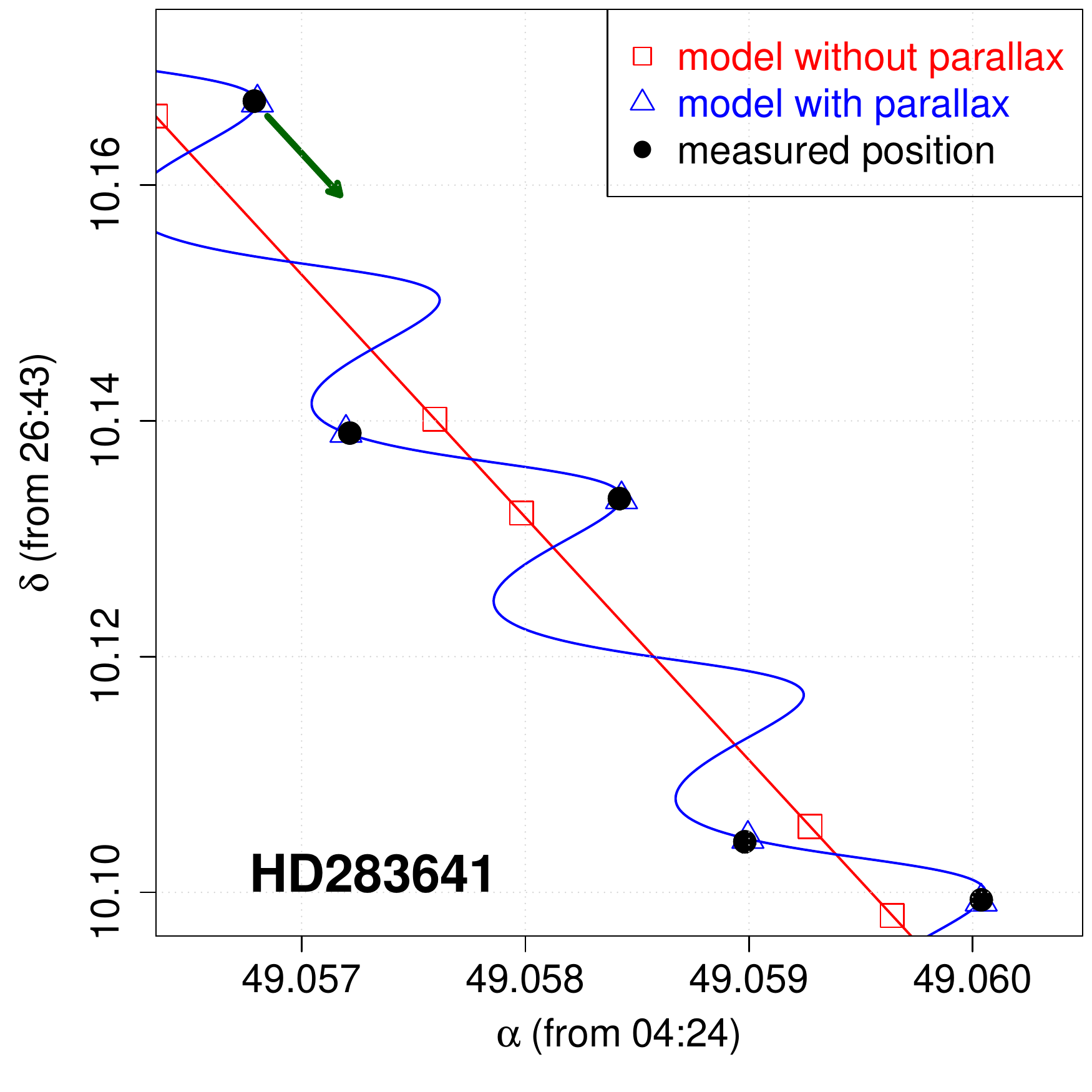}{0.33\textwidth}{}
\fig{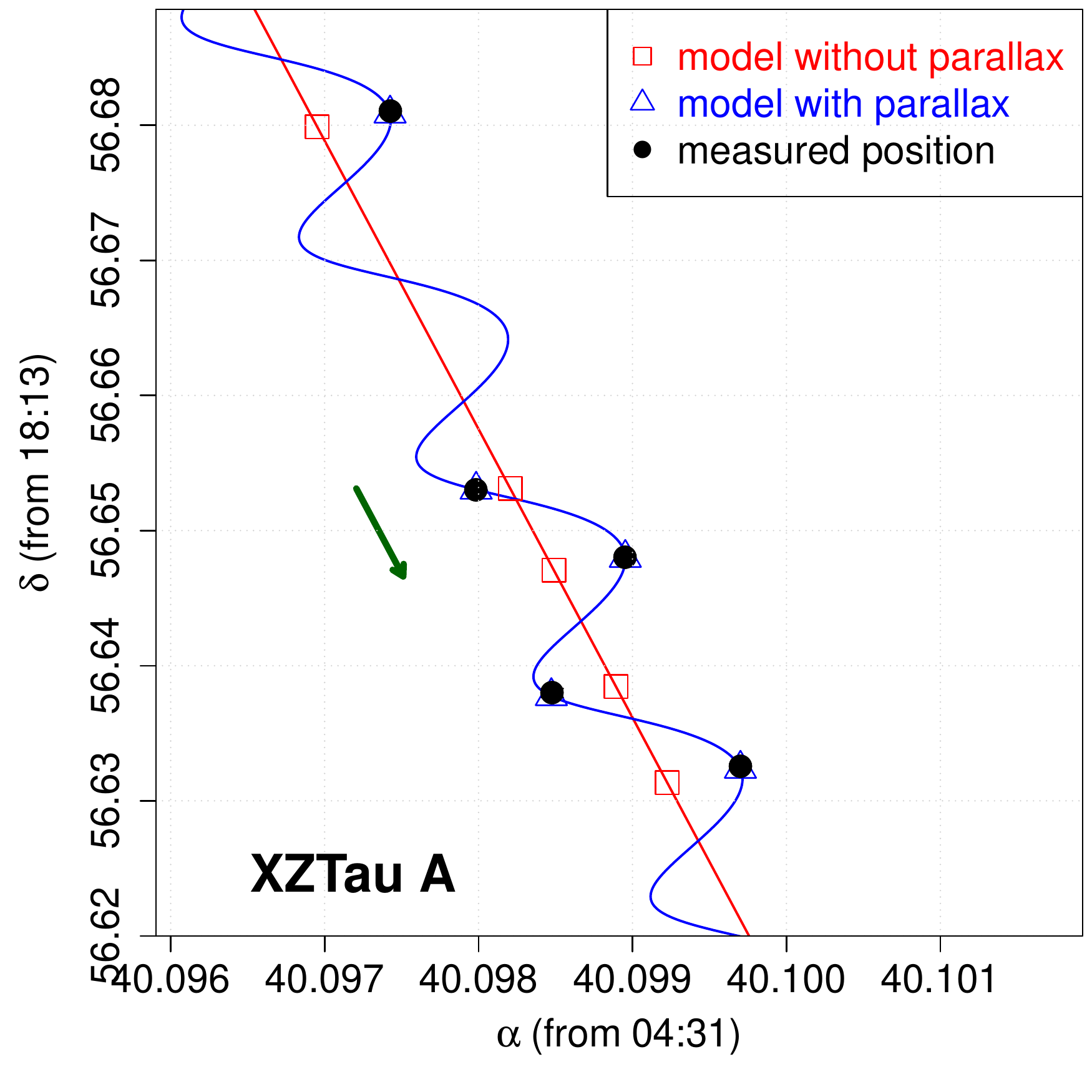}{0.33\textwidth}{}
\fig{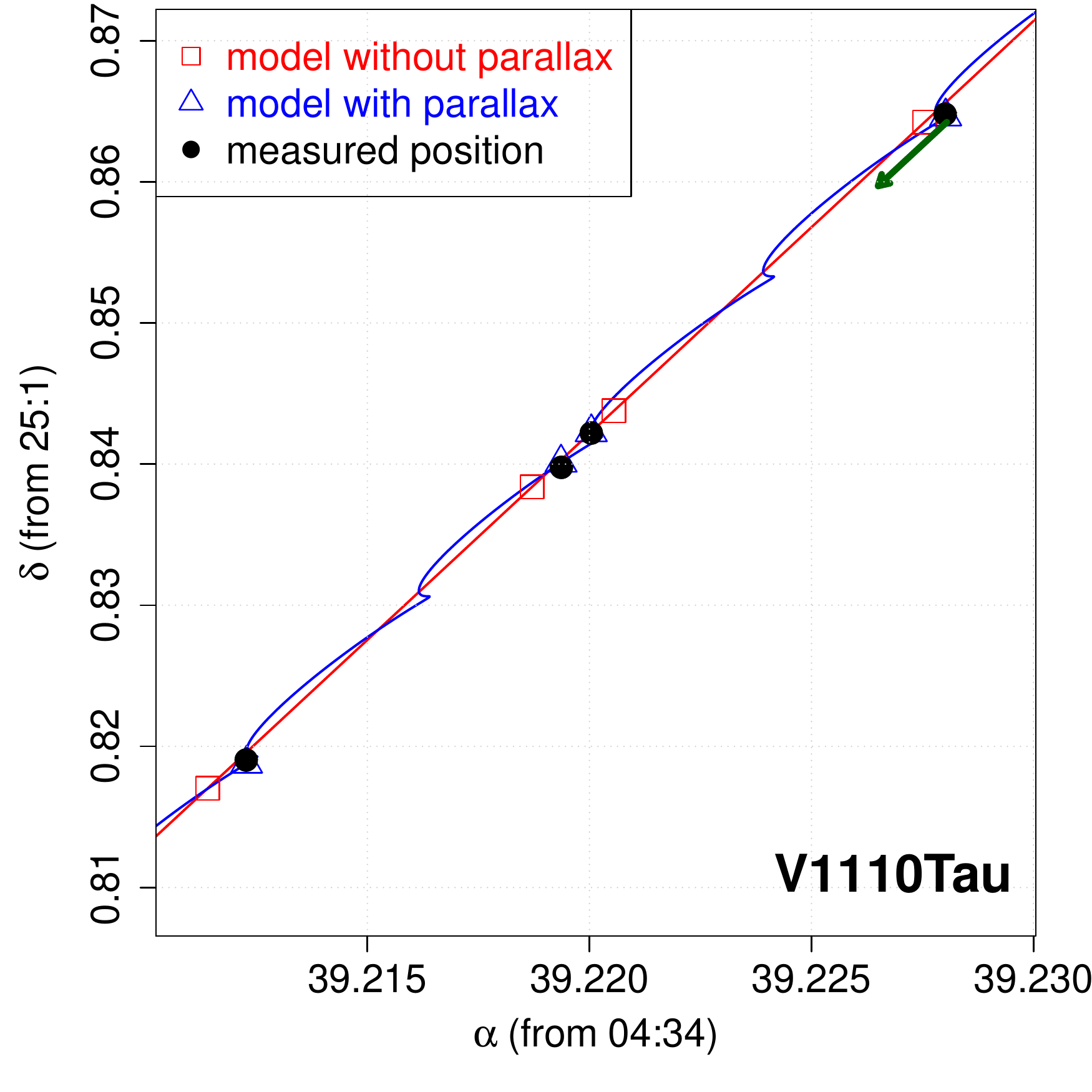}{0.33\textwidth}{}
}
\caption{Astrometry fit for the single and long orbital period YSOs in our sample. Measured positions with the VLBA are shown as black circles and their corresponding error bars (including systematic errors) are mostly smaller than the symbols. The red line indicates the model with the parallax signature removed, and the blue line represents the model including the parallax. Red squares and blue triangles mark the expected positions for the corresponding models. The green arrow shows the direction of the stellar proper motions in the plane of the sky.
\label{fig4}}
\end{figure*}

\setcounter{figure}{3}
\begin{figure*}[]
\gridline{
\fig{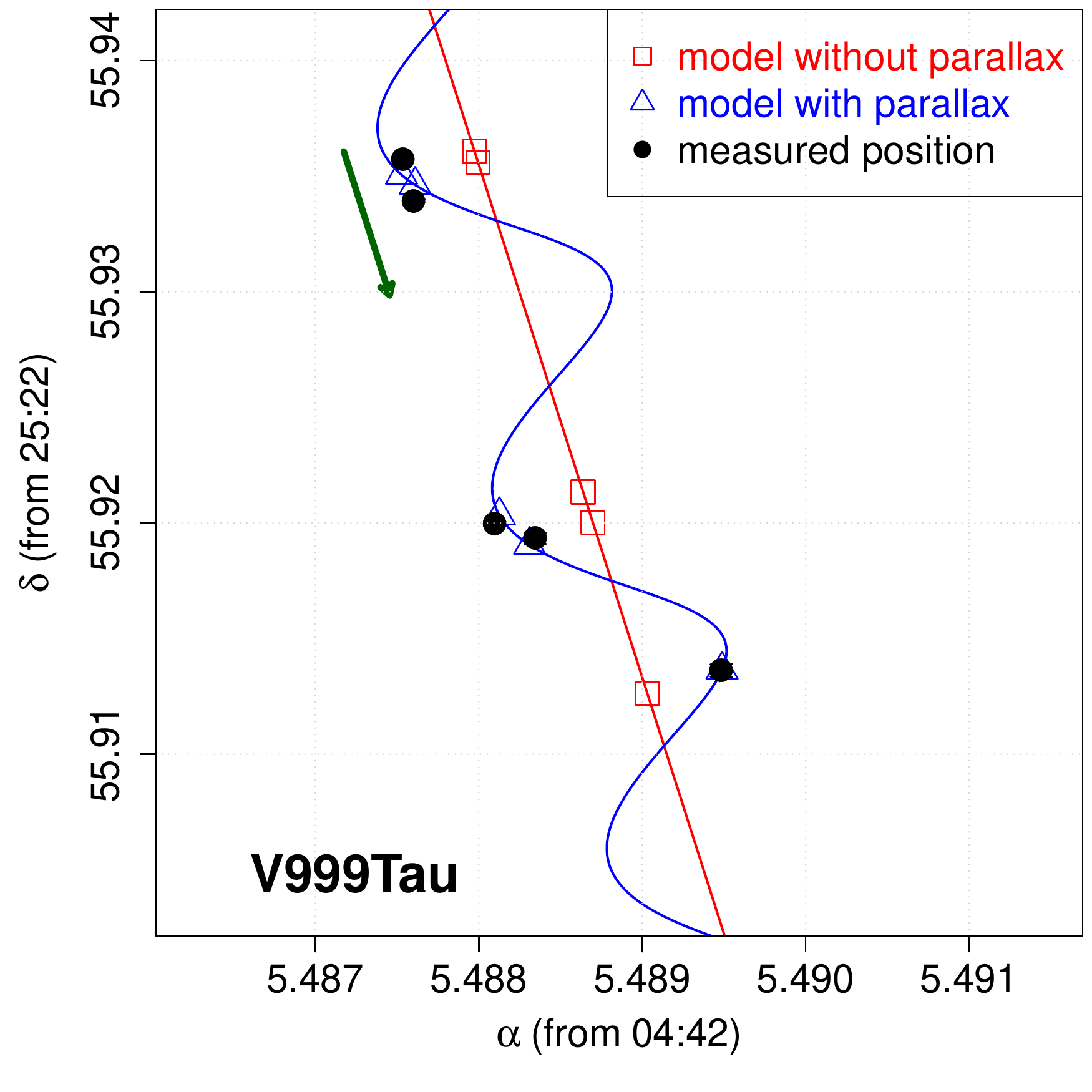}{0.33\textwidth}{}
\fig{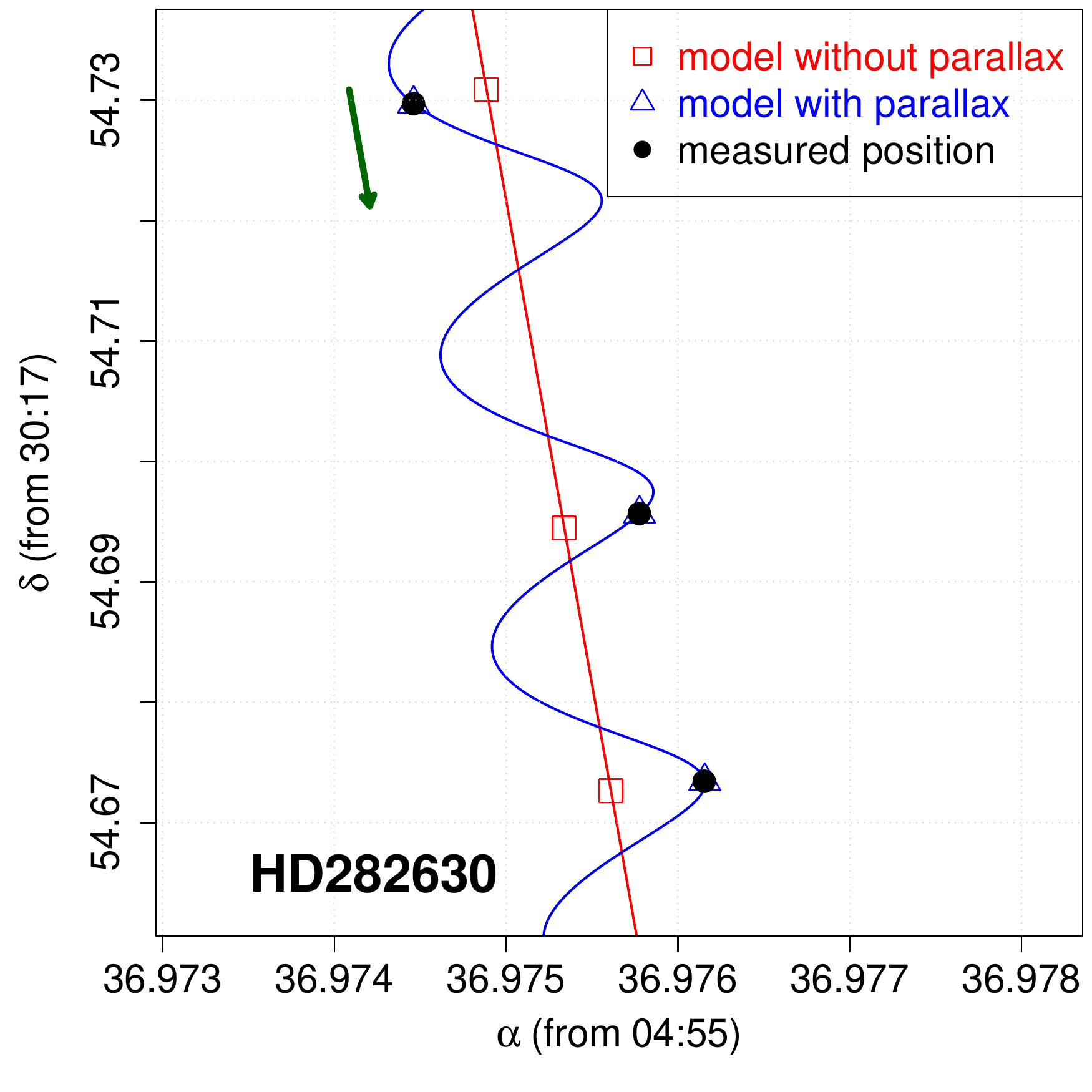}{0.33\textwidth}{}
\fig{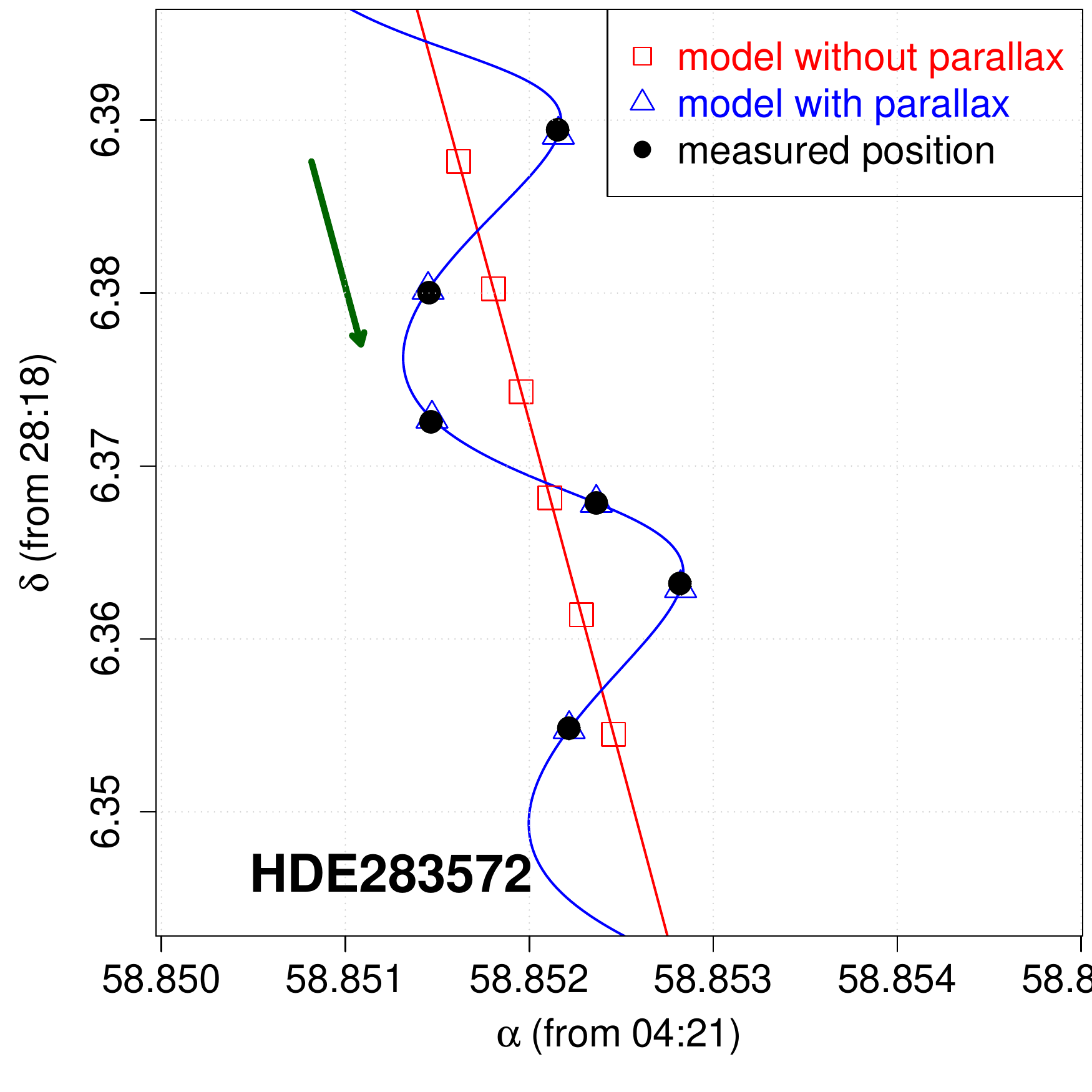}{0.33\textwidth}{}
}
\caption{continued. }
\end{figure*}

\begin{figure*}[]
\gridline{
\fig{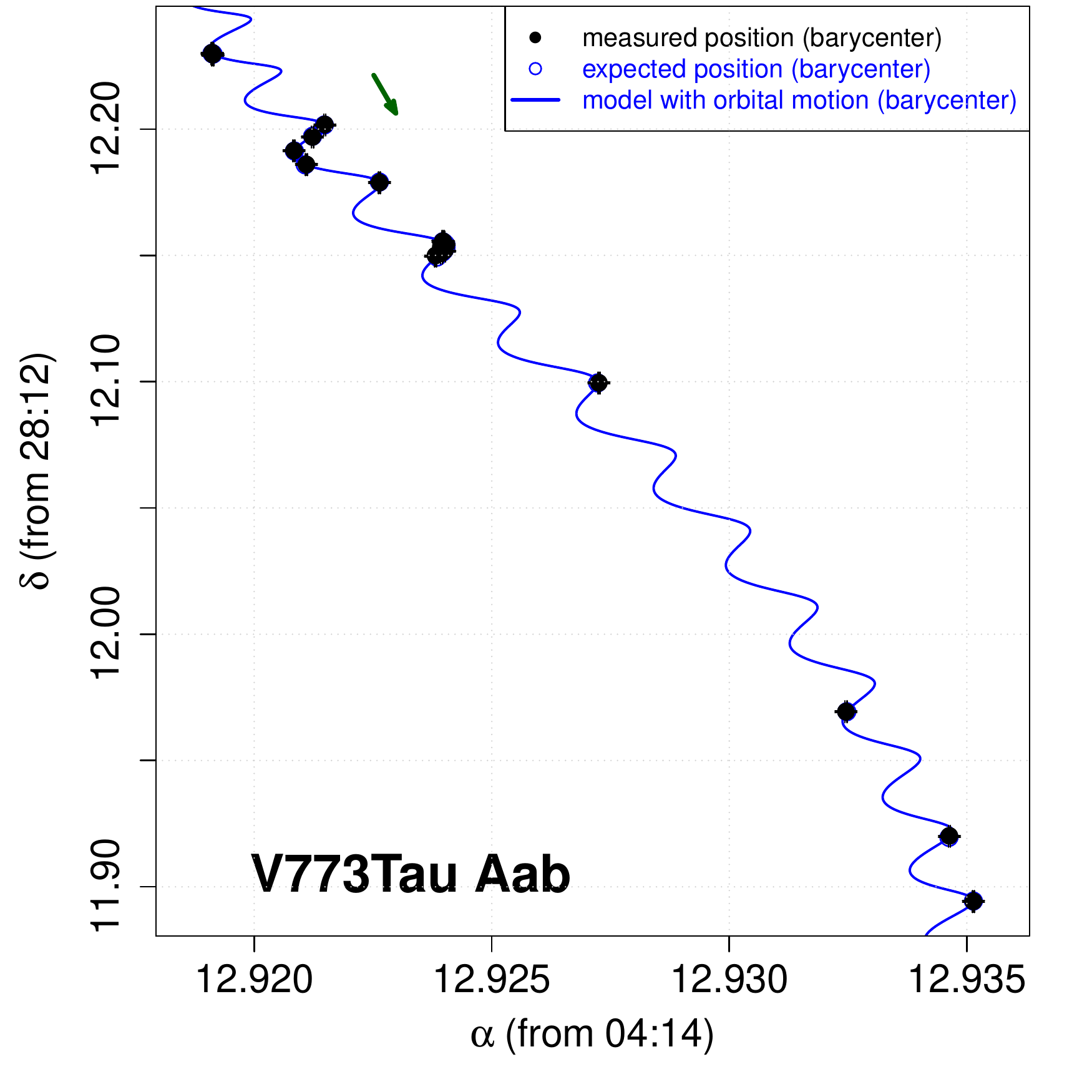}{0.49\textwidth}{}
\fig{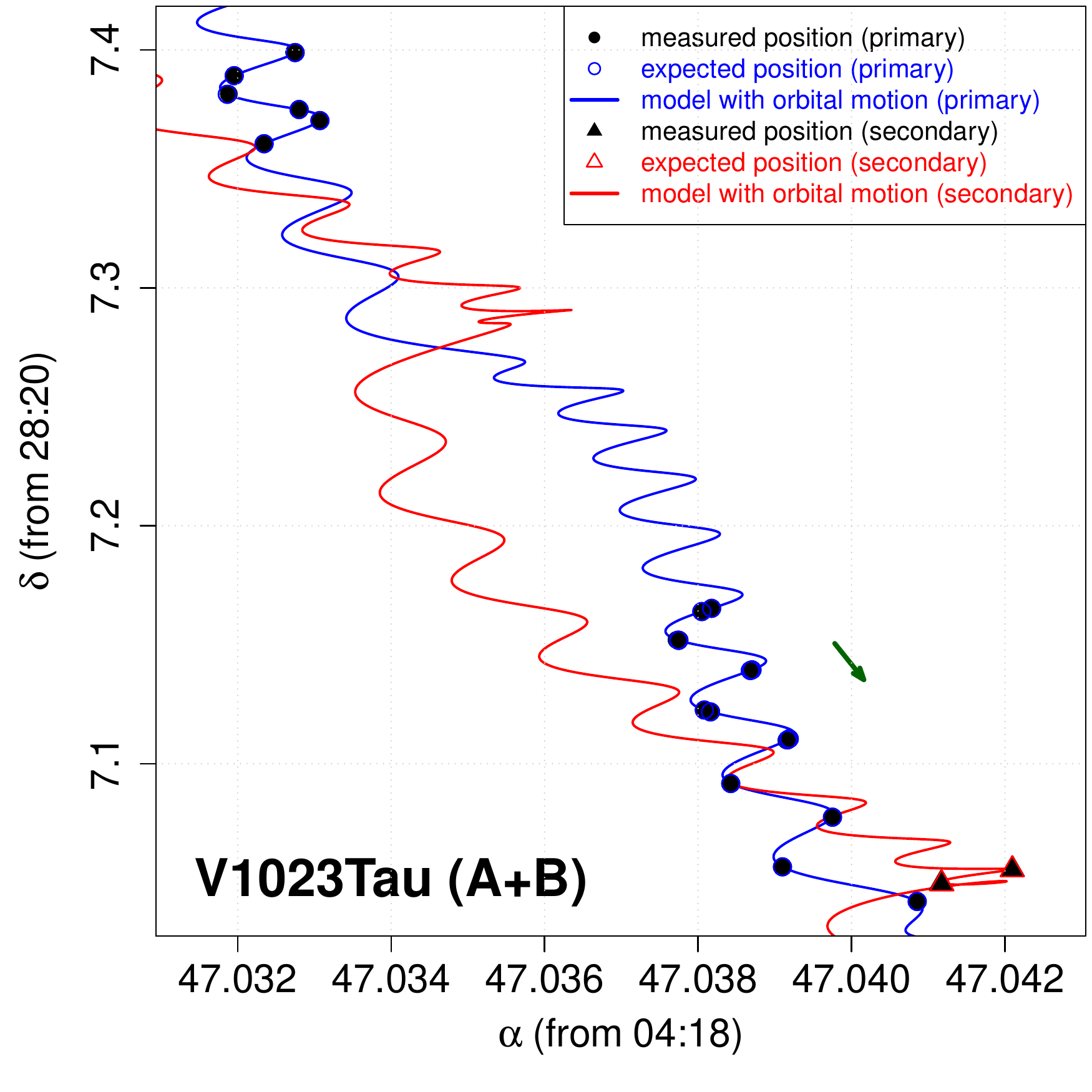}{0.49\textwidth}{}
}
\caption{Astrometry fit for binaries and multiple systems in our sample including the orbital motion of the system in the analysis (see Sect.~\ref{section3.2}). The multiple components of each system detected with the VLBA in our observing campaign are shown in different colors.  The green arrow shows the direction of the stellar proper motions in the plane of the sky. \vspace{2cm}
\label{fig5}}
\end{figure*}

\setcounter{figure}{4}
\begin{figure*}[]
\gridline{
\fig{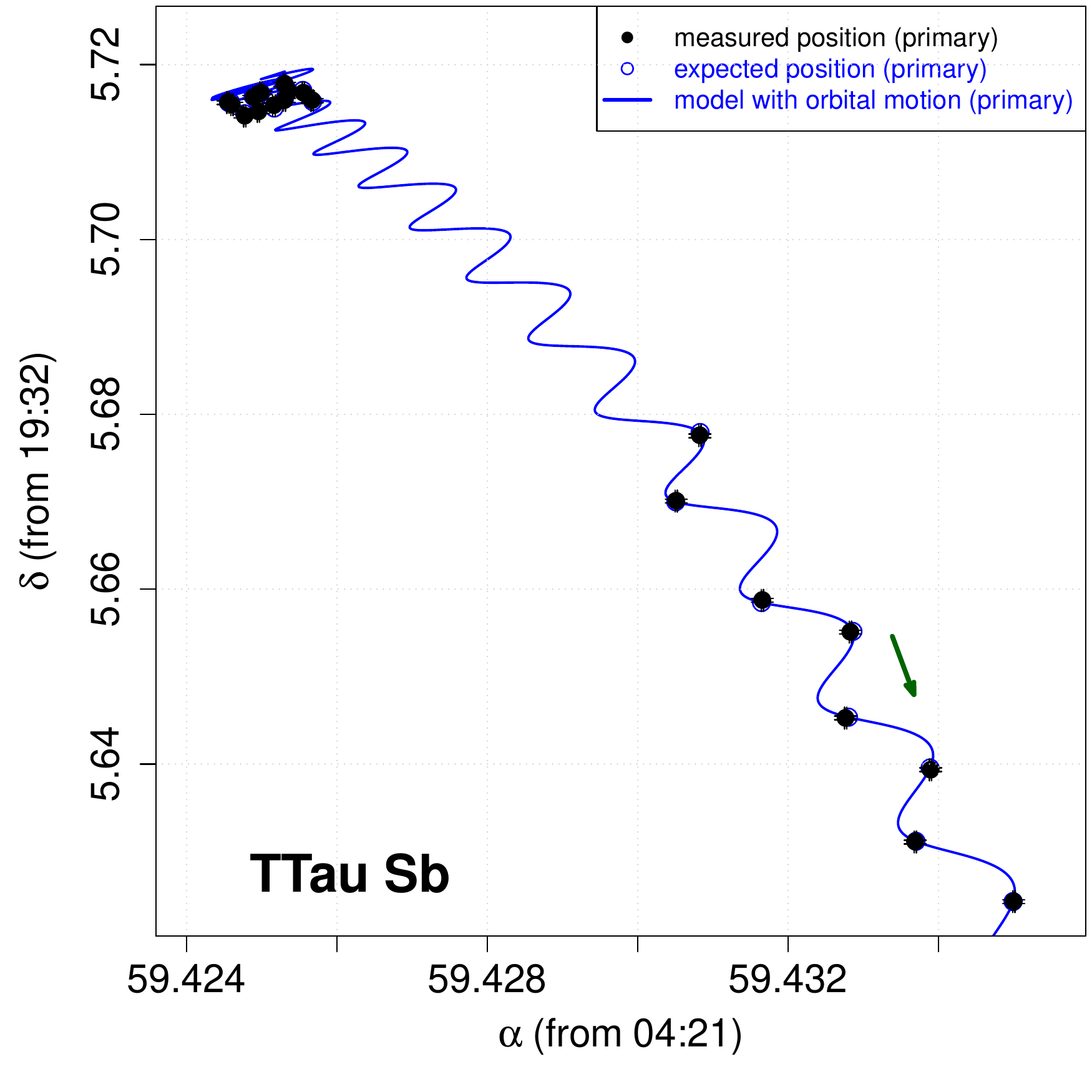}{0.49\textwidth}{}
\fig{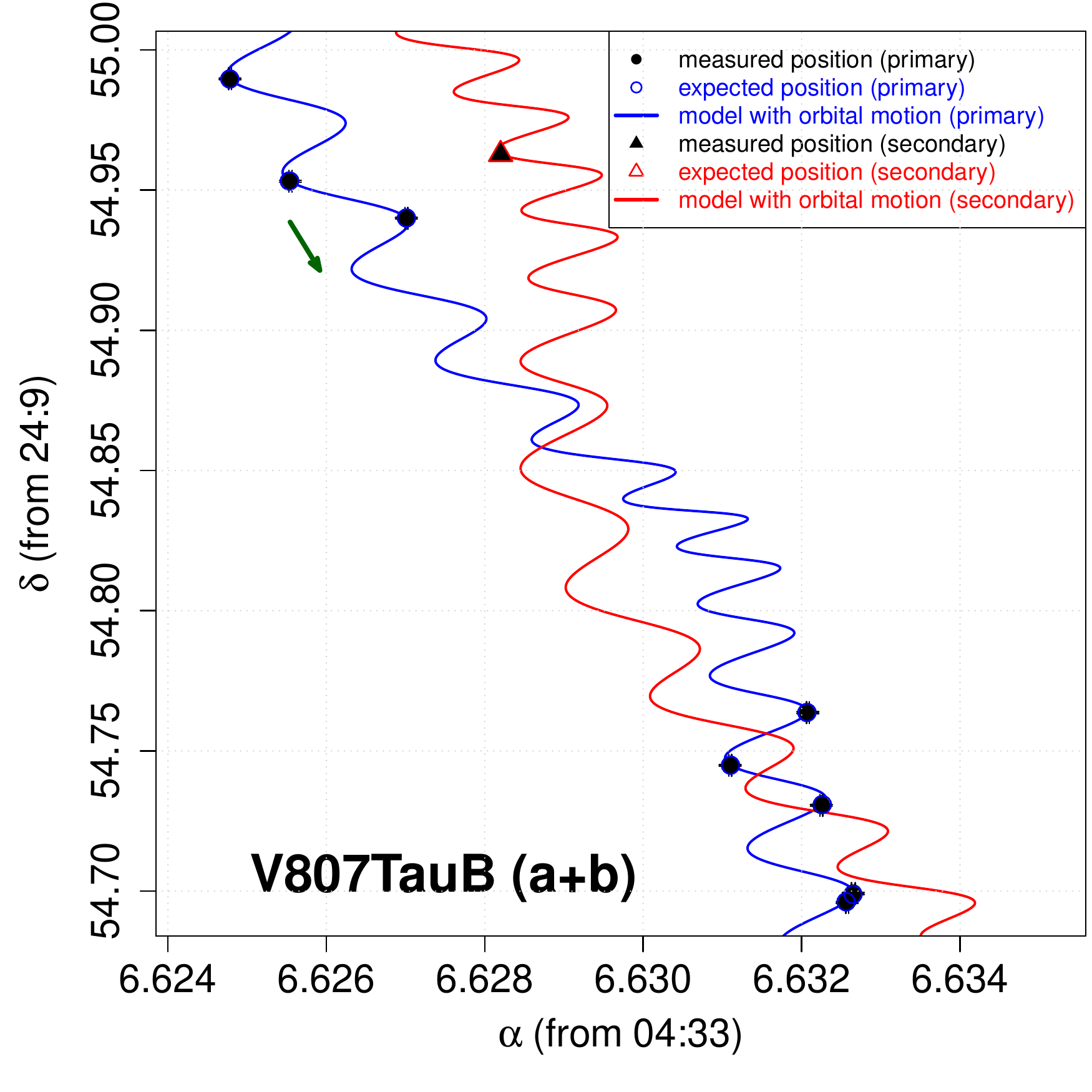}{0.49\textwidth}{}
}
\gridline{
\fig{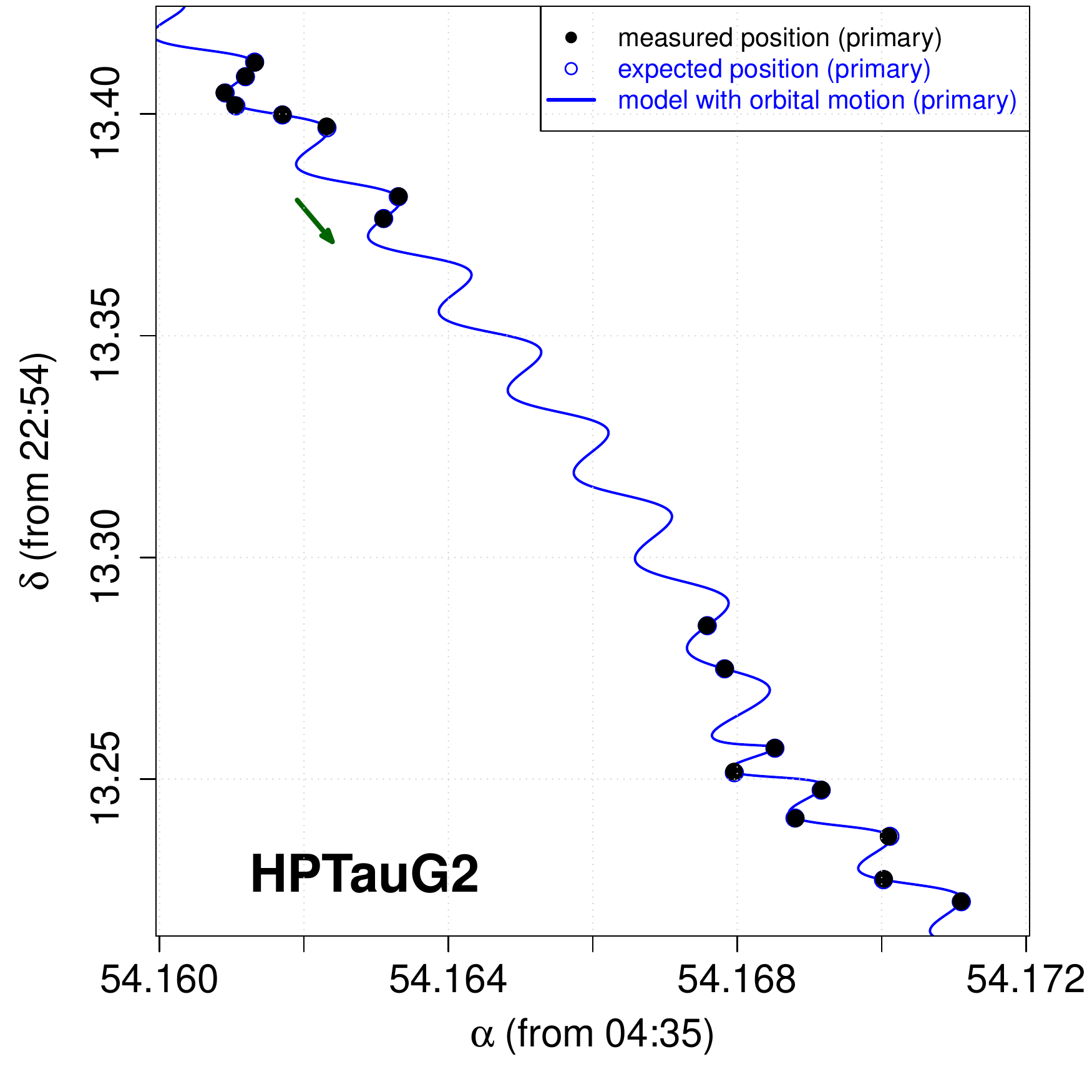}{0.49\textwidth}{}
\fig{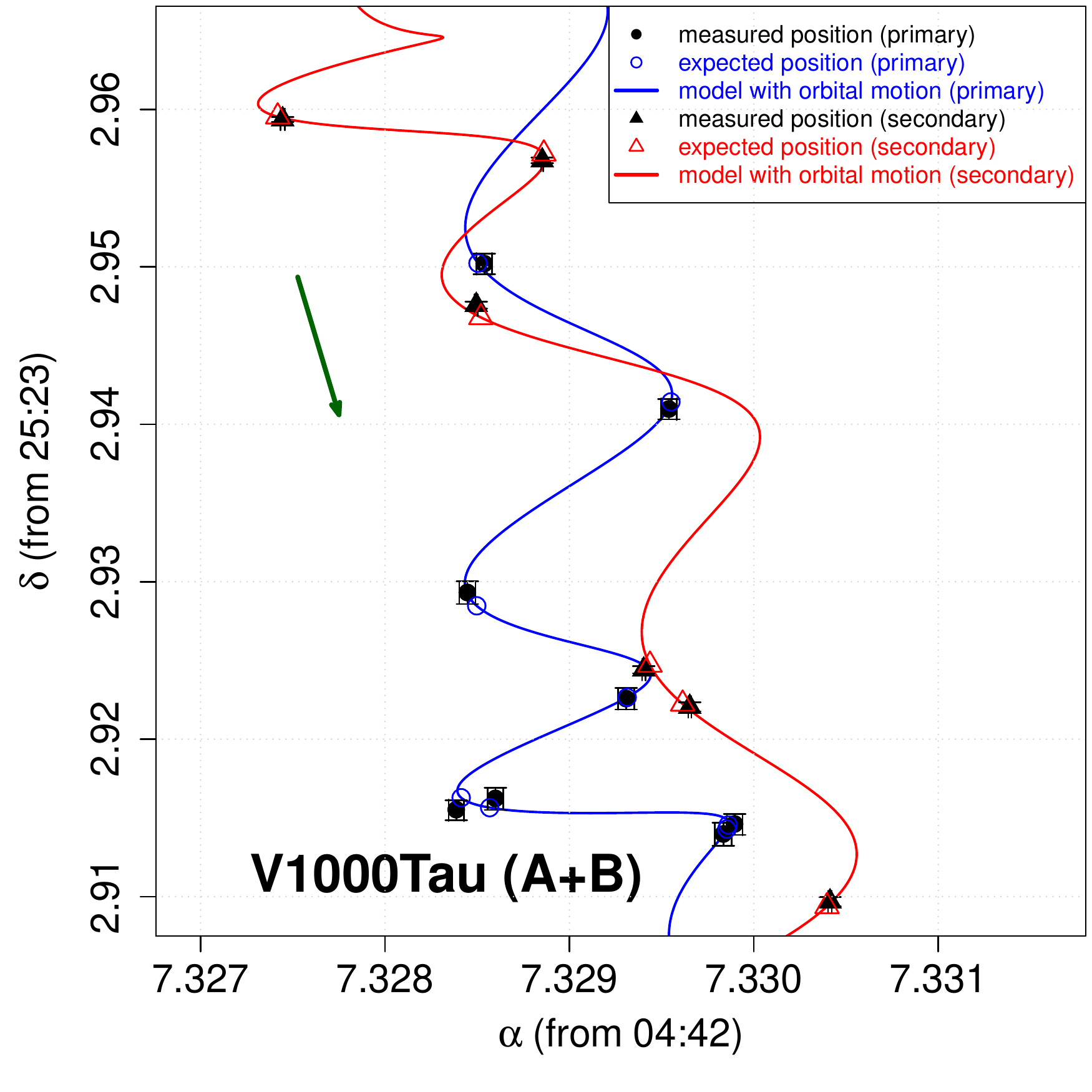}{0.49\textwidth}{}
}
\caption{continued.}
\end{figure*}

\begin{figure*}[]
\gridline{
\fig{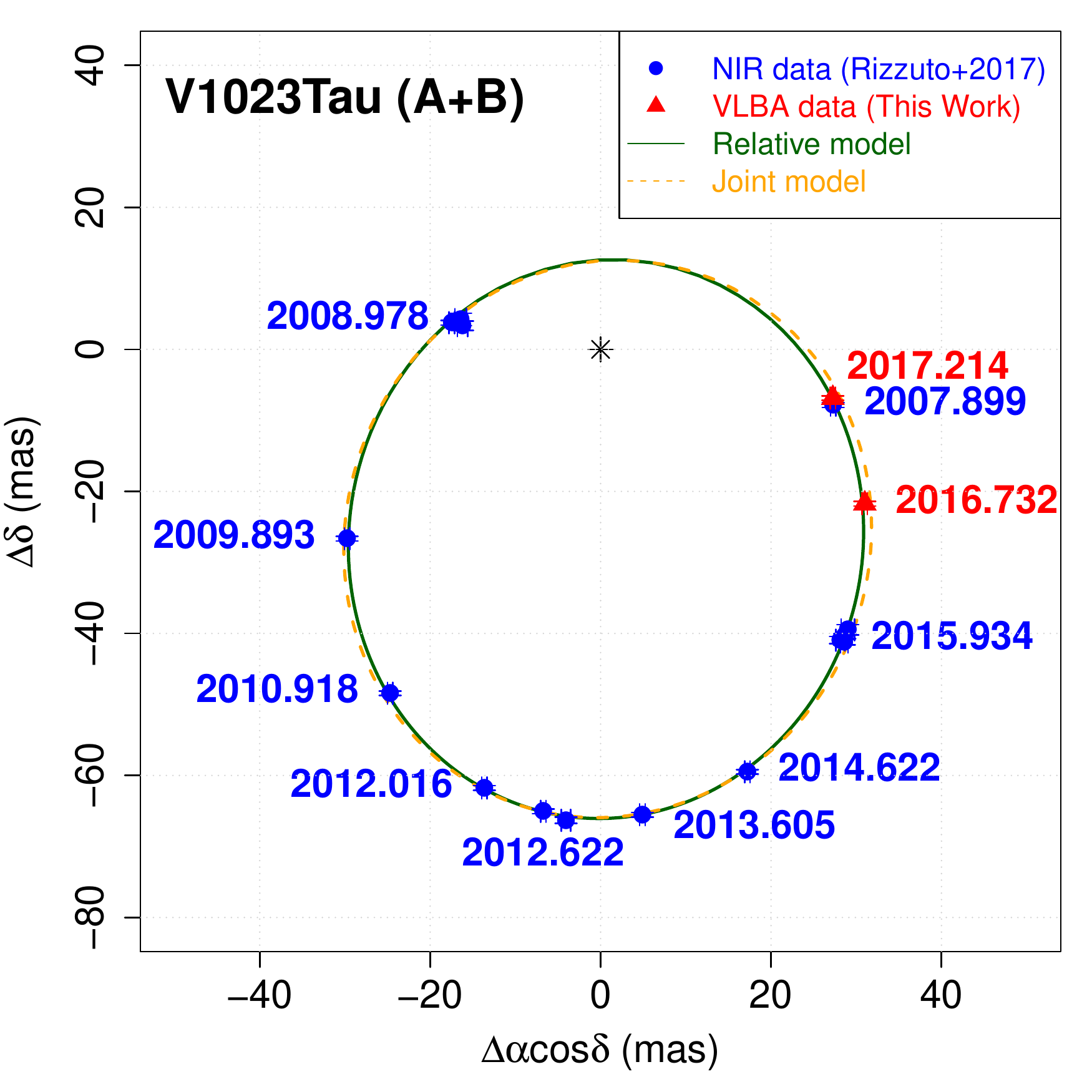}{0.49\textwidth}{}
\fig{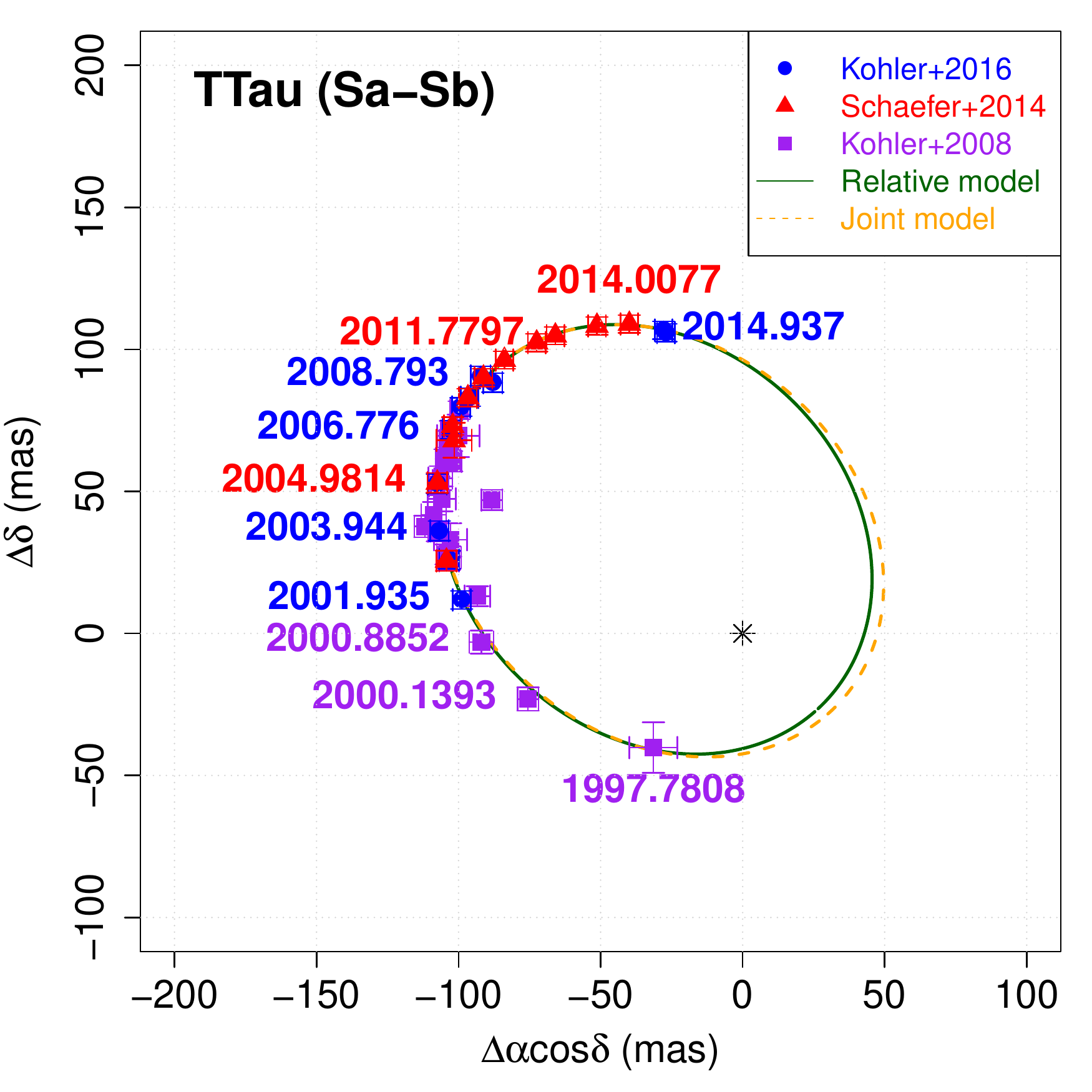}{0.49\textwidth}{}
}
\gridline{
\fig{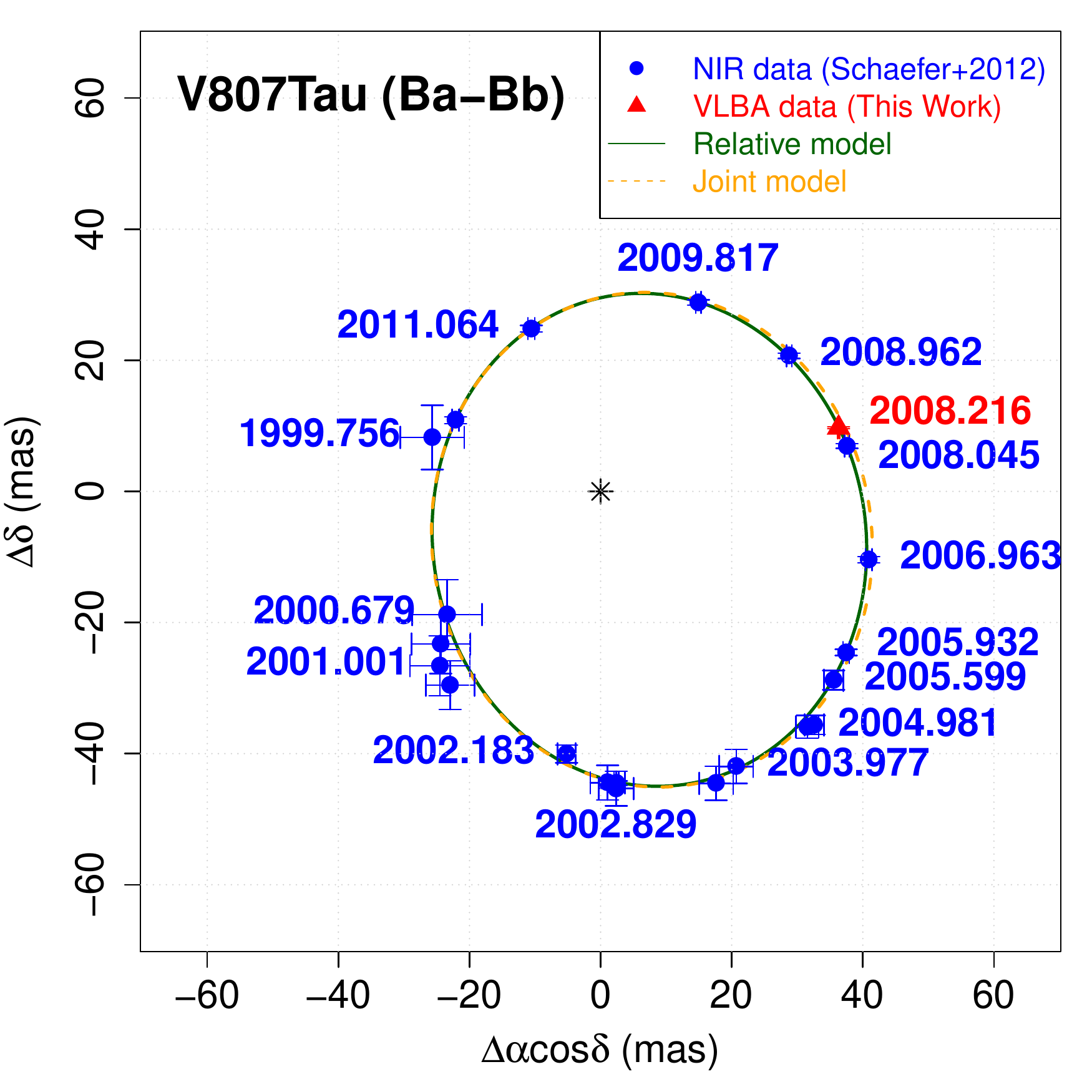}{0.49\textwidth}{}
}
\caption{Relative astrometry of the components for three binary systems in our sample with measured relative positions in the literature. The black asterisk indicates the position of the secondary component (relative to the primary). 
\label{relFIT}}
\end{figure*}

\subsection{V1096~Tau}\label{sectionV1096Tau}

V1096~Tau is a binary system that could be resolved in our observations. Both components were simultaneously detected in only one epoch (project BL175IE) where the primary component of the system V1096~Tau~A exhibits a flux density  of $0.71\pm0.05$~mJy that is almost twice that of the secondary component (V1096~Tau~B) with $0.36\pm0.05$~mJy. We detected the primary and secondary, respectively, in a total of 4 and 3 epochs. However, our observations cover a small fraction of the orbit so that the fit including the orbital motion of the system does not converge. So, we fit the measured positions of the individual components solely for the proper motion and parallax. We included the acceleration term in the astrometry fit for the primary component which indeed represents a better description to the data, but this was not possible for the secondary due to the limited number of detections available. Our result for the secondary component is only indicative and should be regarded with caution as it is likely to be biased by the non-corrected orbital motion of the system. The systematic errors that we add to the stellar positions of V1096~Tau~A and V1096~Tau~B (as described in Sect~\ref{section3.3}) reach up to 1~mas and 3~mas, respectively, and they are significantly larger as compared to the other stars. The weighted mean parallax of the results given in Table~\ref{tab5} for the two components is $\pi=8.04\pm 0.50$~mas. This is consistent with a distance estimate of $d=124.4^{+8.2}_{-7.2}$~pc. Despite the admittedly large errors on our results, this is the first distance determination for this binary system to date. The weighted mean proper motions of the two components is $\mu_{\alpha}\cos\delta=2.612\pm0.691$~mas/yr and $\mu_{\delta}=-17.372\pm0.676$~mas/yr. V1096~Tau continues to be monitored by our team, and the results presented in this paper will be refined when more observations become available.

\subsection{V773~Tau}

V773~Tau is a well-known quadruple system \citep[see e.g.][]{Duchene2003,Woitas2003} and the source detected in our observations is the primary component V773~Tau~A which itself is a tight binary. V773~Tau~A has been detected in a total of 7 epochs during our observing, but a simultaneous detection of both components occurred only in 3 epochs (see Table~\ref{tab4}). Moreover, it has also been observed with the VLBA in 27 additional epochs between March 2004 and September 2009 for projects BM198, BL128, BL146 and BM306 (see Table~\ref{tab3}). We note that project BM198 observed J0403+2600 as the main phase calibrator while the other projects (including GOBELINS) used J0408+3032 instead. We corrected the positions measured in BM198 before combining them with the other projects. The mean position of J0403+2600 measured in projects BM198 is $\alpha=04^{h}03^{m}05.586049^{s}$ and $\delta=26^{\circ}00\arcmin 01.50275\arcsec$. The mean position of J0403+2600 (relative to J0408+3032) in the other projects is $\alpha=04^{h}03^{m}05.586052^{s}$ and $\delta=26^{\circ}00\arcmin 01.50137\arcsec$. So, we applied a correction of $\Delta\alpha=0.000003$~s and $\Delta\delta=-0.00138$\arcsec~to the measured positions in project BM198. Another important point to mention about projects BM198 is that these observations include data collected with both the VLBA and the Effelsberg antenna. But for consistency with the rest of the data used in this work, we decided to remove the Effelsberg antenna from the data reduction. The same applies to the BM306 project which observed with the High Sensitivity Array (HSA). We removed the external antennae and calibrated the observation using only VLBA data. 

The orbital parameters of the short-period binary system V773~Tau~Aa-Ab have already been constrained by \citet{Torres2012} based on the past VLBA observations listed in Table~\ref{tab3} supplemented by radial velocity observations from \citet{Boden2007}. They derived the dynamical masses of the two components ($m_{Aa}=1.55\pm0.11$~$M_{\odot}$ and $m_{Ab}=1.293\pm0.068$~$M_{\odot}$) and a trigonometric parallax of $\pi=7.70\pm0.19$~mas computed from the positions of the barycenter of the system including a uniformly accelerated proper motion in the analysis. As explained in their study, the model including a uniform acceleration (see Sect.~\ref{section3.2}) provided a good description to the data, because their observations covered a small fraction of the orbit of the V773~Tau A-B system. However, this not the case when we include the more recent data from the  GOBELINS project in this analysis which increases the time base of collected observations to about 13~years. We computed the baricentric positions of the V773~Tau~Aa-Ab system for the 3 epochs with simultaneous detections of the two components in our observing campaign and combined them with the re-calibrated baricentric positions reported by \citet{Torres2012}. We verified that the model assuming a uniform acceleration indeed provides a poor fit to the data with the extended time base of our observations. We thus performed a full fit to the data including the orbital motion of the V773~Tau~A-B in the analysis. Doing so, we derive an orbital period of $P=20.3\pm0.8$~yr that is somewhat shorter than the value of $P=26.2\pm1.1$~yr obtained by \citet{Boden2012} based on relative astrometry and radial velocities. The resulting trigonometric parallax of $\pi=7.692\pm 0.085$~mas yields a distance estimate of $d=130.0^{+1.5}_{-1.4}$~pc for the barycenter of V773~Tau~Aa-Ab system that is more accurate and precise than the result of $d=132.8\pm2.3$~pc obtained by \citet{Torres2012}. 

\subsection{V1098~Tau, 2MASS~J04182909+2826191, V999~Tau and HD~282630}\label{sect_V999TAU}  

V1098~Tau, 2MASS~J04182909+2826191 and HD~282630 have been detected in only 3 epochs of our observing campaign. On the other hand, V999~Tau has been detected in 5 different projects, but in practice this corresponds to only 3 epochs due to the short time interval between projects BL175IC/BL175I7 and BL175JW/BL175JY. As a result, their trigonometric parallaxes are less precise as compared to the other stars in our sample with more detections. Because of the small number of detections we computed the additional errors on the stellar positions (see Sect.~\ref{section3.3}) from the methodology outlined in \citet{Pradel2006}. The resulting errors reach up to 0.8~mas depending on the position of the source in the sky and the angular separation to the main phase calibrator.

\subsection{HD~283518}

HD~283518 (V410~Tau) was systematically detected in every epoch of our observing campaign. It is known to be a multiple system \citep[see e.g.][]{Harris2012}, but only one component could be detected in our observations. We do not see the signature of the orbital motion in the stellar positions measured in this work indicating that the gravitational effects of the secondary on the primary are negligible. So, we solved for the astrometric parameters of this source using the methodology described in Sect.~\ref{section3.1} for single stars. We also compared our results with the alternative approach outlined in Sect.~\ref{section3.2} using a uniform acceleration. Both methods return compatible results (with the same level of accuracy and precision) and the derived acceleration terms in the latter approach are consistent with zero. Thus, the results presented in Table~\ref{tab5} refer to the model without acceleration. The trigonometric parallax that we derive for HD~283518 has a relative error of $0.3\%$ making it the most precise result in our sample.   

\subsection{V1023~Tau}

V1023~Tau (Hubble~4) is another binary system that was resolved in our observations. The primary component of this system is a weak-line T~Tauri star of spectral type K7 \citep{Nguyen2012}, and it has been detected in a total of 14 epochs during our observing campaign. On the other hand, the secondary component could be detected in only 2 epochs. In addition, the primary has also been observed between September 2004 and December 2005 for projects BL124 and BL136. These data have already been analyzed and published by \citet{Torres2007}. We combined these observations with the more recent GOBELINS data to fit the measured positions over a timebase of $\sim13$~yr which covers one full orbit of the system. Both data sets used the same main phase calibrator so that no correction has to be applied to the measured positions.

First, we performed a full fit of the orbit using only the VLBA observations reported in this paper and in \citet{Torres2007}. Then, we combined the VLBA observations of projects BL175J2 and BL175JS where both components were simultaneously detected with the NIR relative astrometry from Rizzuto et al. (2018, submitted) to refine the orbital elements of the binary system. Finally, we combined our VLBA absolute positions with the NIR relative positions to perform a joint fit.  Tables~\ref{tab6} and \ref{tab7} compares our results from the different methods. The distance that we derive here from the full model ($d=130.1^{+0.5}_{-0.5}$~pc) is somewhat shorter than the value of $d=132.8\pm0.5$~pc obtained by \citet{Torres2007} using VLBA data from projects BL124 and BL136. We argue that the latter result is likely biased by the non-corrected binarity of the source. By combining the VLBA absolute positions with the NIR relative astrometry we find a distance of $d=129.0^{+2.0}_{-1.9}$~pc that is less precise but fully consistent with our previous result. The weighted mean of both values yields $d=130.0^{+0.5}_{-0.5}$~pc which is the most precise and accurate distance estimate for this source to date. Our analysis also made it possible to determine, for the first time, the dynamical masses of the individual components of this system ($m_{A}=1.234\pm0.023$~M$_{\odot}$ and $m_{B}=0.730\pm0.020$~M$_{\odot}$). This implies a somewhat smaller mass ratio  ($q=0.592\pm0.012$)  than the value of $q=0.73$ previously reported by \citet{Harris2012}. 

\subsection{T~Tau}

T~Tau is a well-known triple system  \citep[see e.g.][]{Duchene2002} and the component detected in our observations with the VLBA is T~Tau~Sb. It has been detected in 8 epochs during our observing campaign for the GOBELINS project. In addition, it was also observed by our team between September 2003 and July 2005 for projects BL118 and BL124 (see Table~\ref{tab3}). However, we note that projects BL118 and BL128 observed J0428+1732 as the main phase calibrator, while GOBELINS uses it as a secondary calibrator (the main calibrator is J0412+1856). The mean position of J0428+1732 measured for projects BL118 and BL124 is $\alpha=04^{h}28^{m}35.633679^{s}$ and $\delta=17^{\circ}32\arcmin23.58803\arcsec$. In the GOBELINS campaign the mean position of J0428+1732 (relative to J0412+1856) is $\alpha=04^{h}28^{m}35.633685^{s}$ and $\delta=17^{\circ}32\arcmin23.58840\arcsec$. Thus, we applied an offset of $\Delta\alpha=0.000006$~s and $\Delta\delta=0.00037\arcsec$ to the measured positions obtained from projects BL118 and BL124. 

We fit the measured positions of T~Tau~Sb based on the full model to solve for the parallax and orbital motion of the binary system. The orbital elements obtained in this work (see Table~\ref{tab6}) refer to the T~Tau~Sa-Sb system. We attempted to include an additional acceleration term due to the T~Tau~N component of the system, but the resulting acceleration parameter was consistent with zero. The results presented in Table~\ref{tab5} refer to our first solution (without acceleration). The distance that we derive in this paper for T~Tau~Sb ($d=148.3^{+2.1}_{-2.1}$~pc) is in good agreement with the result of $d=147.6\pm0.6$~pc published previously by \citet{Loinard2007} using only data from projects BL118 and BL124. However, we consider our result to be more accurate because it takes into account the binarity/multiplicity of the source. Our solution obtained from the joint fit using the relative astrometry of the T~Tau~Sa-Sb published in previous studies yields a somewhat more precise distance estimate for this system ($d=148.7^{+1.0}_{-1.0}$~pc) that confirms our first result. The weighted mean of both results yields a final distance of $d=148.7^{+0.9}_{-0.9}$~pc.  

Other studies have already investigated the orbital motion of the T~Tau~Sa-Sb system based on relative astrometry and different measurements \citep{Kohler2008,Kohler2016,Schaefer2014}. As illustrated in Fig.~\ref{relFIT}, we combined the relative positions from previous works to provide a more refined solution for the orbital parameters of the system. This analysis does not include VLBA data since our observations can only detect one component of the system. We note that most of the resulting orbital elements obtained in this paper including all measurements are more precise than the results reported in the individual studies. However, the errors on the individual parameters are still larger as compared to other stars in our sample. One reason to explain this result is the small coverage of the orbit which requires further monitoring of the T~Tau system.

\subsection{V1201~Tau and HD~283641}

V1201~Tau and HD~283641 are members of the same hierarchical multiple system where both sources themselves are binary systems \citep[see][]{Kohler1998,Mason2001}. V1201~Tau and HD~283641 were also identified as wide binaries with a separation of $13.23\arcsec$ in a recent study conducted by \citet{Andrews2017} based on data from Gaia-DR1. Indeed, their result is consistent with the mean angular separation of $14.53\arcsec$ derived from our observations. The two components of the V1201~Tau system were simultaneously detected in only 2 epochs (projects BL175I1 and BL175KJ). We measured a flux density of $2.40\pm0.04$~mJy in project BL175I1 (February 28, 2016) for the brightest component (hereafter, V1201TauA) that significantly decreased to $0.15\pm 0.05$~mJy in project BL175KJ (September 18, 2017) becoming fainter than the so-defined secondary component V1201~Tau~B. We derived a trigonometric parallax only for V1201~Tau~B that was detected in 6 epochs. We note from Figure~\ref{fig4} that the fit using a uniform acceleration represents a good description to the data yielding a distance estimate of $d=157.2\pm 1.7$~pc. In the case of HD~283641 we detected only one component of the system in our observations. First, we fit the data as described in Sect.~\ref{section3.1}, then we introduced the acceleration. We verified that the inclusion of a uniform acceleration in the model increases the errors on the astrometric parameters and that the derived acceleration terms are consistent with zero. As expected, the distance obtained for HD~283641 ($d=159.1\pm1.8$~pc, without acceleration) is compatible within $1\sigma$ of the distance derived for V1201~Tau. Both sources are currently being monitored by our team, and we will deliver a more refined solution (including the orbital motion of the system) when these observations become available. 

\subsection{XZ~Tau}

The XZ~Tau system is composed of two components (XZ~Tau~A and XZ~Tau~B) with angular separation of about $0.3\arcsec$ \citep{Harris2012,Joncour2017}. \citet{Carrasco-Gonzalez2009} report on the detection of a third component (XZ~Tau~C) in this system separated by $0.09\arcsec$ from XZ~Tau~A making it a triple system. However, a more recent study conducted by \citet{Forgan2014} did not detect XZ~Tau~C and cast doubt on the existence of the third component. Only one component of this system could be detected in our observations and we confirm that it corresponds to XZ~Tau~A. \citet{Frist2008} estimated that the minimum orbital period of the XZ~Tau~A-B system should be about 99~years which greatly exceeds the time base of our observations. Indeed, we see no evidence of binarity in our data and modeling XZ~Tau~A as a single star (as described in Sect.~\ref{section3.1}) provides a good fit to the data (see Fig.~\ref{fig4}). Finally, it is interesting to note that all components of this system were suggested to be thermal radio sources \citep[see][]{Carrasco-Gonzalez2009}, but our study shows that XZ~Tau~A also produces non-thermal emission.

\startlongtable
\begin{deluxetable*}{lcccccc}
\tablecaption{Proper motions, parallaxes and distances derived from VLBA observations. \label{tab5}}
\tablehead{
\colhead{Identifier}&
\colhead{$\mu_{\alpha}\cos\delta$}&
\colhead{$\mu_{\delta}$}&
\colhead{$a_{\alpha}\cos\delta$}&
\colhead{$a_{\delta}$}&
\colhead{Parallax}&
\colhead{Distance}\\
\colhead{}&
\colhead{(mas/yr)}&
\colhead{(mas/yr)}&
\colhead{(mas/yr$^{2}$)}&
\colhead{(mas/yr$^{2}$)}&
\colhead{(mas)}&
\colhead{(pc)}
}
\startdata
V1096~Tau~A	&$	2.089	\pm	0.730	$&$	-16.167	\pm	0.711	$&$	11.640	\pm	2.945	$&$	0.234	\pm	3.710	$&$	8.055	\pm	0.535	$&$	124.1	_{	-7.7	}^{+	8.8	}$\\	
V1096~Tau~B	&$	7.147	\pm	2.149	$&$	-28.765	\pm	2.186	$&	\nodata			&	\nodata			&$	7.924	\pm	1.334	$&$	126.2	_{	-18.2	}^{+	25.5	}$\\	
V773~Tau~A	&$	10.253	\pm	0.8434	$&$	-25.119	\pm	0.301	$&	\nodata			&	\nodata			&$	7.692	\pm	0.085	$&$	130.0	_{	-1.4	}^{+	1.5	}$\\	
V1098~Tau	&$	11.148	\pm	0.175	$&$	-27.327	\pm	0.172	$&	\nodata			&	\nodata			&$	8.070	\pm	0.310	$&$	123.9	_{	-4.6	}^{+	5.0	}$\\	2MASS~J04182909+2826191	&$	8.384	\pm	0.195	$&$	-19.627	\pm	0.217	$&	\nodata			&	\nodata			&$	7.583	\pm	0.389	$&$	131.9	_{	-6.4	}^{+	7.1	}$\\	
HD~283518	&$	8.703	\pm	0.017	$&$	-24.985	\pm	0.020	$&	\nodata			&	\nodata			&$	7.751	\pm	0.027	$&$	129.0	_{	-0.5	}^{+	0.5	}$\\	
V1023~Tau	&$	8.371	\pm	0.020	$&$	-25.490	\pm	0.020	$&	\nodata			&	\nodata			&$	7.686	\pm	0.032	$&$	130.1	_{	-0.5	}^{+	0.5	}$\\	
T~Tau~Sb	&$	6.790	\pm	0.432	$&$	-11.131	\pm	0.444	$&	\nodata			&	\nodata			&$	6.723	\pm	0.046	$&$	148.7	_{	-1.0	}^{+	1.0	}$\\	
V1201~Tau~B	&$	10.839	\pm	0.050	$&$	-13.235	\pm	0.058	$&$	0.335	\pm	0.076	$&$	0.147	\pm	0.071	$&$	6.363	\pm	0.069	$&$	157.2	_{	-1.7	}^{+	1.7	}$\\
HD~283641	&$	10.913	\pm	0.037	$&$	-16.772	\pm	0.044	$&	\nodata			&	\nodata			&$	6.285	\pm	0.070	$&$	159.1	_{	-1.8	}^{+	1.8	}$\\	
XZ~Tau~A	&$	10.858	\pm	0.027	$&$	-16.264	\pm	0.060	$&	\nodata			&	\nodata			&$	6.793	\pm	0.025	$&$	147.2	_{	-0.5	}^{+	0.5	}$\\	
V807~Tau~B	&$	8.573	\pm	0.068	$&$	-28.774	\pm	0.201	$&	\nodata			&	\nodata			&$	7.899	\pm	0.105	$&$	126.6	_{	-1.7	}^{+	1.7	}$\\	
V1110~Tau	&$	-52.705	\pm	0.062	$&$	-11.321	\pm	0.066	$&	\nodata			&	\nodata			&$	11.881	\pm	0.149	$&$	84.2	_{	-1.0	}^{+	1.1	}$\\	
HP~Tau~G2	&$	11.248	\pm	0.022	$&$	-15.686	\pm	0.013	$&	\nodata			&	\nodata			&$	6.145	\pm	0.029	$&$	162.7	_{	-0.8	}^{+	0.8	}$\\	
V999~Tau	&$	9.533	\pm	0.218	$&$	-15.684	\pm	0.198	$&	\nodata			&	\nodata			&$	6.972	\pm	0.197	$&$	143.4	_{	-3.9	}^{+	4.2	}$\\	
V1000~Tau	&$	6.010	\pm	0.235	$&$	-17.720	\pm	0.159	$&	\nodata			&	\nodata			&$	7.324	\pm	0.132	$&$	136.5	_{	-2.4	}^{+	2.5	}$\\	
HD~282630	&$	3.897	\pm	0.113	$&$	-24.210	\pm	0.132	$&	\nodata			&	\nodata			&$	7.061	\pm	0.125	$&$	141.6	_{	-2.5	}^{+	2.6	}$\\	HDE~283572	&$	8.853	\pm	0.096	$&$	-26.491	\pm	0.113	$&	\nodata			&	\nodata			&$	7.722	\pm	0.057	$&$	129.5	_{	-0.9	}^{+	1.0	}$\\	
\enddata
\end{deluxetable*}


\begin{longrotatetable}
\startlongtable
\begin{deluxetable*}{lcccccccccccccc}
\tablewidth{300pt}
\tabletypesize{\scriptsize}
\tablecaption{Orbital elements of the binaries and multiple systems in our sample.  \label{tab6}}
\tablehead{
\colhead{System}&
\colhead{$a_{1}$}&
\colhead{$a_{2}$}&
\colhead{$a$}&
\colhead{P}&
\colhead{e}&
\colhead{$T_{p}$}&
\colhead{$\Omega$}&
\colhead{$\omega$}&
\colhead{i}&
\colhead{$M_{Total}$}&
\colhead{$m_{1}$}&
\colhead{$m_{2}$}\\
\colhead{}&
\colhead{(mas)}&
\colhead{(mas)}&
\colhead{(mas)}&
\colhead{(yr)}&
\colhead{}&
\colhead{(JD)}&
\colhead{($^{\circ}$)}&
\colhead{($^{\circ}$)}&
\colhead{($^{\circ}$)}&
\colhead{($M_{\odot}$)}&
\colhead{($M_{\odot}$)}&
\colhead{($M_{\odot}$)}
}
\startdata
\hline
\multicolumn{15}{l}{\textbf{V773~Tau~A-B}}\\
\hline
Full model&$	44.3	\pm	5.0	$&	\nodata	&	\nodata	&$	20.301	\pm	0.842	$&$	
0.003	\pm	0.012	$&$	2459016	\pm	554	$&$	285.8	\pm	0.5	$&$	94.3	\pm	25.7	$&$	72.2	\pm	0.9	$&	\nodata	&	\nodata	&	\nodata	&\\	
\hline
\multicolumn{15}{l}{\textbf{V1023~Tau~A-B}}\\
\hline
Full model	&$	17.7	\pm	0.3	$&$	30.9	\pm	0.7	$&$	48.6	\pm	0.7	$&$	9.302	\pm	0.045	$&$	0.686	\pm	0.011	$&$	2454692	\pm	17	$&$	82.5	\pm	3.0	$&$	84.0	\pm	2.5	$&$	143.8	\pm	1.8	$&$	2.918	\pm	0.102	$&$	1.855	\pm	0.067	$&$	1.063	\pm	0.049	$&\\	
Relative model	&	\nodata	&	\nodata	&$	41.8	\pm	0.1	$&$	9.301	\pm	0.008	$&$	0.680	\pm	0.002	$&$	2454705	\pm	2	$&$	64.2	\pm	1.8	$&$	67.1	\pm	1.7	$&$	158.4	\pm	0.8	$&$	1.856	\pm	0.005	$&	\nodata	&	\nodata	&\\	
Joint fit	&$	16.0	\pm	0.3	$&$	27.0	\pm	0.3	$&$	43.0	\pm	0.4	$&$	9.329	\pm	0.017	$&$	0.682	\pm	0.003	$&$	2454712	\pm	3	$&$	66.1	\pm	2.3	$&$	70.0	\pm	2.2	$&$	153.8	\pm	1.2	$&$	1.964	\pm	0.033	$&$	1.234	\pm	0.023	$&$	0.730	\pm	0.020	$&\\	
\hline
\multicolumn{15}{l}{\textbf{T~Tau~Sa-Sb}}\\
\hline
Full model	&	\nodata	&$	60.4	\pm	2.2	$&	\nodata	&$	24.744	\pm	0.816	$&$	0.548	\pm	0.031	$&$	2459439	\pm	104	$&$	90.5	\pm	10.8	$&$	45.9	\pm	14.7	$&$	30.1	\pm	3.9	$&	\nodata	&	\nodata	&	\nodata	&\\	
Relative model	&	\nodata	&	\nodata	&$	83.3	\pm	1.9	$&$	26.946	\pm	0.858	$&$	0.546	\pm	0.031	$&$	2459907	\pm	240	$&$	102.0	\pm	39.2	$&$	36.0	\pm	38.3	$&$	13.4	\pm	8.1	$&$	2.620	\pm	0.208	$&	\nodata	&	\nodata	&\\
Joint fit	&$	15.7	\pm	3.0	$&$	70.0	\pm	4.4	$&$	85.7	\pm	5.3	$&$	27.933	\pm	1.156	$&$	0.514	\pm	0.041	$&$	2460214	\pm	333	$&$	112.0	\pm	28.8	$&$	28.7	\pm	29.3	$&$	22.6	\pm	8.9	$&$	2.660	\pm	0.490	$&$	2.172	\pm	0.408	$&$	0.489	\pm	0.134	$&\\		
\hline
\multicolumn{15}{l}{\textbf{V807~Tau~ Ba-Bb}}\\
\hline
Full model&$	17.8	\pm	1.4	$&$	21.5	\pm	2.0	$&$	39.3	\pm	2.4	$&$	12.025	\pm	0.397	$&$	0.339	\pm	0.057	$&$	2456013	\pm	150	$&$	2.7	\pm	10.6	$&$	70.5	\pm	5.7	$&$	146.9	\pm	4.9	$&$	0.849	\pm	0.136	$&$	0.465	\pm	0.078	$&$	0.385	\pm	0.080	$&\\	
Relative model	&	\nodata	&	\nodata	&$	38.6	\pm	0.2	$&$	12.310	\pm	0.060	$&$	0.293	\pm	0.003	$&$	2455732	\pm	7	$&$	1.5	\pm	1.2	$&$	50.4	\pm	1.1	$&$	151.2	\pm	0.9	$&$	0.768	\pm	0.022	$&	\nodata	&	\nodata	&\\	
Joint fit	&$	16.8	\pm	0.3	$&$	22.0	\pm	0.4	$&$	38.8	\pm	0.5	$&$	12.218	\pm	0.050	$&$	0.299	\pm	0.004	$&$	2455738	\pm	7	$&$	3.1	\pm	1.9	$&$	53.3	\pm	1.7	$&$	152.0	\pm	1.0	$&$	0.896	\pm	0.015	$&$	0.507	\pm	0.010	$&$	0.388	\pm	0.013	$&\\	
\hline
\multicolumn{15}{l}{\textbf{HP~Tau~G2-G3}}\\
\hline
Full model&$	12.6	\pm	0.2	$&	\nodata	&	\nodata	&$	11.932	\pm	0.102	$&$	0.691	\pm	0.010	$&$	2456794	\pm	7	$&$	118.4	\pm	1.0	$&$	263.8	\pm	0.8	$&$	46.4	\pm	0.7	$&	\nodata	&	\nodata	&	\nodata	&\\	
\hline
\multicolumn{15}{l}{\textbf{V1000~Tau~A-B}}\\
\hline
Full model	&$	8.3	\pm	0.4	$&$	10.1	\pm	0.5	$&$	18.4	\pm	0.6	$&$	3.616	\pm	0.257	$&$	0.428	\pm	0.050	$&$	2458085	\pm	9	$&$	264.1	\pm	8.2	$&$	268.5	\pm	3.0	$&$	45.6	\pm	3.0	$&$	1.213	\pm	0.181	$&$	0.663	\pm	0.101	$&$	0.550	\pm	0.091	$&\\	\enddata
\end{deluxetable*}
\end{longrotatetable}

\subsection{V807~Tau}

V807~Tau is a hierarchical triple system and the secondary component was resolved by \citet{Simon1995} into two close companions (V807~Tau~Ba/Bb). The secondary has been detected in 5 epochs during the GOBELINS observing campaign. V807~Tau was also observed with the VLBA in the past (from March 2007 to March 2009) by \citet{Schaefer2012} for projects BS171 and BS176. They report on 3 detections of V807~Tau~Ba and one detection of V807~Tau~Bb. Before combining these data with our own observations we decided to download the files from the NRAO archive and reduce them by applying the same calibration procedure used in this work. Both data sets observed  J0426+2327 as the main phase calibrator so that no correction needs to be applied to the positions measured in projects BS171 and BS176. We found that the source detected in the GOBELINS observations corresponds to V807~Tau~Ba. 

The model including a uniform acceleration produces a poor fit to the measured positions for V807~Tau~Ba, because our observations cover almost one full orbit of the Ba-Bb system. We thus performed a full fit including the orbital motion of the close pair in our analysis which indeed represents a better description to the data (see Fig.~\ref{fig5}). The distance that we derive from VLBA observations is $d=126.6\pm1.7$~pc. Then, we used the NIR relative positions for the Ba-Bb system obtained by \citet{Schaefer2012} together with the VLBA observation from project BS176A where both components were simultaneously detected to refine the orbital parameters of the system from the relative model (see Fig.~\ref{relFIT}). By combining the VLBA and NIR data we find a distance of $d=131.8^{+2.4}_{-2.3}$~pc. The larger discrepancy in the distance estimates delivered by the full model and the joint fit (as compared to the other stars in Table~\ref{tab7}) can be explained by the smaller number of data points (i.e., stellar positions) to fit the astrometry. The weighted mean parallax from both methods yields $d=128.5\pm1.4$~pc which is the first distance estimate for V807~Tau~B to date. 

The dynamical masses of the individual components ($m_{Ba}=0.507\pm 0.010$~$M_{\odot}$ and  $m_{Bb}=0.388\pm 0.013$~$M_{\odot}$) that we derive in this paper are more precise than the results obtained by \citet{Schaefer2012}. As discussed in their study, they used the average distance of $140\pm10$~pc to the Taurus region to compute the stellar masses. We have re-scaled the individual masses reported in their work to the distance that we derive in this paper which gives  $m_{Ba}=0.471\pm 0.018$~$M_{\odot}$ and  $m_{Bb}=0.397\pm 0.017$~$M_{\odot}$. However, these numbers are still affected by the systematic error of $0.24M_{\odot}$ which comes from the $\pm10$~pc uncertainty in the distance used in their analysis. Both results are still in good agreement, but we argue that the dynamical masses derived in this paper are more accurate due to the improved accuracy and precision of our distance determination. 

\subsection{V1110~Tau}\label{sectionV1110Tau}

V1110~Tau has been detected in 4 epochs during our observing campaign. The trigonometric parallax ($\pi=11.881\pm0.149$~mas) and proper motion ($\mu_{\alpha}\cos\delta=-52.705\pm0.062$~mas/yr, $\mu_{\delta}=-11.321\pm0.066$~mas/yr) that we derive here clearly confirm it as a foreground star not related to the Taurus star-forming clouds. This is also shown in Fig.~\ref{fig2} where V1110~Tau clearly stands out with a position change rate $>50$~mas/yr. In addition, Gaia-DR1 also confirms this finding yielding a trigonometric parallax of $\pi=12.53\pm 0.61$~mas and proper motion of $\mu_{\alpha}\cos\delta=-54.471\pm1.893$~mas/yr, $\mu_{\delta}=-12.194\pm1.612$~mas/yr. These values are consistent with but are less precise than our results.    

Interestingly, \citet{Martin1994} observed V1110~Tau (Wa~Tau~1) and did not detect the Li~I line in any of the two components of this binary system. \citet{Wahhaj2010} classified V1110~Tau as a weak-line T~Tauri star of spectral type K0 and effective temperature of 5250~K. They used the distance of 145~pc to compute the luminosity of the star (3.04~$L_{\odot}$) and estimate its age (7.3~Myr) based on the \citet{Siess2000} models. We have re-scaled the luminosity of the star to correct for the individual distance ($d=84.2^{+1.1}_{-1.0}$~pc) that we derived in this paper. This yields a luminosity of 1.03~$L_{\odot}$ and age estimate of 24~Myr suggesting that it is much older as indicated in previous studies. More recently, \citet{Xing2010} derived a spectral type of K0 and measured the Li equivalent width of only 39~m{\AA}. In addition, \citet{Xing2012} detected H$\alpha$ in absorption and \citet{Rebull2010} found no significant infrared excess for this source based on Spitzer photometry. Altogether, these properties are consistent with V1110~Tau being a young foreground dwarf not related to the Taurus population of YSOs \citep[see also][]{Briceno1997}. 

\begin{deluxetable*}{lcccc}
\tablecaption{Comparison of the astrometry derived for binaries from different methods. \label{tab7}}
\tablehead{
\colhead{Method}&
\colhead{$\mu_{\alpha}\cos\delta$}&
\colhead{$\mu_{\delta}$}&
\colhead{Parallax}&
\colhead{Distance}\\
\colhead{}&
\colhead{(mas/yr)}&
\colhead{(mas/yr)}&
\colhead{(mas)}&
\colhead{(pc)}
}
\startdata
\hline
\multicolumn{5}{c}{\textbf{V1023~Tau}}\\
\hline
Full model&$	8.371	\pm	0.020	$&$	-25.490	\pm	0.020	$&$	7.686	\pm	0.032	$&$	130.1	^{+	0.5	}_{	-0.5	}$\\
Joint fit	&$	8.393	\pm	0.019	$&$	-25.489	\pm	0.026	$&$	7.752	\pm	0.116	$&$	129.0	^{+	2.0	}_{	-1.9	}$\\
Weighted mean	&$	8.382	\pm	0.014	$&$	-25.490	\pm	0.016	$&$	7.691	\pm	0.031	$&$	130.0	^{+	0.5	}_{	-0.5	}$\\
\hline
\multicolumn{5}{c}{\textbf{T~Tau~Sb}}\\
\hline
Full model	&$	4.674	\pm	0.310	$&$	-10.224	\pm	0.212	$&$	6.743	\pm	0.096	$&$	148.3	^{+	2.1	}_{	-2.1	}$\\
Joint fit&$	4.577	\pm	0.272	$&$	-11.037	\pm	0.250	$&$	6.723	\pm	0.046	$&$	148.7	^{+	1.0	}_{	-1.0	}$\\
Weighted mean	&$	4.619	\pm	0.205	$&$	-10.564	\pm	0.162	$&$	6.727	\pm	0.041	$&$	148.7	^{+	0.9	}_{	-0.9	}$\\
\hline
\multicolumn{5}{c}{\textbf{V807~Tau~B}}\\
\hline
Full model&$	8.573	\pm	0.068	$&$	-28.774	\pm	0.201	$&$	7.899	\pm	0.105	$&$	126.6	^{+	1.7	}_{	-1.7	}$\\
Joint fit	&$	8.544	\pm	0.039	$&$	-28.871	\pm	0.052	$&$	7.588	\pm	0.135	$&$	131.8	^{+	2.4	}_{	-2.3	}$\\
Weighted mean	&$	8.551	\pm	0.034	$&$	-28.865	\pm	0.050	$&$	7.782	\pm	0.083	$&$	128.5	^{+	1.4	}_{	-1.4	}$\\
\enddata
\end{deluxetable*}

\subsection{HP~Tau~G2}

HP~Tau~G2 is a weak-line T Tauri star that belongs to a hierarchical triple system with separation of about $10\arcsec$ from HP~Tau~G3 which is a tight binary system \citep[see e.g.][]{Nguyen2012,Harris2012}. This source has been detected in a total of 9 epochs in this work, and it was also observed in the past from September 2003 to July 2005 for projects BL118 and BL124 (see Table~\ref{tab3}). We corrected the measured positions in these observations before combining them with the more recent GOBELINS data reported in this work. Projects BL118 and BL124 observed J0426+2327 as the main phase calibrator while the GOBELINS observations use it as a secondary calibrator and J0438+2153 as the main calibrator. The mean position of J0426+2327 measured between September 2003 and July 2005 is  $\alpha=04^{h}26^{m}55.734795^{s}$, $\delta=23^{\circ}27\arcmin 39.63371\arcsec$, and the mean position of J0426+2327 (relative to J0438+2153) in the GOBELINS observations is $\alpha=04^{h}26^{m}55.734757^{s}$, $\delta=23^{\circ}27\arcmin 39.63403\arcsec$. Thus, the corresponding offset to correct the measured positions in projects BL118 and BL124 is $\Delta\alpha=-0.000038$~s and $\Delta\delta=0.00032\arcsec$. After combining the two data sets, we performed a full fit to solve for the parallax, proper motion and orbital motion of the HP~Tau~G2-G3 system (see Tables \ref{tab5} and \ref{tab6}). The distance that we derive in this paper ($d=162.7\pm0.8$~pc) is in good agreement with the result of $d=161.2\pm0.9$~pc obtained by \citet{Torres2009} based only on observations collected for projects BL118 and BL124. This confirms that HP~Tau~G2 is indeed farther than other Taurus stars.

\subsection{V1000~Tau}

V1000~Tau has been detected in 7 epochs during our observing campaign, and both components were simultaneously detected in 5 epochs. On March 14, 2017 (project BL175JW) we measured the highest flux density for the primary component ($F_{\nu}=0.96\pm 0.05$~mJy). However, both sources appear significantly fainter in the other observations reported in this work with a flux level below $7\sigma$ which increases the errors on the individual positions of the sources. 

During the calibration process of our observations we noted that two main calibrators have been used for this source during our observing campaign. Projects BL175AE, BL175AO, BL175HD, BL175IC, BL175JW and BL175KV observed J0438+2153 as main phase calibrator while projects BL175D2, BL175I7, BL175IZ, BL175JY and BL175KS used J0429+2724 in this regard. Thus, we corrected the measured positions in the latter projects before combining the two data sets. J0435+2532 was observed as a secondary calibrator in all projected listed before. The mean position of J0435+2532 (relative to J0438+2153) in the first data set is $\alpha=04^{h}35^{m}34.582910^{s}$, $\delta=25^{\circ}32\arcmin 59.69919\arcsec$, and the mean position of J0435+2532 (relative to J0429+2724) in the second data set is  $\alpha=04^{h}35^{m}34.582963^{s}$, $\delta=25^{\circ}32\arcmin 59.69972\arcsec$. Thus, we applied an offset of $\Delta\alpha=-0.000053$~s and $\Delta\delta=-0.00052\arcsec$ to the measured position of the second data set. We applied the same correction to the positions measured for V999~Tau (see Sect.~\ref{sect_V999TAU}) because both sources were observed in the same field and used the same calibrators in all epochs. 
  
We proceeded as follows to calculate the distance to the V1000~Tau system. First, we applied the model with a uniform acceleration to the individual components of the system but it produced a poor fit to the data. We then considered the model including the orbital motion of the system and solved simultaneously for the proper motion, parallax and orbital elements. Doing so, we find a distance of $d=136.5_{-2.4}^{+2.5}$~pc. This approach also allowed us to derive the individual masses of the two components of this system ($m_{A}=0.663\pm 0.101$~M$_{\odot}$ and $m_{B}=0.550\pm 0.091$~M$_{\odot}$). To further investigate our result, we compute the barycenter of the V1000~Tau system using the individual masses derived in this paper and perform a fit including the acceleration terms in our equations. This approach yields a distance estimate of $d=136.4^{+3.4}_{-3.2}$~pc that is in good agreement with our previous result and confirms our solution despite the low detection threshold.  
 
\subsection{V892~Tau}

V892~Tau is the only source in our sample with a minimum of 3 detections for which we do not provide a trigonometric parallax. Unlike other stars in our sample, V892~Tau appears as a faint source in our observations with a detection level of roughly $6\sigma$ in the best case. V892~Tau is a Herbig Ae/Be star with a low-mass companion \citep{Leinert1997}, and at this stage it is not clear whether the positions measured in this work refer to the same component of the system since they provide a poor astrometric fit.

\subsection{HDE~283572}

HDE~283572 was not included in our target list, but it has been observed by \citet{Torres2007} between September 2004 and December 2005. We have re-calibrated and re-analyzed these data by applying the same methods used for the GOBELINS observations and described throughout this paper. The new distance estimate of $d=129.5^{+1.0}_{-0.9}$~pc that we derive here is fully consistent with the previous result of $d=128.5\pm0.6$~pc obtained by  \citet{Torres2007}.


\section{Discussion} \label{section5}

\subsection{Comparison with Gaia-DR1}\label{section5.1}

The first step in our analysis to compare the results obtained in this paper with Gaia-DR1 is to build a list of Taurus stars that have been previously identified in the literature. In a recent study, \citet{Joncour2017} published an updated census with 338 stars (and stellar systems) in this region. We add V1201~Tau, HD~283641 and V1110~Tau to their list which have been investigated in this work and were not included in their compilation.  

\begin{figure}[!h]
\gridline{
\fig{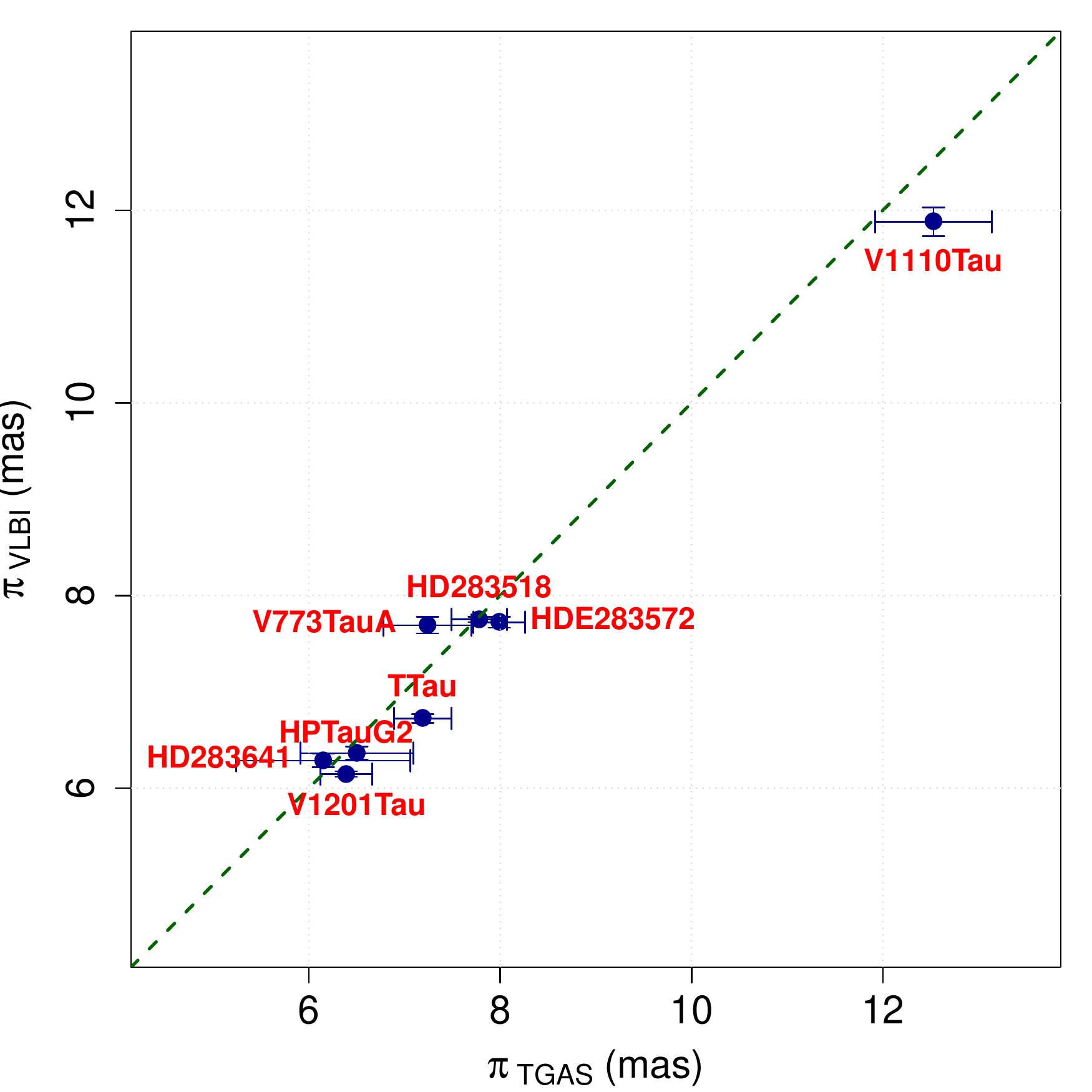}{0.47\textwidth}{}
}
\caption{Comparison of the trigonometric parallaxes derived in this paper with the results delivered by the TGAS catalog from Gaia-DR1 for the stars in common. The green dashed line indicates perfect correlation of both data sets. 
\label{fig_compPLX}}
\end{figure}

The Gaia satellite observed 204~stars in our list of known YSOs in the Taurus region but the \textit{Tycho-Gaia Astrometric Solution} \citep[TGAS,][]{Lindegren2016}  catalog from Gaia-DR1 provides trigonometric parallaxes for only 18 stars (and stellar systems). Eight stars are in common with our sample of 18 stars with measured trigonometric parallaxes (see Table~\ref{tab5}) including V1110~Tau which is more likely a foreground dwarf (see Sect.~\ref{sectionV1110Tau}). Figure~\ref{fig_compPLX} illustrates the comparison of our results with Gaia-DR1. The rms and mean difference between the trigonometric parallaxes derived in both projects are, respectively, 0.38~mas and -0.15~mas (in the sense, ``GOBELINS" minus ``TGAS"). These numbers are smaller than the mean error of $\sigma_{\pi}=0.46$~mas on the trigonometric parallaxes from Gaia-DR1 in the Taurus region. Thus, our results are in good agreement with Gaia-DR1, but the trigonometric parallaxes derived in this work are more precise than the ones given in the TGAS catalog. For example, we measured a trigonometric parallax of $\pi=7.751\pm0.027$~mas for HD~283518 that lies exactly onto the equality line (see Fig.~\ref{fig_compPLX}) and is more precise by almost one order of magnitude than the result of $\pi=7.78\pm0.29$~mas delivered by Gaia-DR1. This confirms the state-of-the-art accuracy and precision that can be obtained from VLBI astrometry, and the good complementarity with the Gaia space mission. 

Two points regarding the comparison of our results with Gaia-DR1 are worth mentioning here. 
First, the trigonometric parallaxes in the TGAS catalog are affected by a systematic error of about 0.30~mas depending, for example, on the position and color of the stars \citep{Lindegren2016}. Thus, we added the value of 0.30~mas quadratically to the formal errors given in the TGAS catalog.  
The second and potentially more serious problem is that most of our sources in common with Gaia-DR1 are binaries (or multiple systems). In this context, it is important to mention that all sources in the TGAS catalog were modeled as single stars, so that the orbital motion in binaries is neglected. This explains the most discrepant results obtained for some sources in Fig.~\ref{fig_compPLX} (e.g. T~Tau~Sb and V773~Tau~A) where this analysis is of ultimate importance and can only be applied with long-term monitoring of the system under investigation. For the reasons mentioned above, we consider the trigonometric parallaxes from the GOBELINS project to be more accurate and precise than the ones given in Gaia-DR1 for the targets in common. 

\begin{figure*}[!]
\gridline{
\fig{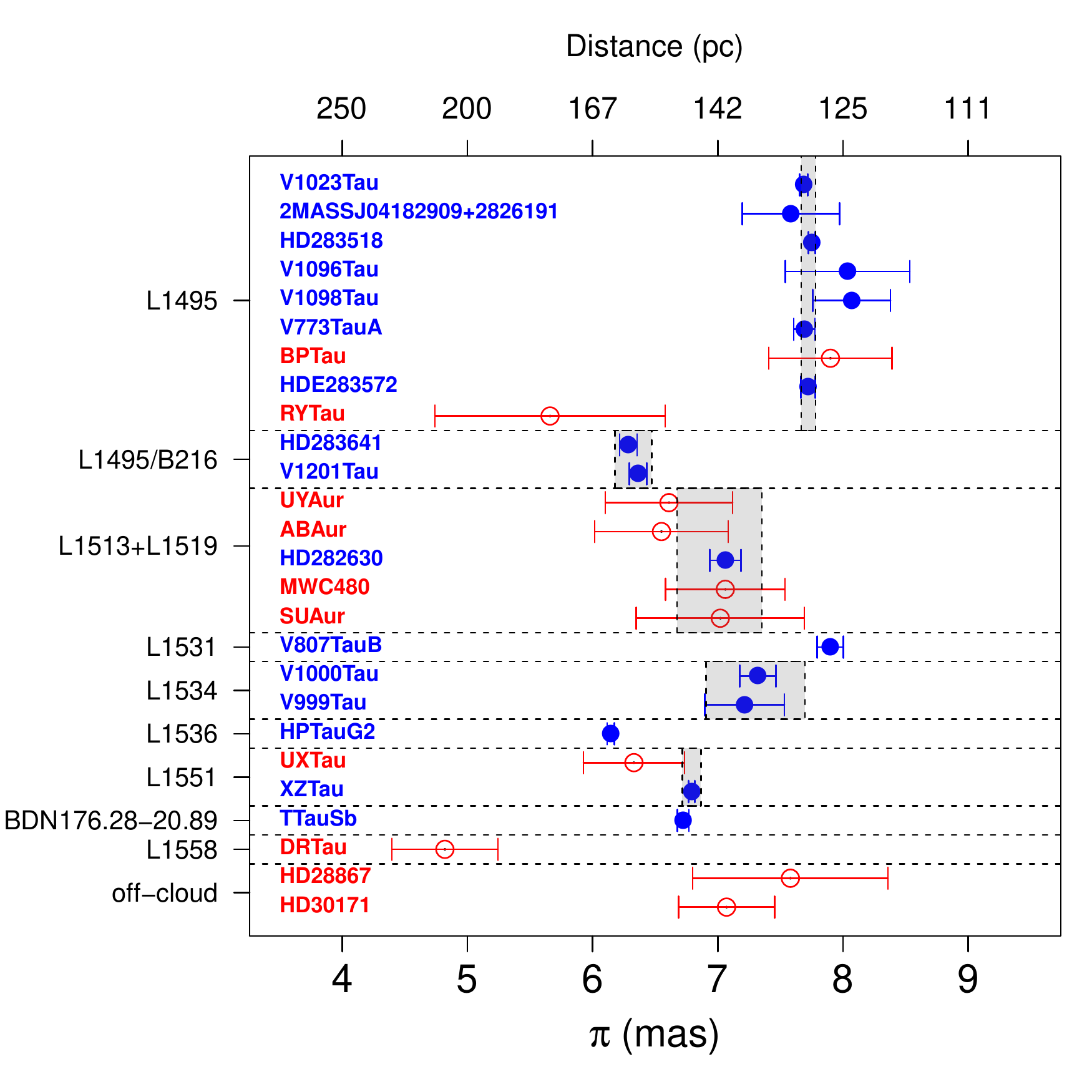}{0.7\textwidth}{}
}
\caption{Summary of the trigonometric parallax measurements for Taurus stars. The sources are grouped according to the star-forming cloud to which they belong. Filled and open symbols indicate, respectively, VLBI trigonometric parallaxes obtained in this paper and TGAS results from Gaia-DR1. The gray vertical bars indicate the weighted mean parallax (at the $3\sigma$-level) of the clouds with more than one member in our sample to illustrate the depth effects.}
\label{fig_summarydistances}
\end{figure*}

\subsection{Distance and spatial velocity of Taurus stars}\label{section5.2}

The effective sample of stars that we use in the forthcoming analysis to discuss the distance and kinematics of the Taurus star-forming region consists of 26 stars (or stellar systems). It includes all stars (or stellar systems) listed in Table~\ref{tab5} (excluding V1110~Tau) and the additional 10 sources with trigonometric parallaxes from Gaia-DR1 (see Sect.~\ref{section5.1}). V1110~Tau is not included in this discussion for the reasons presented in Sect.~\ref{sectionV1110Tau}. We use the proper motions and trigonometric parallaxes derived in this work from VLBI observations, and take the results from the TGAS catalog to complement our sample for stars that were not included in our observing campaign. In the case of V1096~Tau we use the weighted mean parallax and proper motion of the two components given in Sect.~\ref{sectionV1096Tau} that provide a better solution for the system than the individual values listed in Table~\ref{tab5}. Figure~\ref{fig_summarydistances} summarizes the trigonometric parallaxes of the stars in our sample, and Figures~\ref{fig_L1495}, \ref{fig_L1519}, \ref{fig_L1534} and \ref{fig_L1551} illustrate the structure in and around the various star-forming clouds investigated in our analysis based on the extinction maps from \citet{Dobashi2005}. 

We also searched the literature for the radial velocity of the stars in our sample using the data mining tools available in the CDS databases \citep{Wenger2000}. Our search for radial velocities is based on \citet{Hartmann1986}, \citet{Herbig1988}, \citet{Gontcharov2006}, \cite{Kharchenko2007} and \citet{Nguyen2012}. The properties of  GOBELINS and Gaia-DR1 stars in our sample are collectively listed in Table~\ref{tab8}. Then, we converted the observed trigonometric parallaxes into distances, and used the radial velocities to calculate the three-dimensional Galactic spatial velocities from the procedure described by \citet{Johnson1987}. In Table~\ref{tab9} we present the UVW spatial velocity and the peculiar velocity of individual stars in our sample after correcting for the velocity of the Sun with respect to the local standard of rest (LSR). For this correction we use the solar motion obtained by \citet{Schoenrich2010}. We also present in Table~\ref{tab9} the radial velocities of the stars converted to the LSR which will be used in the forthcoming discussion to compare with the velocity field of the CO molecular gas in this region produced by the Five College Radio Astronomy Observatory (FCRAO). In this case we used the older IAU standard solar motion to convert the radial velocities to the LSR for consistency with the FCRAO maps \citep[see][]{Jackson2006}. In the following we comment on the distance and kinematics of the individual clouds of the Taurus complex. The properties of the various star-forming clouds discussed below are also summarized in Table~\ref{tab10}. 

\subsubsection{Lynds 1495}\label{sectionL1495}

\begin{figure*}[!]
\gridline{
\fig{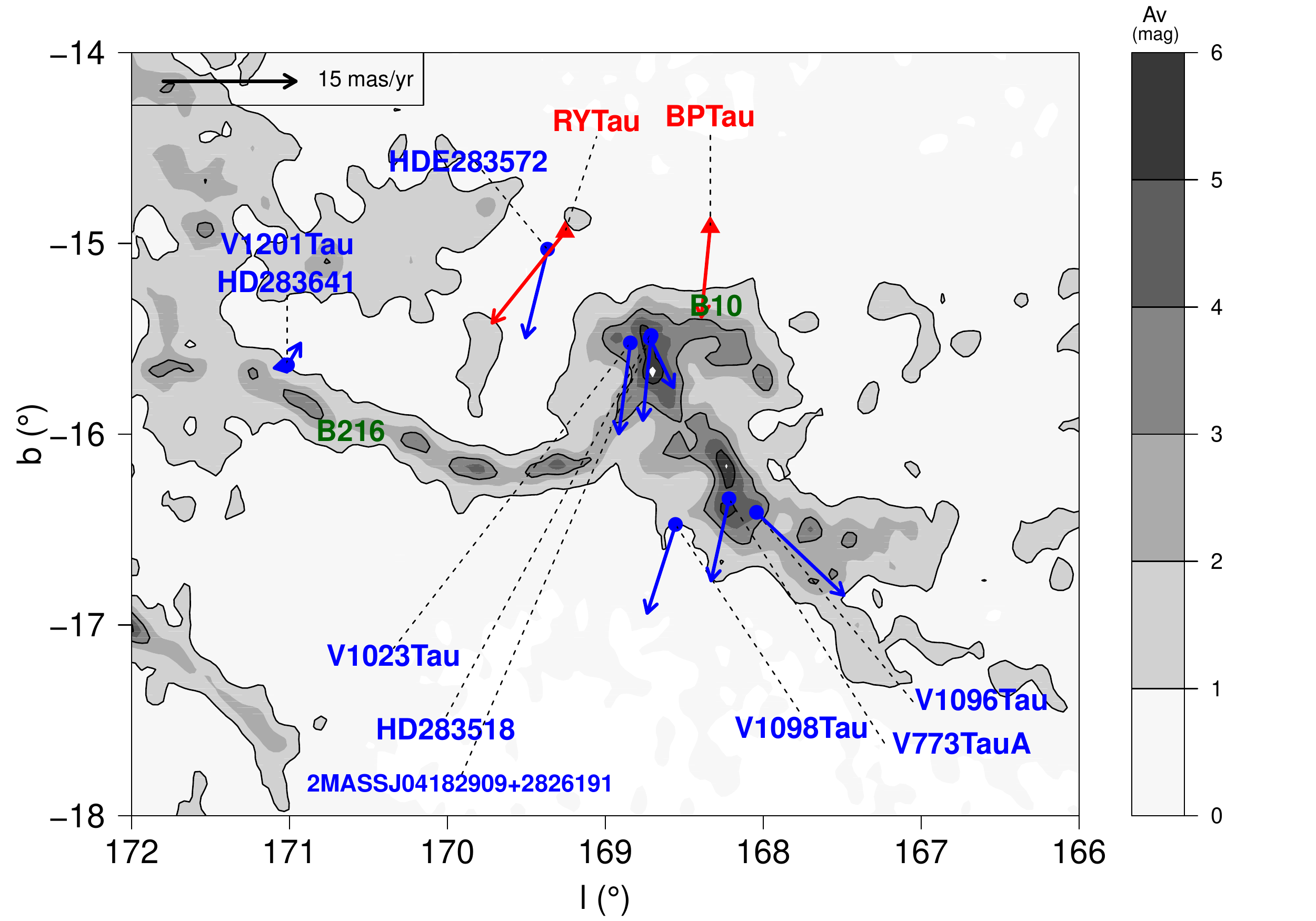}{0.65\textwidth}{}
}
\caption{Structure of the L~1495 cloud ($d=129.5^{+0.3}_{-0.3}$~pc) overlaid on the extinction map from \citet{Dobashi2005}. Blue circles and red triangles denote, respectively, the stars with VLBI and TGAS trigonometric parallax. The vectors indicate the stellar proper motions from Table~\ref{tab8} converted to the Galactic reference system and corrected for the Solar motion \citep{Schoenrich2010} using the formalism described by \citet{Abad2005}}.
\label{fig_L1495}
\end{figure*}

Lynds~1495 \citep[L1495,][]{Lynds1962} is the main star-forming site of the Taurus complex and the most important structure to discuss in this work, because it contains about 40\% of the stars in our sample. \citet{Schmalzl2010} divided it into five clumps (B~211, B~213, B~216, B~217 and B~218) which form the filament projected on the plane of the sky, and the central part B~10. We note that V1023~Tau, HD~283518 and 2MASS~J04182909+2826191 are located in the northern part of B~10 (see Fig.~\ref{fig_L1495}). The weighted mean parallax of these sources is $\pi=7.724\pm 0.021$~mas which is consistent with a distance estimate of $d=129.5^{+0.4}_{-0.3}$~pc. On the other hand, V1096~Tau, V1098~Tau and V773~Tau are projected towards the southern part of the B~10 clump with a weighted mean parallax of $\pi=7.727\pm 0.081$~mas and distance of $d=129.4^{+1.4}_{-1.3}$~pc. Thus, we conclude that both substructures are located at the same distance. As discussed in Sect.~4, the trigonometric parallaxes obtained in this paper for V1096~Tau, V1098~Tau and 2MASS~J04182909+28261 require further improvement, but based on the current results we find no evidence of significant depth effects within the B~10 clump of the L~1495 cloud. The weighted mean parallax of the six stars mentioned before is $\pi=7.724\pm 0.020$~mas. This is consistent with a distance estimate of $d=129.5^{+0.3}_{-0.3}$~pc which is the most precise and accurate present-day distance determination of L~1495.

Figure~\ref{fig_L1495} reveals 3 stars in our sample (HDE~283572, RY~Tau and BP~Tau) in the outskirts of L~1495. We note that the trigonometric parallaxes of HDE~283572 ($\pi=7.722\pm0.057$~mas) and BP~Tau ($\pi=7.900\pm0.492$~mas) are  consistent with the results obtained in this paper for the other stars in the B~10 clump. However, the trigonometric parallax of RY~Tau given in the TGAS catalog ($\pi=5.660\pm0.920$~mas) is in obvious disagreement with the other stars suggesting either a problem with the result delivered by Gaia-DR1 (e.g. non-corrected binarity or low precision/accuracy) or that RY~Tau is indeed not related to the L~1495 cloud. Moreover, we note that its radial velocity exceeds the mean radial velocity of the other stars in this region by about 9~km/s (see Table~\ref{tab8}). The revised weighted mean parallax of L~1495 (including HDE~283572 and BP~Tau) is $\pi=7.724\pm 0.019$~mas and yields a distance of $d=129.5^{+0.3}_{-0.3}$~pc confirming our previous result. 

One interesting finding of our analysis is that V1201~Tau and HD~283641, which are projected towards the B~216 clump in the filamentary structure of L~1495, are located at a different distance as compared to the other stars in L~1495. The mean parallax of these two sources is $\pi=6.325\pm 0.049$~mas and yields a distance of $d=158.1^{+1.2}_{-1.2}$~pc. This distance estimate differs by almost 30~pc from the result mentioned before and reveals the existence of important depth effects between the central part of the cloud and the filament. In addition, the spatial velocity that we derive for HD~283641 also suggests that both structures exhibit different kinematic properties (see Table~\ref{tab10}). 

\subsubsection{Lynds~1513 and 1519}

\begin{figure*}[!]
\gridline{
\fig{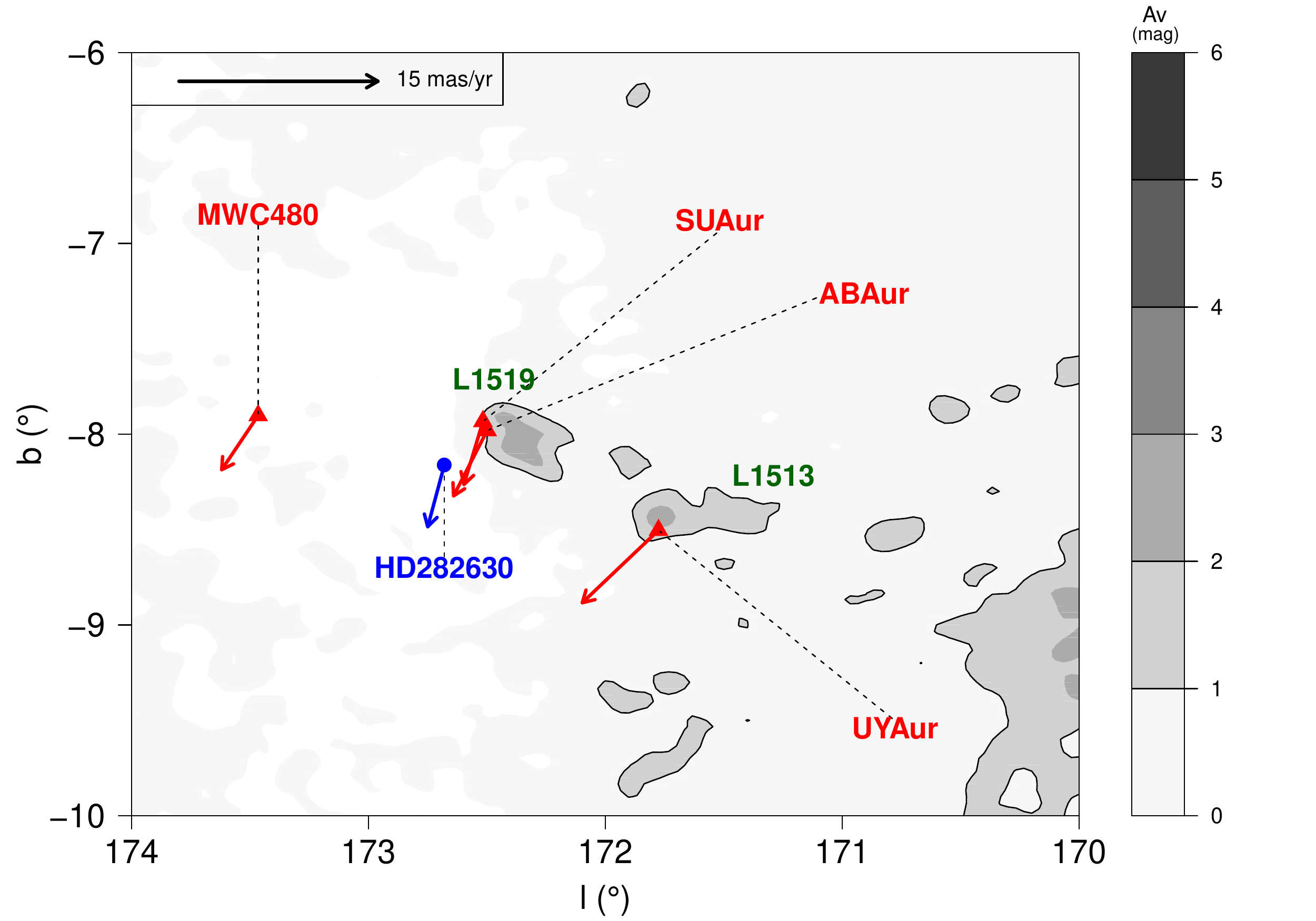}{0.65\textwidth}{}
}
\caption{Structure of the L~1513  ($d=151^{+13}_{-11}$~pc) and L~1519  ($d=142.1^{+2.4}_{-2.3}$~pc) clouds overlaid on the extinction map from \citet{Dobashi2005}. Blue circles and red triangles denote, respectively, the stars with VLBI and TGAS trigonometric parallax. The vectors indicate the stellar proper motions from Table~\ref{tab8} converted to the Galactic reference system and corrected for the Solar motion \citep{Schoenrich2010} using the formalism described by \citet{Abad2005}}.
\label{fig_L1519}
\end{figure*}

\begin{figure*}[!]
\gridline{
\fig{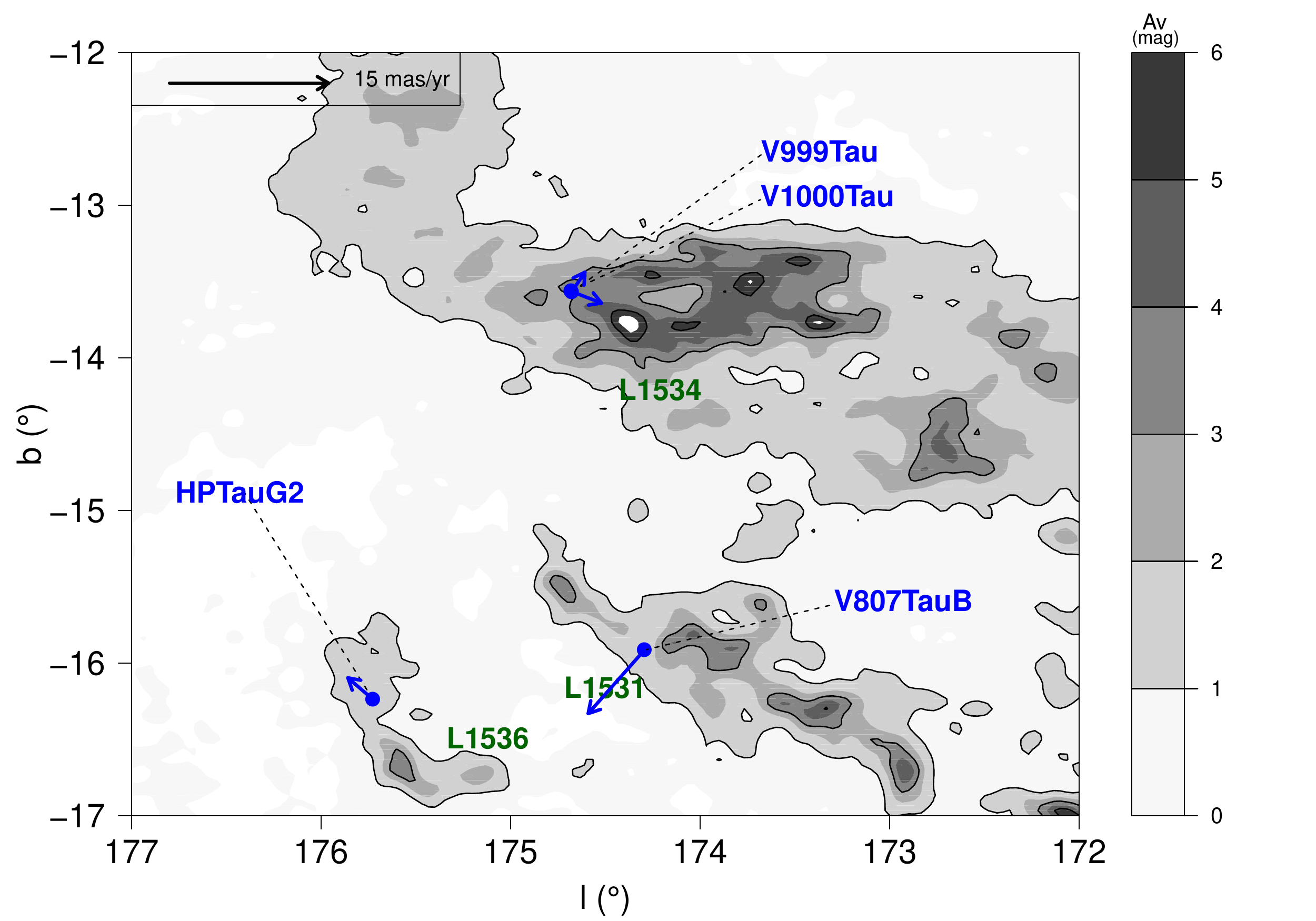}{0.65\textwidth}{}
}
\caption{Structure of the L~1531 ($d=126.6^{+1.7}_{-1.7}$~pc), L~1534  ($d=138.6^{+2.1}_{-2.1}$~pc) and L~1536  ($d=162.7^{+0.8}_{-0.8}$~pc) clouds overlaid on the extinction map from \citet{Dobashi2005}. Blue circles and red triangles denote, respectively, the stars with VLBI and TGAS trigonometric parallax. The vectors indicate the stellar proper motions from Table~\ref{tab8} converted to the Galactic reference system and corrected for the Solar motion \citep{Schoenrich2010} using the formalism described by \citet{Abad2005}}.
\label{fig_L1534}
\end{figure*}

UY~Aur is projected towards the Lynds~1513 cloud \citep[L1513,][]{Lynds1962} with a trigonometric parallax of $\pi=6.610\pm0.508$~mas. This result is in good agreement (at the $1\sigma$ level) with the trigonometric parallaxes of AB~Aur ($\pi=6.550\pm0.532$~mas) and SU~Aur ($\pi=7.020\pm0.671$~mas) which are located in Lynds~1519 \citep[L1519,][]{Lynds1962}. Another two stars (HD~282630 and MWC480) are located in the surroundings of L~1519 (see Fig.\ref{fig_L1519}) and their trigonometric parallaxes are also consistent with UY~Aur, AB~Aur and SU~Aur. The trigonometric parallaxes and spatial velocities of these sources are all consistent between themselves within the admittedly large errors. The weighted mean parallax is $\pi=7.014\pm 0.113$~mas and yields a distance of $d=142.6^{+2.3}_{-2.3}$~pc that we consider at this stage to be representative of both L~1513 and L~1519.

\begin{figure*}[!]
\gridline{
\fig{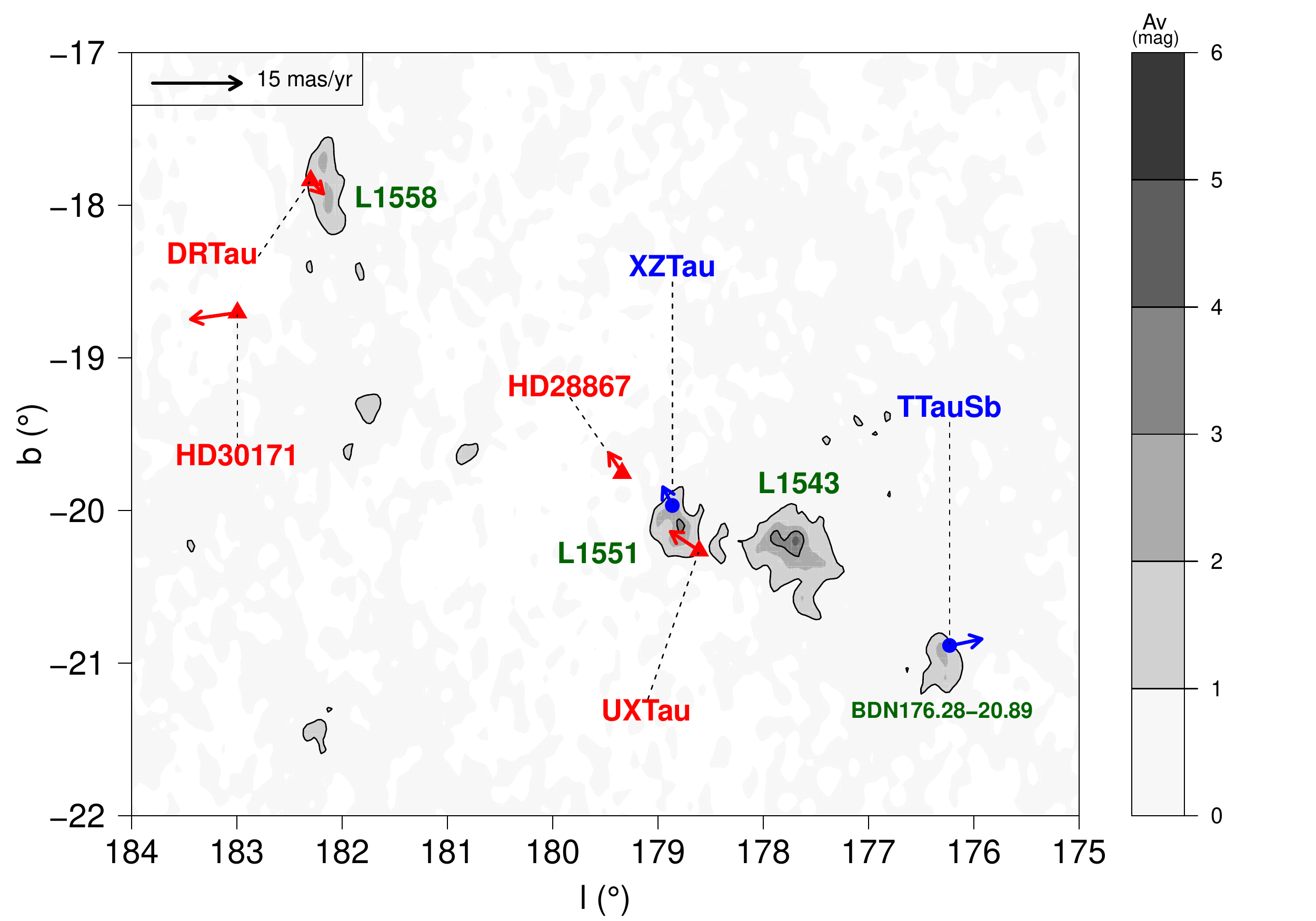}{0.65\textwidth}{}
}
\caption{Structure of the L~1551 ($d=147.3^{+0.5}_{-0.5}$~pc) and L~1558  ($d=208^{+20}_{-17}$~pc) clouds overlaid on the extinction map from \citet{Dobashi2005}. Blue circles and red triangles denote, respectively, the stars with VLBI and TGAS trigonometric parallax. The vectors indicate the stellar proper motions from Table~\ref{tab8} converted to the Galactic reference system and corrected for the Solar motion \citep{Schoenrich2010} using the formalism described by \citet{Abad2005}}.
\label{fig_L1551}
\end{figure*}

\clearpage
\begin{longrotatetable}
\startlongtable
\begin{deluxetable*}{llcccccccccccc}
\tablewidth{75pt}
\tabletypesize{\scriptsize}
\tablecaption{Proper motions, trigonometric parallaxes, distances and radial velocities of the GOBELINS-Gaia sample in Taurus.  \label{tab8}}
\tablehead{
\colhead{Star}&
\colhead{2MASS Identifier}&
\colhead{$\mu_{\alpha}\cos\delta$}&
\colhead{$\mu_{\delta}$}&
\colhead{$\pi$}&
\colhead{$d$}&
\colhead{Ref.}&
\colhead{$V_{r}$}&
\colhead{Ref.}\\
\colhead{}&
\colhead{}&
\colhead{(mas/yr)}&
\colhead{(mas/yr)}&
\colhead{(mas)}&
\colhead{(pc)}&
\colhead{}&
\colhead{(km/s)}&
\colhead{}&
}
\startdata
V1096~Tau & J04132722+2816247 &$ 2.612 \pm 0.691 $&$ -17.372 \pm 0.676 $&$ 8.037 \pm 0.497 $&$ 124.4 ^{+ 8.2 }_{ -7.2 }$& ThisWork &$ 12.00 \pm 5.00 $& 1 \\
V773~Tau~A & J04141291+2812124 &$ 10.253 \pm 0.843 $&$ -25.119 \pm 0.301 $&$ 7.692 \pm 0.085 $&$ 130.0 ^{+ 1.5 }_{ -1.4 }$& ThisWork &$ 16.00 \pm 2.50 $& 2 \\
HD~283518 & J04183110+2827162 &$ 8.703 \pm 0.017 $&$ -24.985 \pm 0.020 $&$ 7.751 \pm 0.027 $&$ 129.0 ^{+ 0.5 }_{ -0.4 }$& ThisWork &$ 19.90 \pm 0.30 $& 3 \\
V1023~Tau & J04184703+2820073 &$ 8.371 \pm 0.020 $&$ -25.490 \pm 0.020 $&$ 7.686 \pm 0.032 $&$ 130.1 ^{+ 0.5 }_{ -0.5 }$& ThisWork &$ 15.00 \pm 1.70 $& 4 \\
BP~Tau & J04191583+2906269 &$ 8.683 \pm 0.355 $&$ -25.846 \pm 0.239 $&$ 7.900 \pm 0.492 $&$ 126.6 ^{+ 8.4 }_{ -7.4 }$& Gaia-DR1 &$ 15.24 \pm 0.04 $& 3 \\
RY~Tau & J04215740+2826355 &$ 9.100 \pm 0.140 $&$ -25.859 \pm 0.091 $&$ 5.660 \pm 0.920 $&$ 176.7 ^{+ 34.3 }_{ -24.7 }$& Gaia-DR1 &$ 24.30 \pm 1.90 $& 2 \\
HDE~283572 & J04215884+2818066 &$ 8.853 \pm 0.096 $&$ -26.491 \pm 0.113 $&$ 7.722 \pm 0.057 $&$ 129.5 ^{+ 1.0 }_{ -0.9 }$& ThisWork &$ 14.22 \pm 0.08 $& 3 \\
T~Tau~Sb & J04215943+1932063 &$ 6.790 \pm 0.432 $&$ -11.131 \pm 0.444 $&$ 6.723 \pm 0.046 $&$ 148.7 ^{+ 1.0 }_{ -1.0 }$& ThisWork &$ 19.23 \pm 0.02 $& 3 \\
HD~283641 & J04244904+2643104  &$ 10.913 \pm 0.037 $&$ -16.772 \pm 0.044 $&$ 6.285 \pm 0.070 $&$ 159.1 ^{+ 1.8 }_{ -1.8 }$& ThisWork &$ 16.23 \pm 0.04 $& 3 \\
UX~Tau & J04300399+1813493 &$ 12.641 \pm 0.435 $&$ -17.552 \pm 0.257 $&$ 6.330 \pm 0.404 $&$ 158.0 ^{+ 10.8 }_{ -9.5 }$& Gaia-DR1 &$ 15.45 \pm 0.02 $& 3 \\
XZ~Tau & J04314007+1813571 &$ 10.858 \pm 0.027 $&$ -16.264 \pm 0.060 $&$ 6.793 \pm 0.025 $&$ 147.2 ^{+ 0.5 }_{ -0.5 }$& ThisWork &$ 18.30 \pm 0.04 $& 3 \\
V807~Tau~B & J04330664+2409549 &$ 8.573 \pm 0.068 $&$ -28.774 \pm 0.201 $&$ 7.899 \pm 0.105 $&$ 126.6 ^{+ 1.7 }_{ -1.7 }$& ThisWork &$ 16.85 \pm 0.03 $& 3 \\
HD~28867 & J04333297+1801004 &$ 12.017 \pm 0.047 $&$ -18.635 \pm 0.030 $&$ 7.580 \pm 0.780 $&$ 131.9 ^{+ 15.1 }_{ -12.3 }$& Gaia-DR1 &$ -14.70 \pm 7.40 $& 5 \\
HP~Tau~G2 & J04355415+2254134 &$ 11.248 \pm 0.022 $&$ -15.686 \pm 0.013 $&$ 6.145 \pm 0.029 $&$ 162.7 ^{+ 0.8 }_{ -0.8 }$& ThisWork &$ 16.60 \pm 1.00 $& 3 \\
V999~Tau & J04420548+2522562 &$ 9.533 \pm 0.218 $&$ -15.684 \pm 0.198 $&$ 6.972 \pm 0.197 $&$ 143.4 ^{+ 4.2 }_{ -3.9 }$& ThisWork&$ 14.70 \pm 2.00 $& 1 \\
HD~30171 & J04455129+1555496 &$ 9.824 \pm 1.675 $&$ -24.288 \pm 0.994 $&$ 7.070 \pm 0.384 $&$ 141.4 ^{+ 8.1 }_{ -7.3 }$& Gaia-DR1 &$ 21.13 \pm 0.17 $& 3 \\
DR~Tau & J04470620+1658428 &$ 1.372 \pm 2.747 $&$ -11.225 \pm 1.785 $&$ 4.820 \pm 0.424 $&$ 207.5 ^{+ 20.0 }_{ -16.8 }$& Gaia-DR1 &$ 21.10 \pm 0.04 $& 3 \\
UY~Aur & J04514737+3047134 &$ 6.099 \pm 2.824 $&$ -27.027 \pm 1.840 $&$ 6.610 \pm 0.508 $&$ 151.3 ^{+ 12.6 }_{ -10.8 }$& Gaia-DR1 &$ 13.92 \pm 0.07 $& 3 \\
HD~282630 & J04553695+3017553 &$ 3.897 \pm 0.113 $&$ -24.210 \pm 0.132 $&$ 7.061 \pm 0.125 $&$ 141.6 ^{+ 2.6 }_{ -2.5 }$& ThisWork &$ 13.58 \pm 0.01 $& 3 \\
AB~Aur & J04554582+3033043 &$ 3.889 \pm 0.056 $&$ -24.050 \pm 0.039 $&$ 6.550 \pm 0.533 $&$ 152.7 ^{+ 13.5 }_{ -11.5 }$& Gaia-DR1 &$ 8.90 \pm 0.90 $& 2 \\
SU~Aur & J04555938+3034015 &$ 3.857 \pm 0.126 $&$ -24.367 \pm 0.088 $&$ 7.020 \pm 0.671 $&$ 142.5 ^{+ 15.1 }_{ -12.4 }$& Gaia-DR1 &$ 14.26 \pm 0.05 $& 3 \\
\hline
V1098~Tau & J04144797+2752346 &$ 11.148 \pm 0.175 $&$ -27.327 \pm 0.172 $&$ 8.070 \pm 0.310 $&$ 123.9 ^{+ 5.0 }_{ -4.6 }$& ThisWork &\nodata&\nodata\\
2MASS~J04182909+2826191 & J04182909+2826191 &$ 8.384 \pm 0.195 $&$ -19.627 \pm 0.217 $&$ 7.583 \pm 0.389 $&$ 131.9 ^{+ 7.1 }_{ -6.4 }$& ThisWork &\nodata&\nodata\\
V1201~Tau & J04244815+2643161  &$ 10.839 \pm 0.050 $&$ -13.235 \pm 0.058 $&$ 6.363 \pm 0.069 $&$ 157.2 ^{+ 1.7 }_{ -1.7 }$& ThisWork &\nodata&\nodata\\
V1000~Tau & J04420732+2523032 &$ 6.010 \pm 0.235 $&$ -17.720 \pm 0.159 $&$ 7.324 \pm 0.132 $&$ 136.5 ^{+ 2.5 }_{ -2.4 }$& ThisWork &\nodata&\nodata\\
MWC480 & J04584626+2950370 &$ 4.790 \pm 0.081 $&$ -25.044 \pm 0.049 $&$ 7.060 \pm 0.476 $&$ 141.6 ^{+ 10.2 }_{ -9.0 }$& Gaia-DR1 &\nodata&\nodata\\
\enddata
\tablecomments{References for radial velocities: (1)~\citet{Herbig1988}; (2)~\citet{Gontcharov2006}; (3)~\citet{Nguyen2012}; (4)~\citet{Hartmann1986}; (5)~\citet{Kharchenko2007}.}
\end{deluxetable*}
\end{longrotatetable}


\begin{longrotatetable}
\startlongtable
\begin{deluxetable*}{llccccccccc}
\tablewidth{75pt}
\tabletypesize{\scriptsize}
\tablecaption{Spatial velocity for Taurus stars with measured trigonometric parallaxes and radial velocities.  \label{tab9}}
\tablehead{
\colhead{Star}&
\colhead{2MASS Identifier}&
\colhead{$U$}&
\colhead{$V$}&
\colhead{$W$}&
\colhead{$V_{space}$}&
\colhead{$u$}&
\colhead{$v$}&
\colhead{$w$}&
\colhead{$V_{pec}$}&
\colhead{$V_{r}^{LSR}$}\\
\colhead{}&
\colhead{}&
\colhead{(km/s)}&
\colhead{(km/s)}&
\colhead{(km/s)}&
\colhead{(km/s)}&
\colhead{(km/s)}&
\colhead{(km/s)}&
\colhead{(km/s)}&
\colhead{(km/s)}&
\colhead{(km/s)}\\
}
\colnumbers
\startdata
V1096Tau & J04132722+2816247 &$ -11.4 ^{+ 4.9 }_{ -4.9 }$&$ -6.3 ^{+ 2.0 }_{ -2.1 }$&$ -9.1 ^{+ 2.4 }_{ -2.4 }$&$ 15.9 ^{+ 3.9 }_{ -3.9 }$&$ -0.3 ^{+ 5.0 }_{ -5.0 }$&$ 6.0 ^{+ 2.0 }_{ -2.2 }$&$ -1.8 ^{+ 2.4 }_{ -2.5 }$&$ 6.2 ^{+ 2.1 }_{ -2.2 }$&$ 6.29 \pm 5.00 $\\
V773TauA & J04141291+2812124 &$ -16.5 ^{+ 2.6 }_{ -2.6 }$&$ -12.5 ^{+ 1.1 }_{ -1.1 }$&$ -10.3 ^{+ 1.3 }_{ -1.3 }$&$ 23.1 ^{+ 2.0 }_{ -2.0 }$&$ -5.4 ^{+ 2.7 }_{ -2.7 }$&$ -0.3 ^{+ 1.2 }_{ -1.2 }$&$ -3.1 ^{+ 1.4 }_{ -1.4 }$&$ 6.2 ^{+ 2.4 }_{ -2.4 }$&$ 10.27 \pm 2.50 $\\
HD283518 & J04183110+2827162 &$ -20.0 ^{+ 0.3 }_{ -0.3 }$&$ -11.2 ^{+ 0.1 }_{ -0.1 }$&$ -11.5 ^{+ 0.1 }_{ -0.1 }$&$ 25.6 ^{+ 0.2 }_{ -0.2 }$&$ -8.9 ^{+ 0.8 }_{ -0.8 }$&$ 1.1 ^{+ 0.5 }_{ -0.5 }$&$ -4.3 ^{+ 0.4 }_{ -0.4 }$&$ 9.9 ^{+ 0.7 }_{ -0.7 }$&$ 14.13 \pm 0.30 $\\
V1023Tau & J04184703+2820073 &$ -15.3 ^{+ 1.6 }_{ -1.6 }$&$ -12.4 ^{+ 0.4 }_{ -0.4 }$&$ -10.6 ^{+ 0.5 }_{ -0.5 }$&$ 22.3 ^{+ 1.2 }_{ -1.2 }$&$ -4.2 ^{+ 1.8 }_{ -1.8 }$&$ -0.1 ^{+ 0.6 }_{ -0.6 }$&$ -3.4 ^{+ 0.6 }_{ -0.6 }$&$ 5.4 ^{+ 1.4 }_{ -1.4 }$&$ 9.20 \pm 1.70 $\\
BPTau & J04191583+2906269 &$ -15.7 ^{+ 0.2 }_{ -0.2 }$&$ -12.0 ^{+ 1.1 }_{ -1.3 }$&$ -10.4 ^{+ 1.1 }_{ -1.2 }$&$ 22.4 ^{+ 0.8 }_{ -0.9 }$&$ -4.6 ^{+ 0.7 }_{ -0.8 }$&$ 0.2 ^{+ 1.2 }_{ -1.3 }$&$ -3.2 ^{+ 1.2 }_{ -1.2 }$&$ 5.6 ^{+ 0.9 }_{ -0.9 }$&$ 9.57 \pm 0.04 $\\
RYTau & J04215740+2826355 &$ -24.7 ^{+ 2.3 }_{ -2.4 }$&$ -16.9 ^{+ 3.4 }_{ -4.6 }$&$ -14.8 ^{+ 3.7 }_{ -4.1 }$&$ 33.4 ^{+ 2.9 }_{ -3.5 }$&$ -13.6 ^{+ 2.4 }_{ -2.5 }$&$ -4.7 ^{+ 3.4 }_{ -4.6 }$&$ -7.6 ^{+ 3.7 }_{ -4.1 }$&$ 16.3 ^{+ 2.8 }_{ -3.2 }$&$ 18.46 \pm 1.90 $\\
HDE283572 & J04215884+2818066 &$ -14.6 ^{+ 0.1 }_{ -0.1 }$&$ -13.2 ^{+ 0.2 }_{ -0.2 }$&$ -10.4 ^{+ 0.2 }_{ -0.2 }$&$ 22.3 ^{+ 0.2 }_{ -0.2 }$&$ -3.5 ^{+ 0.7 }_{ -0.8 }$&$ -1.0 ^{+ 0.5 }_{ -0.5 }$&$ -3.1 ^{+ 0.4 }_{ -0.4 }$&$ 4.8 ^{+ 0.6 }_{ -0.6 }$&$ 8.36 \pm 0.08 $\\
TTauSb & J04215943+1932063 &$ -18.0 ^{+ 0.2 }_{ -0.2 }$&$ -8.0 ^{+ 0.5 }_{ -0.5 }$&$ -8.1 ^{+ 0.5 }_{ -0.5 }$&$ 21.3 ^{+ 0.3 }_{ -0.3 }$&$ -6.9 ^{+ 0.7 }_{ -0.8 }$&$ 4.2 ^{+ 0.7 }_{ -0.7 }$&$ -0.8 ^{+ 0.6 }_{ -0.6 }$&$ 8.2 ^{+ 0.7 }_{ -0.7 }$&$ 11.83 \pm 0.02 $\\
HD283641 & J04244904+2643104  &$ -17.1 ^{+ 0.1 }_{ -0.1 }$&$ -12.5 ^{+ 0.2 }_{ -0.2 }$&$ -6.5 ^{+ 0.2 }_{ -0.2 }$&$ 22.2 ^{+ 0.1 }_{ -0.2 }$&$ -6.0 ^{+ 0.7 }_{ -0.8 }$&$ -0.2 ^{+ 0.5 }_{ -0.5 }$&$ 0.8 ^{+ 0.4 }_{ -0.4 }$&$ 6.1 ^{+ 0.7 }_{ -0.8 }$&$ 10.02 \pm 0.04 $\\
UXTau & J04300399+1813493 &$ -14.5 ^{+ 0.5 }_{ -0.5 }$&$ -15.9 ^{+ 1.3 }_{ -1.5 }$&$ -6.1 ^{+ 1.3 }_{ -1.3 }$&$ 22.4 ^{+ 1.0 }_{ -1.2 }$&$ -3.4 ^{+ 0.8 }_{ -0.9 }$&$ -3.7 ^{+ 1.4 }_{ -1.6 }$&$ 1.1 ^{+ 1.4 }_{ -1.3 }$&$ 5.1 ^{+ 1.2 }_{ -1.3 }$&$ 7.68 \pm 0.02 $\\
XZTau & J04314007+1813571 &$ -17.0 ^{+ 0.1 }_{ -0.1 }$&$ -13.4 ^{+ 0.1 }_{ -0.1 }$&$ -7.3 ^{+ 0.1 }_{ -0.1 }$&$ 22.8 ^{+ 0.1 }_{ -0.1 }$&$ -5.9 ^{+ 0.7 }_{ -0.8 }$&$ -1.1 ^{+ 0.5 }_{ -0.5 }$&$ 0.0 ^{+ 0.4 }_{ -0.4 }$&$ 6.0 ^{+ 0.7 }_{ -0.7 }$&$ 10.50 \pm 0.04 $\\
V807TauB & J04330664+2409549 &$ -15.8 ^{+ 0.1 }_{ -0.1 }$&$ -15.1 ^{+ 0.3 }_{ -0.3 }$&$ -11.4 ^{+ 0.3 }_{ -0.3 }$&$ 24.7 ^{+ 0.3 }_{ -0.3 }$&$ -4.7 ^{+ 0.7 }_{ -0.8 }$&$ -2.9 ^{+ 0.6 }_{ -0.6 }$&$ -4.2 ^{+ 0.5 }_{ -0.5 }$&$ 6.9 ^{+ 0.6 }_{ -0.6 }$&$ 10.04 \pm 0.03 $\\
HD28867 & J04333297+1801004 &$ 14.3 ^{+ 7.5 }_{ -7.4 }$&$ -13.9 ^{+ 1.4 }_{ -1.7 }$&$ 3.5 ^{+ 3.7 }_{ -3.8 }$&$ 20.2 ^{+ 5.4 }_{ -5.4 }$&$ 25.4 ^{+ 7.5 }_{ -7.5 }$&$ -1.6 ^{+ 1.4 }_{ -1.7 }$&$ 10.7 ^{+ 3.8 }_{ -3.8 }$&$ 27.6 ^{+ 7.0 }_{ -7.0 }$&$ -22.57 \pm 7.40 $\\
HPTauG2 & J04355415+2254134 &$ -16.7 ^{+ 1.0 }_{ -1.0 }$&$ -13.8 ^{+ 0.2 }_{ -0.2 }$&$ -5.5 ^{+ 0.4 }_{ -0.4 }$&$ 22.3 ^{+ 0.7 }_{ -0.7 }$&$ -5.6 ^{+ 1.2 }_{ -1.2 }$&$ -1.5 ^{+ 0.5 }_{ -0.5 }$&$ 1.8 ^{+ 0.5 }_{ -0.5 }$&$ 6.1 ^{+ 1.1 }_{ -1.2 }$&$ 9.52 \pm 1.00 $\\
V999Tau & J04420548+2522562 &$ -15.0 ^{+ 2.0 }_{ -2.1 }$&$ -11.1 ^{+ 0.7 }_{ -0.7 }$&$ -5.0 ^{+ 1.0 }_{ -1.0 }$&$ 19.3 ^{+ 1.7 }_{ -1.7 }$&$ -3.9 ^{+ 2.2 }_{ -2.2 }$&$ 1.1 ^{+ 0.8 }_{ -0.9 }$&$ 2.3 ^{+ 1.0 }_{ -1.0 }$&$ 4.6 ^{+ 1.9 }_{ -1.9 }$&$ 7.96 \pm 2.00 $\\
HD30171 & J04455129+1555496 &$ -17.7 ^{+ 0.8 }_{ -0.8 }$&$ -18.1 ^{+ 2.0 }_{ -2.3 }$&$ -10.8 ^{+ 2.1 }_{ -2.1 }$&$ 27.5 ^{+ 1.7 }_{ -1.8 }$&$ -6.6 ^{+ 1.1 }_{ -1.1 }$&$ -5.8 ^{+ 2.1 }_{ -2.3 }$&$ -3.5 ^{+ 2.1 }_{ -2.1 }$&$ 9.5 ^{+ 1.7 }_{ -1.8 }$&$ 12.71 \pm 0.17 $\\
DRTau & J04470620+1658428 &$ -18.0 ^{+ 1.3 }_{ -1.2 }$&$ -10.5 ^{+ 3.7 }_{ -4.3 }$&$ -11.5 ^{+ 3.8 }_{ -3.9 }$&$ 23.9 ^{+ 2.6 }_{ -2.8 }$&$ -6.9 ^{+ 1.4 }_{ -1.4 }$&$ 1.7 ^{+ 3.7 }_{ -4.3 }$&$ -4.3 ^{+ 3.8 }_{ -3.9 }$&$ 8.3 ^{+ 2.4 }_{ -2.5 }$&$ 12.83 \pm 0.04 $\\
UYAur & J04514737+3047134 &$ -14.9 ^{+ 0.5 }_{ -0.6 }$&$ -16.0 ^{+ 3.4 }_{ -3.9 }$&$ -10.5 ^{+ 3.6 }_{ -3.6 }$&$ 24.3 ^{+ 2.7 }_{ -3.1 }$&$ -3.8 ^{+ 0.9 }_{ -1.0 }$&$ -3.7 ^{+ 3.4 }_{ -4.0 }$&$ -3.3 ^{+ 3.6 }_{ -3.6 }$&$ 6.3 ^{+ 2.8 }_{ -3.1 }$&$ 8.05 \pm 0.07 $\\
HD282630 & J04553695+3017553 &$ -14.1 ^{+ 0.1 }_{ -0.1 }$&$ -12.8 ^{+ 0.4 }_{ -0.4 }$&$ -9.7 ^{+ 0.3 }_{ -0.3 }$&$ 21.3 ^{+ 0.3 }_{ -0.3 }$&$ -3.0 ^{+ 0.7 }_{ -0.8 }$&$ -0.6 ^{+ 0.6 }_{ -0.6 }$&$ -2.4 ^{+ 0.5 }_{ -0.5 }$&$ 3.9 ^{+ 0.6 }_{ -0.7 }$&$ 7.57 \pm 0.01 $\\
ABAur & J04554582+3033043 &$ -9.6 ^{+ 1.0 }_{ -1.0 }$&$ -14.4 ^{+ 1.3 }_{ -1.5 }$&$ -9.6 ^{+ 1.2 }_{ -1.3 }$&$ 19.7 ^{+ 1.2 }_{ -1.4 }$&$ 1.5 ^{+ 1.2 }_{ -1.2 }$&$ -2.1 ^{+ 1.4 }_{ -1.6 }$&$ -2.3 ^{+ 1.2 }_{ -1.3 }$&$ 3.5 ^{+ 1.3 }_{ -1.4 }$&$ 2.94 \pm 0.90 $\\
SUAur & J04555938+3034015 &$ -14.8 ^{+ 0.1 }_{ -0.1 }$&$ -12.8 ^{+ 1.4 }_{ -1.7 }$&$ -9.8 ^{+ 1.2 }_{ -1.3 }$&$ 21.9 ^{+ 1.0 }_{ -1.1 }$&$ -3.7 ^{+ 0.7 }_{ -0.8 }$&$ -0.6 ^{+ 1.4 }_{ -1.7 }$&$ -2.6 ^{+ 1.3 }_{ -1.4 }$&$ 4.6 ^{+ 0.9 }_{ -1.0 }$&$ 8.30 \pm 0.05 $\\
\enddata
\tablecomments{Columns (3)-(6) provide the stellar spatial velocity not corrected for the Solar motion. The peculiar velocity of the stars after correcting for the velocity of the Sun with respect to the LSR are given in columns (7)-(10). The radial velocity of the stars (see Table~\ref{tab8}) converted to the LSR are given in column (11).} 
\end{deluxetable*}\end{longrotatetable}

\subsubsection{Lynds~1531, 1534 and 1536}

V999~Tau and V1000~Tau are projected towards Lynds~1534 \citep[L1534,][]{Lynds1962}. The weighted mean parallax of these two sources is $\pi=7.215\pm 0.110$~mas. This yields a distance estimate of $d=138.6^{+2.1}_{-2.1}$~pc. Our results also reveal that the nearby (in the plane of the sky) star-forming clumps Lynds~1531 (L~1531) and Lynds~1536 (L~1536) are located at different distances. The distance that we derive in this work for L~1531 and L~1536 is based solely on the trigonometric parallaxes of V807~Tau~B and HP~Tau~G2, and represents one first distance determination to these clouds that will be refined when more data for the remaining cloud members becomes available. We find a distance of $ d=126.6^{+1.7}_{-1.7}$~pc and $ d=162.7^{+0.8}_{-0.8}$~pc, respectively, for L~1531 and L~1536. This reveals a difference of about 36~pc between these two clouds along the line of sight. 

Based on the VLBI trigonometric parallaxes derived in this paper we find that V807~Tau~B is the closest star ($d=126.6\pm1.7$~pc) in Taurus. The nominal distances obtained for V1096~Tau and V1098~Tau in L1495 indicate that they are somewhat closer than V807~Tau~B, but  the larger errors given in our solution due to the non-corrected binarity of V1096~Tau and the small number of detections for V1098~Tau make our results for these sources rather uncertain. On the other hand, we confirm HP~Tau~G2 as the remotest star ($d=162.7\pm0.8$~pc) in the complex. Despite the different distances, we note that V807~Tau~B and HP~Tau~G2 move with the same speed. However, their velocity vectors differ significantly as the $w$-component of HP~Tau~G2 points to a different direction than most stars in our sample (see Table~\ref{tab9}).  

\subsubsection{Lynds~1551 and L1558}

XZ~Tau is projected towards Lynds~1551 \citep[L1551,][]{Lynds1962} with a trigonometric parallax of 
$\pi=6.793\pm 0.025$~mas while UX~Tau located on the border of L~1551 (see Fig.~\ref{fig_L1551}) has a trigonometric parallax of $\pi=6.330\pm 0.404$~mas. The weighted mean of these values yields $\pi=6.791\pm 0.025$~mas, and a distance estimate of $d=147.3^{+0.5}_{-0.5}$~pc. We note that this result is in good agreement with the distance of $d=148.7^{+0.9}_{-0.9}$~pc obtained in this work for T~Tau~Sb which is projected towards the BDN~176.28-20.89 cloud \citep[see][]{Dobashi2005}. We thus conclude that L~1551 and BDN~176.28-20.89 are located at the same distance. Interestingly, this result also constrains the distance to the recently imaged HL~Tau star \citep{ALMA-HLTAU} which is also projected towards L~1551 and located $<0.5\arcmin$ from XZ~Tau.  

On the other hand, we note that DR~Tau, which is projected towards Lynds~1558  \citep[L~1558,][]{Lynds1962} has a trigonometric parallax of $\pi=4.820\pm 0.424$~mas in the TGAS catalog. If we assume that the result delivered by Gaia-DR1 for this star is accurate enough, this will put DR~Tau (and L~1558) in the background of the Taurus star-forming complex at a distance of $d=208^{+20}_{-17}$~pc. More study is clearly warranted in this regard, and the upcoming (and more precise) trigonometric parallaxes from Gaia-DR2 will allow us to confirm this scenario. 

In addition to the stars mentioned in this section, we note that HD~28867 and HD~30171 are located in the vicinity of L~1551 and L~1558. HD~28867 has a radial velocity of $V_{r}=-14.70\pm 7.40$~km/s which is in obvious disagreement with the observed radial velocity of other group members (see Table~\ref{tab8}). On the other hand, HD~30171 has a trigonometric parallax and spatial velocity that are more consistent with the properties of the L~1551 cloud (see Tables~\ref{tab8} and \ref{tab9}) despite the closer proximity in the plane of the sky with L~1558 (see also Fig.~\ref{fig_L1551}). In a recent study, \citet{Kraus2017} performed a global reassessment of the membership status of known YSOs in Taurus, and suggested the existence of a distributed older population of stars. HD~28867 was not included in their study and HD~30171 was classified as a YSO candidate member confirming that its membership status is rather uncertain. For these reasons, we have assigned HD~28867 and HD~30171 to an ``off-cloud" population that will require further investigation with Gaia-DR2.

\begin{figure*}[!]
\gridline{
\fig{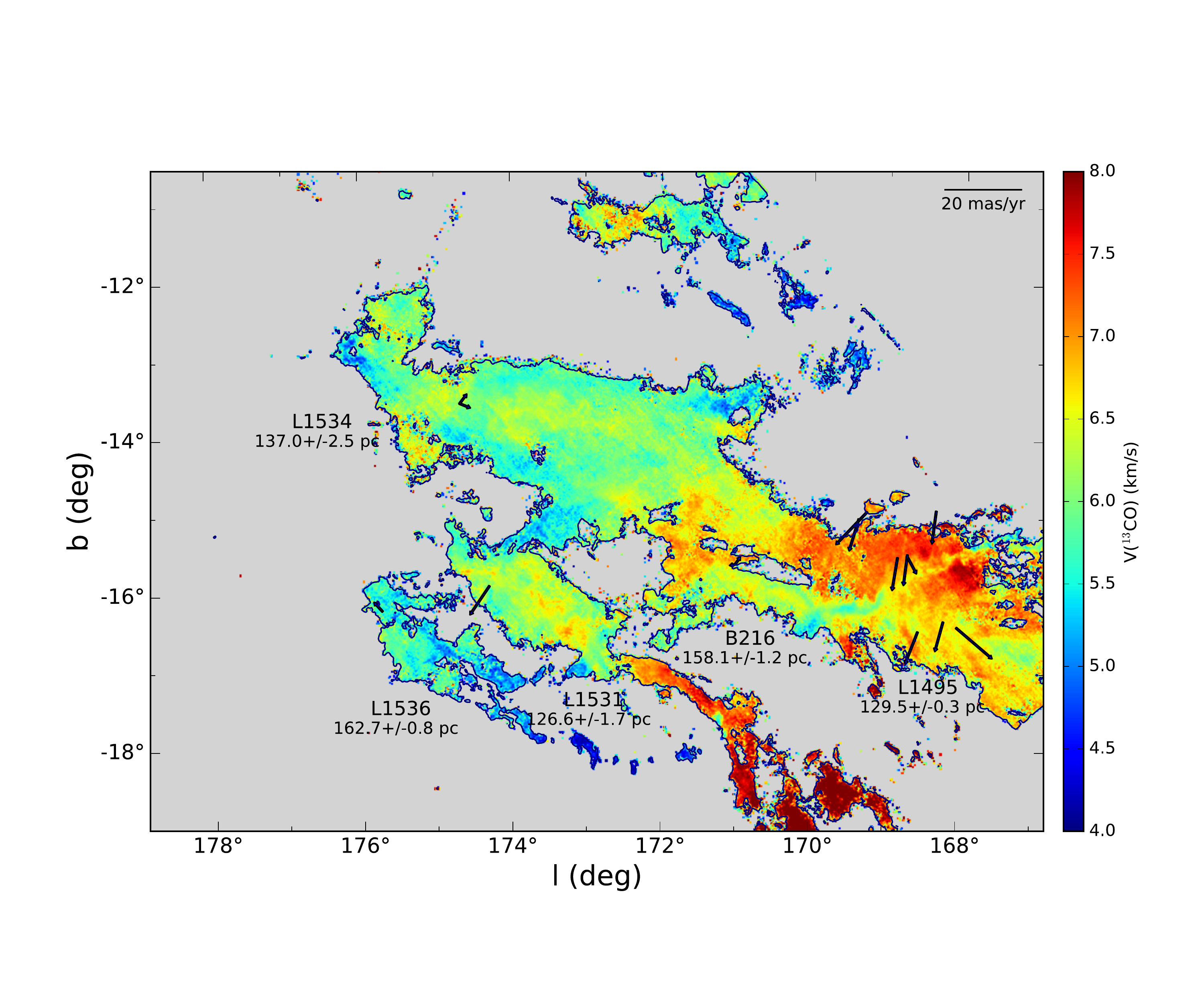}{0.82\textwidth}{}
}
\caption{Location of the stars in our sample overlaid on the $^{13}$CO map from \citet{Goldsmith2008}. The vectors represent the stellar proper motions corrected for the Solar motion \citep{Schoenrich2010} using the formalism described by \citet{Abad2005}. The most prominent star-forming clouds in the central portion of the complex and their distances are indicated in this diagram.}
\label{fig_COmap}
\end{figure*}

\begin{deluxetable*}{lcccccccc}
\tablecaption{Distance and spatial velocity of the various clouds in Taurus. \label{tab10}}
\tablehead{
\colhead{Cloud}&
\colhead{$N_{1}$}&
\colhead{$N_{2}$}&
\colhead{$\pi$}&
\colhead{$d$}&
\colhead{$U$}&
\colhead{$V$}&
\colhead{$W$}&
\colhead{$V_{space}$}\\
\colhead{}&
\colhead{}&
\colhead{}&
\colhead{(mas)}&
\colhead{(pc)}&
\colhead{(km/s)}&
\colhead{(km/s)}&
\colhead{(km/s)}&
\colhead{(km/s)}
}

\startdata
L~1495&8&6&$7.724\pm0.019$&$129.5^{+0.3}_{-0.3}$&$-15.4\pm0.1$&$-11.7\pm0.1$&$-11.1\pm0.1$&$23.4\pm0.1$\\
L~1495 (B~216)&2&1&$6.325\pm0.049$&$158.1^{+1.2}_{-1.2}$&$-17.1\pm0.1$&$-12.5\pm0.2$&$-6.5\pm0.2$&$22.2\pm0.1$\\
L~1513&1&1&$6.610\pm0.508$&$151^{+13}_{-11}$&$-14.9\pm0.6$&$-16.0\pm3.7$&$-10.5\pm3.6$&$24.3\pm2.9$\\
L~1519&4&3&$7.035\pm0.116$&$142.1^{+2.4}_{-2.3}$&$-14.1\pm0.1$&$-12.9\pm0.4$&$-9.7\pm0.3$&$21.3\pm0.3$\\
L~1531&1&1&$7.899\pm0.105$&$126.6^{+1.7}_{-1.7}$&$-15.8\pm0.1$&$-15.1\pm0.3$&$-11.4\pm0.3$&$24.7\pm0.3$\\
L~1534&2&1&$7.215\pm0.110$&$138.6^{+2.1}_{-2.1}$&$-15.0\pm2.0$&$-11.1\pm0.7$&$-5.0\pm1.0$&$19.3\pm1.7$\\
L~1536&1&1&$6.145\pm0.029$&$162.7^{+0.8}_{-0.8}$&$-16.7\pm1.0$&$-13.8\pm0.2$&$-5.5\pm0.4$&$22.3\pm0.7$\\
L~1551&2&2&$6.791\pm0.025$&$147.3^{+0.5}_{-0.5}$&$-17.0\pm0.1$&$-13.4\pm0.1$&$-7.2\pm 0.1$&$22.8\pm0.1$\\
L~1558&1&1&$4.820\pm0.424$&$208^{+20}_{-17}$&$-18.0\pm1.2$&$-10.5\pm4.0$&$-11.5\pm 3.8$&$23.9\pm2.7$\\
\hline
Taurus (all stars)&23&18&$7.054\pm0.012$&$141.8^{+0.2}_{-0.2}$&$-15.2\pm0.1$&$-12.8\pm0.1$&$-8.7\pm0.1$&$22.8\pm0.1$\\
\enddata
\tablecomments{We provide for each subgroup the number of stars with known trigonometric parallax ($N_{1}$) and radial velocity ($N_{2}$), the weighted mean parallax with the corresponding distance, and the weighted mean spatial velocity.  }
\end{deluxetable*}

\subsubsection{Taurus (all stars)}

In Table~\ref{tab10} we list the mean distance and spatial velocity derived for all stars in our sample (V1110~Tau, HD~28867, HD~30171 and RY~Tau are excluded from this analysis for the reasons discussed before). The mean distance of $d=141.8\pm0.2$~pc that we derive from our analysis is still consistent with the canonical distance estimate of $d=140\pm10$~pc \citep{Kenyon1994} which is commonly used in the literature for the Taurus region.  However, the resulting distance is only representative of a few clouds in the Taurus complex, and it does not reveal the important depth effects that exist in this region as demonstrated in our study. Interestingly, we conclude from Table~\ref{tab10} that the B~216 clump in the filamentary structure of L~1495 is moving at $(\Delta U,\Delta V,\Delta W)=(-1.7,-0.8,4.9)\pm(0.1,0.2,0.2)$~km/s with respect to the central part of the cloud, which implies a relative bulk motion of about $5.2\pm0.2$~km/s between both structures. It is also interesting to note that L~1551, which is the most southern cloud in our sample, is moving at $(\Delta U,\Delta V,\Delta W)=(-1.6,-1.7,4.2)\pm(0.1,0.1,0.1)$~km/s with respect to L~1495 and has a relative motion of $4.8\pm0.1$~km/s. On the other hand, we see from Table~\ref{tab10} that the spatial velocity for L~1519 is fully consistent within 1~km/s of the mean spatial velocity computed for all stars in our sample.

We find that the dispersion of the spatial velocities among the various clouds is $(\sigma_{U},\sigma_{V},\sigma_{W})=(2.4,2.5,2.1)$~km/s. For comparison, the velocity dispersion derived from the proper motions converted to tangential velocities in right ascension and declination are 2.4~km/s and 3.1~km/s, respectively. This implies that the one-dimensional velocity dispersion in Taurus is somewhat higher than the value of 1~km/s adopted by \citet{Luhman2009}, and smaller than the value of 6~km/s estimated by \citet{Bertout2006}. This value is also similar to the one-dimensional velocity dispersion obtained by \citet{Dzib2017} and \citet{Kounkel2017} for YSOs in the Orion Nebula Cluster.

One interesting question that arises from our study is whether the stars and the molecular gas in this region exhibit the same kinematic properties. In this context, \citet{Goldsmith2008} performed a large-scale survey of the $^{12}$CO  and $^{13}$CO molecular gas in Taurus which we use here to further discuss our results. Figure~\ref{fig_COmap} summarizes the distance of the various clouds in the central portion of the Taurus complex that we derive in this paper overlaid on the $^{13}$CO velocity field produced in that survey. We extracted the $^{12}$CO  and $^{13}$CO spectra at the position of the stars in our sample in a velocity interval from 3 to 13~km/s, computed the centroid velocity of the molecular gas in each case and estimated their errors from the r.m.s. of the individual spectra. We note that for some stars in our sample there was no apparent signal in one of the two spectra extracted from the  $^{12}$CO  and  $^{13}$CO maps. So, we decided to restrict our analysis to the stars with measured centroid velocities from both spectra and took the weighted mean of the computed values as our final estimate for the velocity of the gas at the position of a given star. It is important to mention, that this analysis is restricted to only 9 stars in our sample with measured radial velocities in the literature (see Table~\ref{tab8}) that are included in the region surveyed by \citet{Goldsmith2008} and fulfill this condition.

\begin{figure}[!h]
\gridline{
\fig{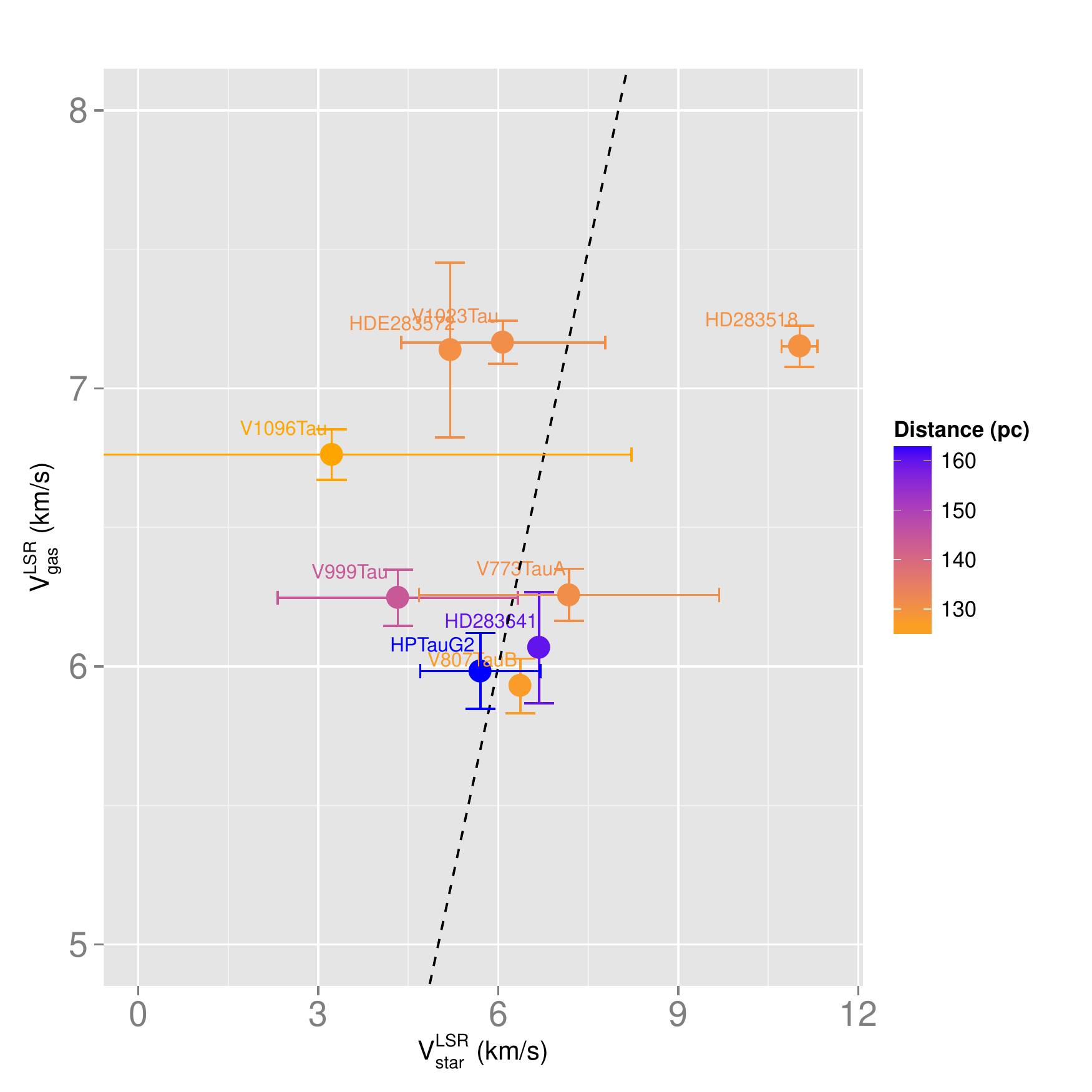}{0.49\textwidth}{}
}
\caption{Comparison of the stellar radial velocity with the velocity of the $^{12}$CO and $^{13}$CO molecular gas measured at the position of each star. The colors indicate the distance of each star obtained in this study from VLBI astrometry. The dashed line indicates perfect correlation between the velocity of the gas and the radial velocity of the stars. }
\label{fig_RV_gasVLSR}
\end{figure}

In Figure~\ref{fig_RV_gasVLSR} we compare (i) the velocity of the molecular gas with the radial velocity of the stars (both are given with respect to the LSR), and (ii) the distance of the stars with the velocity of the associated gaseous clouds. First, we note that the velocity of the molecular gas (measured at the position of our sources) and the spectroscopic radial velocity of the stars are mostly consistent confirming that the stars are indeed associated with the underlying gaseous clouds. It is important to mention that most  stars used in this analysis are binaries which explains both the existence of discrepant values (e.g. HD~283518) and the large errors on the radial velocities given in the literature (e.g. V1096~Tau). Second, we note that the velocity of the gas ranges from 6.5 to 7.5~km/s for most sources projected towards the central part of the L~1495 cloud ($d=129.5^{+0.3}_{-0.3}$~pc), and it varies from 6.0 to 6.5~km/s at the position of the remotest stars in this sample: V999~Tau ($d=143.4^{+4.2}_{-3.9}$~pc), HD~283641 ($d=159.1^{+1.8}_{-1.8}$~pc) and HP~Tau~G2 ($d=162.7^{+0.8}_{-0.8}$~pc). Although a prefect correlation between the distance of the stars and the velocity of the gas is not straightforward from Fig.~\ref{fig_RV_gasVLSR} (see e.g. V773Tau~A and V807~Tau~B), we found evidences that the observed velocity of the molecular gas at the position of our targets decreases with the increasing distance of the star. This finding and the trigonometric parallaxes derived in this paper support our conclusion that the different structures of the Taurus complex are located at different distances. We will continue to investigate this issue using the upcoming parallaxes from Gaia-DR2 and a more significant number of stars to provide an accurate picture of the gas and stellar kinematics in this region. 

Finally, we note that the angular size of the Taurus complex in the plane of sky is about 12$^{\circ}$ in both Galactic longitude and Galactic latitude (see e.g. Fig.~\ref{fig1}) which roughly corresponds to 30~pc using the mean distance given in Table~\ref{tab10}. From the closest (L~1495) and remotest (L~1536) molecular cloud with measured VLBI trigonometric parallaxes in this study (this excludes L~1558) we estimate the depth of 33~pc. Thus, we conclude that the distance range in the plane of the sky and in the line of sight are equivalent.

\section{Conclusions}\label{section6}

In this study we reported on multi-epoch VLBI radio observations taken as part of the GOBELINS project in the Taurus star-forming region. We detected 26 YSOs with the VLBA, and presented the astrometry of 18 stars (or stellar systems) in our sample. The absolute positions measured in this work were modeled to derive the trigonometric parallaxes and proper motions of both single stars and binaries to a few percent of accuracy. By combining our observations with data from previous studies in the literature we were able to solve simultaneously for the astrometry and orbital motion of the sources in binary systems over an extended time base, and provide a more accurate solution for the trigonometric parallax. Thus, our results are more accurate than the trigonometric parallaxes from Gaia-DR1 for both single stars and binaries where the orbital motion of such systems was not taken into account. The VLBI trigonometric parallaxes presented in this paper are also more precise than the results from Gaia-DR1 by almost one order of magnitude. Our analysis also made it possible to determine the dynamical masses of the individual components in four systems (V1023~Tau, T~Tau~S, V807~Tau~B and V1000~Tau). 

We converted the trigonometric parallaxes derived in this study into stellar distances and investigated the three-dimensional structure of the Taurus complex. We confirm the existence of significant depth effects and concluded that the various star-forming clouds of the complex are located at different distances. We found a mean distance of $129.5\pm 0.3$~pc to the central part of the dark cloud L~1495, and report on the distance of $158.1\pm 1.2$~pc towards the B~216 clump in the filamentary structure of this cloud. Based on our VLBI observations we conclude that V807~Tau~B, which is projected towards L~1531, is the closest star in our sample located at $126.6\pm 1.7$~pc. On the other hand, HP~Tau~G2 projected towards the nearby (in the plane of the sky) L~1536 cloud is the farthest star in the complex located at  $162.7\pm 0.8$~pc. Altogether, this implies a depth of about 36~pc based solely on the distances derived from VLBI trigonometric parallaxes. In particular, we note that one of the clouds for which we derive a distance (Lynds~1551) contains the young star HL~Tau that has recently been subject of many studies upon the ALMA imaging of its protoplanetary disk. We argue that the distance derived here ($d=147.3\pm0.5$~pc) should be used for any future study of that specific source.

Finally, we combined the stellar distances obtained in this paper with published radial velocities to compute the spatial velocities of Taurus stars. We verified that the one-dimensional velocity dispersion among the various clouds in the complex amounts to 2-3~km/s. Moreover, we showed that the velocity of the molecular gas structures is somewhat smaller for the remotest  stars is our sample ($d\simeq160$~pc) as compared the closest stars projected towards L~1495 ($d\simeq130$~pc). 

The distances produced by the GOBELINS project in Taurus represent one important step to map the three-dimensional structure of the complex with unprecedented accuracy and precision. In addition, they also provide us with an independent consistency check of the upcoming trigonometric parallaxes from Gaia-DR2 for the targets in common. We anticipate that we will soon be able to deliver more results for other targets in our sample (including binaries and multiple systems) that are currently being monitored by our team and we will use Gaia-DR2 trigonometric parallaxes to provide a more complete picture of the Taurus region. 

\acknowledgments

P.A.B.G. acknowledges financial support from the S\~ao Paulo Research Foundation (FAPESP) through grants 2013/04934-8 and 2015/14696-2. L.L. acknowledges the financial support of DGAPA, UNAM (project IN112417), and CONACyT, Mexico. G.-N.O.L acknowledges support from the Alexander von Humboldt Foundation in the form of a Humboldt Fellowship. M.K. acknowledges support provided by the NSF through grant AST-1449476, and from the Research Corporation via a Time Domain Astrophysics Scialog award (\#24217).


%

\vspace{5mm}
\facilities{
VLBA (NRAO) - The National Radio Astronomy Observatory
is operated by Associated Universities, Inc., under cooperative
agreement with the National Science Foundation. The Long Baseline Observatory is a facility of the National Science Foundation operated under cooperative agreement by Associated Universities, Inc. DiFX correlator - This work made use of the Swinburne University of Technology software correlator, developed as part of the Australian Major National Research Facilities Programme and operated under licence. This work has made use of the computing facilities of the Laboratory of Astroinformatics (IAG/USP, NAT/Unicsul), whose purchase was made possible by the Brazilian agency FAPESP (grant 2009/54006-4) and the INCT-A.
}


\software{
AIPS \citep{Greisen2003},
\texttt{emcee} \citep{Foreman2012},
\texttt{astropy} \citep{astropy}, 
\texttt{novas} \citep{NOVAS}.
}


\newpage
\bibliography{references.bib}

\end{document}